\documentclass[12pt,a4paper%
 ]{amsbook}

\makeatletter
\def\thempfootnote{{\it\@alph\c@mpfootnote}}
\makeatother

\usepackage{amsmath}
\usepackage{amsfonts}
\usepackage[mathcal]{euscript}
\usepackage{amsthm}
\usepackage{multicol}
\usepackage{graphics}
\usepackage{tabularx}
\usepackage{amssymb}
\usepackage[dvips]{graphicx}
\usepackage[dvips]{epsfig}
\usepackage{epigraph}

\usepackage{DALdefs1}

\makeindex

{
}

\begin{document}

\graphicspath{{./figs/}}

\title{The thermodynamic approach\\ to market}

\author{{\bf Victor Sergeev}\\
translated from the Russian and edited by 
Dimitry~Leites}




\date{}

\maketitle

\begin{abstract}
The book gives an explanation of several intriguing phenomena,
providing new insights and answers to some deeply vexing
questions.

Why the economic \lq\lq shock therapy" implemented in Eastern
Europe was doomed to a~failure whereas the approach adopted by
China and Vietnam should inevitably lead to economic growth
(damped, perhaps, by corruption and inconsistencies)?

Why some restrictions imposed on markets are dangerous?
Politicians (and laymen) usually believe that the more
restrictions you impose on the society, the easier it is to
govern. The reality seems to be more subtle. Some restrictions are
necessary for the markets to function. However, restrictions on
salary, for example, seem to always result in unemployment, as
a~simple \lq\lq spin" model shows. Evidently, optimizing freedom
is an equilibrium problem, and none of the extrema is devoid of
danger.

Why should crooked dealings be prevented? While honesty has
a~price, the the dishonest market collapses.

What is a~reasonable rate of taxation, if any such exists?

Other questions abound, raising many points of interest.

Appendices contain an essay which informally can be entitled
\lq\lq Why mathematics and physics major should study economics"
and a~deep mathematical paper summarizing a~century long study of
nonholonomic systems (such as ideal gas or market economy) in the
quest for an analog of the curvature tensor in nonholonomic
setting.

\end{abstract}


\tableofcontents

\chapter*{Preface}

Scientific notions possess an intrinsic power. For this
author, it is, of course, essential to understand when and under what
circumstances
are born the particular ideas that forced him to exercise the considerable
effort to write this book. But for the process of creation and accumulation of
knowledge, this is hardly essential.

Nevertheless, ideas are seldom conceived in isolation.
Therefore this author is morally indebted to the colleagues who,
perhaps unawares, helped him to crystallize these notions and
mold them into a~coherent dissertation.

It seems that at the very beginning the idea that there exists an
affinity between economic processes and phenomena studied in the
statistical physics appeared as a~result of the author's repeated
discussions of parallels between the biological and social
processes with Ya.~Dorfman\index{Dorfman Ya.} between 1980 and
1985. Unfortunately, we were unable to develop these ideas because
of the tragic death of Ya.~Dorfman in the summer of 1986.

My interest in this problem was further stimulated by the
inspiring atmosphere of A.~Chernavsky's seminar at the Institute
of Problems of Information Transmission in \hbox{1985--86} and our joint
work with Yu.~Sandler between 1987 and 1990 on the application of
neural networks to the study of collective decision-making
processes. During the same period and somewhat later the author
intensively discussed the possibilities of neural-net modelling of
economic processes with A.~Vedenov.\index{Vedenov A.}

Still, the final formation of a~thermodynamic approach to
equilibrium problems in economics happened during informal
discussions of the economics of science organized by V.~Olshansky in
one of the Moscow biological institutes and subsequent periodic
discussions of this type of problems with V.~Pavlov and O.~Bushmanov.
I am sincerely thankful to all the above.

The author is also thankful to his foreign colleagues: Prof.
O.~Andersson\index{Andersson O.} for the numerous discussions of
the problems of theoretical economics and the possibility to
discuss the first result of this work at the Conference on
Economics of Eastern Asia in Beijing in October 1996; Professors
B.~Arthur, M.~Bekman and Zhang Weibin for stimulating discussions
of the problems of economic equilibrium.

Naturally, none of the above people take responsibility neither
for the views and conclusions in this book nor for the possible
mistakes.

Finally, the author is deeply thankful to the Russian Foundation for
Fundamental Research for the financial support of the Russian edition
of this book. The first three chapters are written under the support
of RFFR grant No 96-06-80625.

Moscow, November 1998.

\section*{Preface to the English edition}

The book was
translated and edited by D.~Leites, Sophus-Lie-Professor at the
Max-Planck-Institute for Mathematics in Sciences, Leipzig. He also
appended the book with translations of excerpts from two of my
papers published recently in the editions difficult to find (his
deliberations on feasibility of a~qualitative study of stability
of market economy based on the approach given in this book are
published separately\footnote{D.~Leites, On computer-aided solving
differential equations and stability studies of markets. In:
Vershik~A. (ed.)\index{Vershik A. M.}  Teorija predstavlenij,
dinamicheskie sistemy. [Representation theory and dynamical
systems] XI. (Proc. of the L.~V.~Kantorovich memorial conference
\lq\lq Mathematics and economics: Old problems and new
approaches", Jan. 8--13, Euler Mathematica International Institute
(EIMI), 2004.) Zapiski nauchnyh seminarov POMI, tom 312,
S.-Peterburg, 2004, 165--187.}). I am deeply grateful to D.~Leites
for the time and efforts he devoted to the process of translation
as well as his numerous comments and suggestions which improved
the text.

I am also deeply grateful to the  Max-Planck-Institute for
Mathematics in Sciences, Leipzig, which hosted me in October 2005
and its director, Prof. Dr. Juergen Jost, for an opportunity to
finish together with D.~ Leites the work on the English text of
the book.

Access to originals of several references when the book was being
written has become difficult to me. In order not to overburden the
translator with the effort of tracking precise quotations, we
confined ourselves with a~translation of a~quotation back from the
Russian in most places, usually omitting the quotation marks in
such cases. Except for some minor improvements, the main text is
basically the same as in the Russian edition. To comment the
recent developments in the subject would be to write a~completely
new book.

I'd like to make some additional  acknowledgements: I am very
grateful to David Chassin for various comments that improved the
exposition of the book and his invitation to the panels he shared
at the 2nd International Conference on Computing, Communications
and Control Technologies: CCCT'04. August 14--17, 2004~--- Austin,
Texas, USA and at the AAAS conference at Washington, February
2005, where I was able, among other things, to express the ideas
presented in this book.

I am very grateful to all the people who participated in
discussions of the subject of this book during my stay as a~Senior
International Fellow at the Santa Fe Institute during 2002--04. My
stay at the Santa Fe Institute was extremely helpful from the
point of view of my further studies of the applications of
thermodynamic approach to economics and for an opportunity to
publish a~working paper of the Santa Fe Institute based on parts
of this book. This made at least some of its results readily
available to the English speaking audience. I am also indebted to
the members of the Santa Fe Institute who provided me with an
opportunity to participate in the vivid scientific life of the
Institute. I am especially grateful to Prof.~Elen Goldberg, the
then director of the Institute for the first invitation to Santa
Fe during the fall of 2001. I also thank Prof.~Hayward Alker, Jr.
who connected me for the first time with the Santa Fe Institute. I
am particularly indebted to Professors~Doyne Farmer and Duncan
Foley for discussions and their interest in this work. I
appreciate very much their interest to my work. I thank Prof.
M.~Gell Mann for his interest in my work.

I am thankful to D.~Leites and A.~Vershik for their contributions.

After this book had been published in Russian in 1999, a~lot of
papers and several books were published on application of ideas of
thermodynamics and statistical physics to problems of economics.
They are being taken into account in a~sequel to this book I am
writing now.

\vskip 0.5 cm

October 2005, MPIMiS, Leipzig \vskip 0.5 cm

\footnotesize {\bf About the author}.\index{Sergeev V.} MA in
physics from Moscow Power Engineering (1967) and MA in mathematics
from Department of Mechanical engineering and Mathematics (\lq\lq
mekh-mat") of Moscow State University (1970), Ph.D. in Mathematics
and Physics from Moscow Institute for High Temperature (1974); Dr.
Sci. in History; thesis \lq\lq Genesis of Democracy in Russia and
the Tradition of Sobornost'"(1990).

Hon. Research Prof. Leeds University, UK (1995-); Member of the
Russian Academy of Natural Sciences (1997).

Author of more than 150 published works, in particular, monographs

(2003)  {\em Perspective of Russian Oil}  (with V.~Petrov, G.~
Polyakov, T.~ Poliakova) Fazis, Moscow (in Russian)

(2003)   {\em  Economic Reconstruction: the comparison of post war
Europe and Post-Communist Russia}  (with R.~Griffiths). Letnii
Sad, Moscow (in Russian)

(1999)   {\em Democracy as a~Negotiation Process}. MONF, Moscow
(in Russian)

(1999)    {\em The Limits of Rationality: Termodynamic Approach to
the Problem of Economic Equilibrium}, Fasis, Moscow  (in Russian)

(1997)  {\em Central Asia After the Empire} (with Yu.~ Kulchik and
A.~Fadin) Pluto Press, London

(1997)  {\em Russian Politics in Transition}. (with
N.~Biryukov)\index{Biryukov N.} Aldershot, Ashgae Publishing Ltd.,
Brookfield, USA, Singapore, Sydney.

(1997)   {\em The Wild East}. M.~E. Sharpe, N. Y.

(1993)  {\em Russia's Road to Democracy: Parliament, Communism and
Traditional Culture} (with N.~Biryukov) Aldershot, Edward Elgar
Publishing Ltd., Hampshire

(1987) Editor of {\em Language and Modelling of Social
Interaction} Progress, Moscow (in Russian)

(1986)    {\em Space Weapons: A Security Dilemma} (with
A.~G.~Arbatov et al., Mir, Moscow (in Russian and in English)

\normalsize

\newpage

\section*{Editor's preface}

When I was 15, my chemistry teacher gave me {\it Feynman Lectures on
Physics} just translated into Russian. Although, in Yu.~Manin's
words, these lectures represent physics approximately as much as La
Fontaine's fables represent reality, or, perhaps, just thanks to
this feature of {\it Lectures}, they are most absorbing reading.
Feynman's charismatic presentation of the material is felt even on
paper and I was overwhelmed. For decades I read no equally
impressive scientific book until I came across Sergeev's book {\it
The Limits of Rationality}\footnote{For the English edition the book
is expounded and the title appropriately changed to \lq\lq The
thermodynamic approach to market".}.

The most intriguing words in Feynman's {\it Lectures} were those
about the mysterious notion of {\it force}. At about the same time
I read {\it Lectures}, I learned about Hertz's attempts to
exorcize the notion of {\it force} from the set of basic physical
notions. These ideas were related with (ascending to Riemann)
notion of ``geometrodynamic'', namely the idea that there are no
{\it forces}, just the curved space.

Later on, when as a~researcher I tried to understand how
supergravity equations should look, I have realized how to write
them: one should use Hertz's definition of constrained dynamics and
define an analog of the Riemann curvature tensor for such
constrained dynamics (more precisely, with {\it nonintegrable}
constraints on velocities; phase spaces of dynamic systems with such
constraints are said to be {\it nonholonomic}); the supergravity
equations should be a~vanishing condition on certain components of
this tensor.\footnote{Some researchers argue that (super)gravity
equations may have a different origin.}

Having realized this much, I was on the look out for examples of
nonholonomic manifolds. My interest in the problem of mathematical
description of economic systems given in this book is related with
a particular way of this description, namely, in terms of
nonintegrable distributions\footnote{In this book, the term \lq\lq
distribution" means two different things, obvious from the
context: either a~probabilistic distribution or a~subbundle of the
tangent or cotangent bundle.} or nonholonomic manifolds (a {\it
distribution}\index{distribution, nonholonomic}\index{manifold,
nonholonomic}\index{nonholonomic manifold, distribution, dynamics}
is singled out in the tangent bundle to a~manifold by a~system of
Pfaff equations,\index{Pfaff equation}\index{equation, Pfaff}
i.e., is the common set of zeros of differential
1\defis forms).\footnote{There are two interpretations of the notions
which bare the same name \lq\lq Pfaff equation". One is given
above and in the main text (a solution of a~system of Pfaff
equations in this sense is a~common set of zeros of a~collection
of differential 1\defis forms), whereas here we assume that the
differentials of the independent variables are independent, and
solve the equation $\alpha(X)=0$ for unknown vector fields $X$;
such fields $X$ constitute the distribution in question for the
given collection of differential 1\defis forms $\alpha$.}

No wonder then that Sergeev's {\it Limits} made an impact on me
comparable with Feynman's {\it Lectures}: I was psychologically
prepared for its message. Later I read several more of Sergeev's
books and highly recommend them, especially {\it The Wild East}
and his books on democracy; regrettably not all of them are yet
published in English.

{\bf For whom is this book}. It was inexplicable to me at first
why {\it Limits} was not a~bestseller, at least, among physicists
and economists, but also among other people: it does not use
mathematics more involved than the notion of partial derivative.

Eventually I have understood why {\it Limits} did not enjoy the
phenomenal success I'd expected it to have, at least no less than
Feynman's {\it Lectures}. Actually, I'd expect much wider spread
of {\it Lectures} since the deductions of {\it Limits} concern
everybody's pocket, are easier to grasp than {\it Lectures};
besides, {\it Limits} is very short.

Although the topic of {\it Limits} is vital for everybody, and
practical applications of {\it Limits} are rather clearly
expressed, its first half (Chapters 1--4), is manifestly addressed
to economists and sociologists and is difficult for mathematicians
and physicists to whom the names even of Nobel laureates in
economics do not mean much, and who (as I witnessed, to my
regret), having read about what is not obvious to them, easily get
too tired to pass to the second half of the book
--- the thermodynamics expressed in terms of market economy (or
the other way round, the market economy expressed in terms of
thermodynamics).

{\bf Economists}, on the other hand, were scared by the second half
by allegedly difficult notions from physics. Therefore I would like
to inform the reader: {\bf no mathematics above the level of basics
on partial differential equations is needed to understand the main
text}. Conclusions do not require any mathematics at all and should
be clear even to government officials.

{\bf Mathematicians and physicists} should, perhaps, begin with
Chapter 5 and turn to earlier chapters for motivations addressed,
mainly, to experts in economics, as well as for general background
and history.

Appendices require rudiments of differential geometry. The fact
that the book mainly appeals to common sense (and uses only most
basic mathematics) is its extra attraction to me.

The present book is not just a~translation into English of its
Russian counterpart. It is about twice enlarged version, with
elucidations and comments.

The main body of the book is appended with an essay \lq\lq{\it
Physics as a~tool in sociology}'' which can be considered as an
extended summary with its own message: {\sl it gives an answer to
the question \lq\lq Why mathematics and physics major should study
sociology or economy"}. This Appendix can be interpreted as
follows: whereas the main body of the book is a~paraphrase of the
first few chapters of any good text-book on Statistical Physics in
terms of market economy, to translate other notions and theorems
is an interesting open problem, a~challenging problem for Ph.D.
{\bf students majoring in theoretical physics and seeking real
life applications}.

This book not only indicates a~method for answering various vital
questions of economics but also widens horizons of applications
for {\bf mathematics major students}. Similar miracles (when
a~branch of mathematics becomes an indispensable tool or language
in a branch of another natural science) were known (applications
of the theory of Hilbert space operators in quantum theory);
Sergeev makes economics a~field of application of representation
theory of Lie algebras (these are needed, for example, to compute
the nonholonomic analog of the curvature tensor, see Vershik's
Appendix and ref. \cite{L} in it).

\normalsize

{\bf Literature on econophysics}. Since the first pioneer papers by
Rozonoer\index{Rozonoer L.} published in 1973 (commented on in the
main text) and Sergeev's {\it Limits} written in 1996 the ideas to
apply statistical physics to economy became widespread.

Ten years after that first bold approach, in collaboration with
A.~Tsyrlin, Rozonoer published another three papers in the same
journal. This time no analogies between thermodynamics and
economics were mentioned but a~very impressive theory of optimal
control in thermodynamic processes was suggested. Possibly, the
cause of such a~turn away from economics was related with general
political atmosphere in Soviet science, not encouraging for
advanced studies in market economics. These new ideas of optimal
control over thermodynamic processes gave a~start to a~flow of
articles in this direction, but this activity was not related
directly with studies of economics. Recently S.~Amelkin,
K.~Martinash and A.~Tsirlin\footnote{Amelkin, S. A.; Martinash,
K.; Tsirlin, A. M. Problems of the optimal control of irreversible
processes in thermodynamics and microeconomics. (Russian) Appendix
1 by L. I. Rozonoer. Avtomat. i Telemekh. 2002, , no. 4, 3--25;
translation in Autom. Remote Control 63 (2002), no. 4, 519--539.}
returned to the idea of applying the methods of optimal control in
thermodynamics to economic problems.

Here is the list of latest monographs on econophysics or related
to it:

McCauley, J. L. {\em Dynamics of markets. Econophysics and
finance}. Cambridge University Press, Cambridge, 2004. xvi+209 pp.

Kleinert, H., {\em Path integrals in quantum mechanics,
statistics, polymer physics, and financial markets}. Third
edition. World Scientific Publishing Co., Inc., River Edge, NJ,
2004. xxxvi+1468 pp.

Sornette, D., {\em Critical phenomena in natural sciences. Chaos,
fractals, selforganization and disorder: concepts and tools}.
Second edition. Springer Series in Synergetics. Springer-Verlag,
Berlin, 2004. xxii+528 pp.

Schweitzer, F., {\em Brownian agents and active particles.
Collective dynamics in the natural and social sciences}. With a
foreword by J. Doyne Farmer. Springer Series in Synergetics.
Springer-Verlag, Berlin, 2003. xvi+420 pp.

Schulz, M., {\em  Statistical physics and economics. Concepts,
tools, and applications}. Springer Tracts in Modern Physics, 184.
Springer-Verlag, New York, 2003. xii+244 pp.

Voit, J., {\em The statistical mechanics of financial markets.}
Texts and Monographs in Physics. Springer-Verlag, Berlin, 2001.
xii+220 pp.

Paul, W.; Baschnagel, J. {\em  Stochastic processes. From physics
to finance}. Springer-Verlag, Berlin, 1999. xiv+231 pp.

\medskip

None of the above listed rival works covers the ideas, approaches
and results presented in the book you are about to read.

 \vfill
\newpage

\part{The Limits of Rationality}

\chapter*{Introduction}

This book is a~result of my thoughts on problems of construction
of models of social phenomena. In the European scientific
tradition starting from Descartes\index{Descartes} and
Bacon,\index{Bacon} a~thick line was drawn between the study of
Nature and the study of Man. This line became a~separating
truncheon about a~hundred years ago when Dilthey,\index{Dilthey
W.} Windelbandt\index{Windelbandt W.} and Rickert\index{Rickert
H.} divided the domains of science into the ones where the essence
of the study is the search for the laws that connect the notions
(physics is a~prime example here) and ``sciences of the Spirit'',
i.e., the domains of science where the essence of the study is the
description of the individual and which are not reproducible
(history is a~prime example here).

But even without accepting the neo-Kantian position, it is
difficult not to concede that the faceless world of Nature should
be essentially different from the world of the Man colored by his
personality and possessing at least empirical freedom of choice
and abilities to understand the world.

To what extent do this freedom and ability to understand (which an
individual is supposed to possess) influence the properties of
communities?

What happens when the interaction between individuals becomes so
essential that, studying the individual, we cannot actually say
anything about the whole?

This borderline situation requires a~special consideration. Are the
methods of the study developed for the description of Nature
applicable in this case? Or the individuality is, in the end,
non-removable and one has to take it into account, at least through
the values, i.e., through the criteria that determine the choices of
individuals? To what extent these criteria themselves are individual?

Basically, the study of human societies in the various incarnations
--- economic, political, social~--- is a~``no man's land'' between the
sciences of Nature and sciences of Spirit. It was already in XVIII
century starting with F.~Quesnay\index{Quesnay A.} and
A.~Smith\index{Smith A.} that it became clear that certain domains
of this ``no man's land'' are much better suited to be subjected
to a~formal ``logical analysis'' than the other ones and that
economics is a~most promising domain for the construction of
rigorous models thanks to the domination of the
interests\index{interest} (in principle, accessible to a~logical
analysis) in the economic behavior of the people.

In this way, we determine, on the individual level, a~force that
governs the behavior of society as a~whole.

Such arguments range in a~simple and neat scheme~--- the
individual choice of behavioral alternatives is determined on the
base of a~personal interest\index{interest}~--- (in the
mathematical models such a~choice is expressed through the
maximization of the ``utility\index{utility} function'') and the
collective behavior is considered as the sum of individual
behavioral acts. On the choice of a~human performance certain
auxiliary conditions may be imposed that restrict this choice, and
therefore the problem to determine the ``conditional maximum''
becomes the center around which the mathematical methods of
theoretical economics are formed.

The efforts of several generations of theoreticians resulted in an
impressive apparatus that enables one to explain the existence of
a market equilibrium\index{equilibrium, market} under the
conditions of an ideal competition and to study the properties of
an equilibrium state of economy. The successes of mathematical
economics far surpass the achievements in the construction of
models in the social and political studies. It was already
A.~Smith\index{Smith A.} who pointed out at the cause of this
situation by discovering the existence of a~general regulative
principle in economics~--- ``the invisible hand''\index{invisible
hand} of the market that transforms the individual aspirations
into a~common order.

The successes of theoretical and mathematical models cannot,
however, hide a~serious crisis in this domain of science; a~crisis
that developed slowly and became noticed only recently.

The gap between the mathematical economics, i.e., the discipline with
its own quite interesting and meaningful object of the study that
develops more and more abstract methods and partly became an esoteric
domain of mathematics that studies certain particular types of
functionals in partly ordered topological vector spaces and the
applied studies whose problems turned out to be weakly related to the
existence problems and the properties of the market equilibrium became
wider and wider.

In practice, it is much more important to know how the system reaches
an equilibrium under various constraints than in the non-constrained
situation.

This problem became manifest in the study of ``economies in
transition'', for which the abstract existence theorems concerning
equilibrium are of little value. In the study of economies in
transition, it is important to understand where this equilibrium
will be reached under initial non-equilibrium conditions.

The deepening of the theoretical studies of the market as such
also led to serious questions that put the theoretical orthodoxy
to doubt. F.~Hayek\index{Hayek F.} gave serious arguments in favor
of the fact that the market as a~social institute exists not as a~corollary of an unbounded strive for full satisfaction of personal
interests\index{interest} but rather thanks to the opposite~---
thanks to the existence of a~system of a~rigid moral rules that
bound this strive.

In this case, the initial paradigm that pillared the mathematical
models~--- the maximization of the utility\index{utility} function
as the base for determining equilibrium~--- should be questioned.

The purpose of this book is to develop a~new approach to the study
of Nature and the mathematical apparatus of the theory of economic
equilibrium; the point of view conceptually lying in the trend of
ideas of F.~Hayek and the Austrian school\index{Austrian school}
and the evolutionary approach to economics.

The idiosyncrasy towards econometrics shared by representatives of
the Austrian school\index{Austrian school} is well known. The
purpose of this work is to show that the mathematical apparatus
that enables one to rigorously formulate certain basic ideas of
C.~Menger,\index{Menger C.} L.~von Mises\index{von Mises L.} and
F.~Hayek\index{Hayek F.} actually exists and, moreover, is well
known.

This apparatus is the statistical thermodynamics\index{statistical
thermodynamics} which, as follows from the studies of C.~
Shannon,\index{Shannon C.} L.~Brillouin\index{Brillouin L.} and
N.~Wiener\index{Wiener N.} and preceded in 1920s by
L.~Szilard's\index{Szilard L.} ideas, has applications far wider
than statistical physics.

Statistical thermodynamics\index{statistical thermodynamics} and
information theory\index{information theory} coincide,
essentially. And what is the market if not the search for
information and the continuous stream of decision-making?

``Information'' here is understood in its most formal theoretical
meaning as a~means to diminish the number of alternatives for
choice.

That is precisely how Hayek\index{Hayek F.} interpreted the
institute of market and it is precisely this interpretation that
provides, as we will show, a~possibility to construct a~qualitative theory of market economy. This theory possesses, we
think, far greater possibilities than the orthodox theory based on
ideas of Walras\index{Walras L.} and Samuelson.\index{Samuelson
P.}

Several final comments. Our discussion of the existing economic theory is by
no means a~review or
overview. We consider only theories and works which were of interest
to us in the context of the formulated problem~--- the development of
a new approach to the study of economic equilibrium.

One more remark concerns the rigor of the mathematical tools
applied. We stick to the standards developed in theoretical
physics, i.e., consider the functions differentiable as many times
as is needed and freely replace the sums by integrals (but doing
so we naturally remember the constraints that should be imposed on
the mathematical objects involved).

\chapter{Models and their role in economics}

In which sense should one understand models in economics?

Is the economic modeling an uncovering of certain ``common laws'' of
economics or this is a~mere instrument that helps to evaluate
a~concrete economic situation?

To answer these questions, we have to overview the types of models
used in economics and, what is most important, establish what type
of knowledge these models allow one to deduce from the analysis of
economic situations.\footnote{The discussion on various types of
models in economics has a~long history. For separation of the
``pure'' theoretical law and an historical study of empirical
forms of economic activity, see \cite{Me}. For the present state
of discussion on the type of economics, see \cite{Bl}.}

The role of economic models in the estimation of practical decisions
is obvious. How to make use of resources optimally? How to invest
with the least risk or with the highest yield? What assortment of
goods should the firm produce at the current market situation?

The answers to the multitude of similar questions constitute the
base of management and in this sense the mathematical models are
of undoubted value as a~means for solving the routine problems of
control of economic activity. The main concept in the formulation
of such problems and obtaining the answers is the notion of
optimization~--- in the process of the search for solutions a~certain
essential parameter of the problem is tested for the
maximum or minimum value.\footnote{The optimization problems in
economics are discussed in very broad literature, see, e.g.,
\cite{La} solidly based on the ideas of
L.~Kantorovich\index{Kantorovich L.} and J.~von~Neumann.\index{von
Neumann J.} To a~great extent this interest was also related with
the general interest in the problem of the study of operations
gained a~great importance during WWII. The methods of the study of
operations based on optimization of solutions demonstrated their
effectiveness for a~very broad type of problems, see, e.g.,
\cite{Saa}.}

A typical problem of this type is the problem of replacement in
the neoclassical theory\index{theory, neoclassical} of demand in
the style of Slutsky-Hicks:\index{theorem, Slutsky and Hicks} How
do variations of the price of one of the goods influence the
demand? (See \cite{AEA}.)

This is an optimization problem with one parameter whose solution is
easy to obtain by means of the usual methods of Calculus.

Far more serious problems appear when optimization takes into
account two or more parameters: say, maximize the yield and
minimize the risk. These problems would have required a~more
sophisticated mathematical apparatus~--- game theory\index{game
theory}
--- but, nevertheless, observe that problems with multiple
criteria remain open to a~considerable extent.

Far from always it is clear how to solve them and the reason here
lies not only in mathematics but mainly in the ``theory of
choice'' since it is not clear, in general, how to adjust and make
commensurable criteria of distinct nature.\footnote{See, for
example, the discussion of this problem in Section 4 of Chapter 1
in the book by J.~von~Neumann\index{von Neumann J.} and
O.~Morgenstern\index{Morgenstern O.} \cite{NM}.}

How to correlate the beauty of the object with its price? Or the
price and the quality?

It is hardly possible to construct universal means for evaluating
beauty or practical value because they depend too much on the
individuality of the evaluator. This is exactly why certain trends
of economic thought, in particular, the Austrian
school,\index{Austrian school} very skeptically considered the
possibility to universalize the estimates. Such an approach
certainly undermines the trust into the practical significance of
the optimization problems that use the notion of
``utility''.\index{utility}

There are, however, a~considerable number of optimization problems
where there is no need to introduce the
``utility''.\index{utility} Such are for example, various
transport problems where the subject of optimization is the time
or the price of transportation.

Optimization problems provide innumerable possibilities for the
skilled application of mathematical apparatus~---  linear, integer
and dynamical programming, various versions of game theory: games
with zero sum, games with non-cooperation, differential games and
so~on.

All these problems, however, possess a~common peculiarity: these
are problems with a~goal. Models of solutions of such problems can
be called  ``normative'' since they are based on the evaluation of
decisions.

The ontology of goal-reaching is organically in-built in these
problems.

Generally speaking, the subject is utterly unnecessary for similar
optimization problems. For example, there is a~broad variety of
variational problems in physics without any subject if, of course,
one does not consider that variational principles\index{principle,
variational, see variational principle} in mechanics are
testimonials of God's existence (such a~possibility was subject to
an active discussion immediately after the discovery of
variational principles\footnote{Thus, one of the authors of the
principle of least action,\index{principle of least action}
P.~Maupertuis,\index{Maupertuis P.} assumed that the laws deduced
from this principle are applicable not only to the world of
mechanics but also to the world of animated nature, see
\cite{Mau}.}).

Nevertheless, both practical economic problems on optimization and
``normative models'', despite the similarity of their tools,
differ from the variational problems of physics: the economic
theory assumes the existence of a~subject with\index{interest}
interests, a~subject able to formulate goals.

In ``practical economics'', optimization problems are the tools
that help the subject to achieve a~goal rather than ``nature's
laws'', which variational problems of mechanics are supposed to
be.

Nothing similar to Hamilton's variational
principle\index{variational principle} can be seen in the
``normative models'' for decision-making.

{\bf To what extent then can one expect the appearance of similar
universal principles in the theory of economics?}

This question was the center of a~theoretical discourse in
economics in the second half of XIX century. Such economists as
C.~Menger,\index{Menger C.} L.~Walras,\index{Walras L.}
W.~Jevons\index{Jevons W.} insisted on the possibility of the
existence of an abstract theory, see \cite{Me, Wa, Je}.

Those who defended the possibility of existence of an abstract
theory in economics based on relations between ``pure'', ``ideal''
notions were actually defending theoretical economics as an
independent sphere of activity, independent of successes of the
practical recommendations in the realm of management.

This discussion was also connected with the existence of another
danger for the newborn theoretical economics~--- the possibility
of its engulfing by economic history, see \cite{Me}. In case of
such an engulfing the theoretical studies would have been reduced
to the documentation of the uncountable variety of various
``national economics''.

An approach formulated in the works of L.~Walras,\index{Walras L.}
W.~Jevons and C.~Menger and later called
``neoclassical''\index{approach to economic,
neoclassical}\index{neoclassical approach to economic} in
distinction from the ``classical'' formulations of
A.~Smith\index{Smith A.} assumed the possibility to obtain precise
answers to ``general questions'' and explanation of such
phenomena as, for example, the surge of prices under the increase
of the supply of goods, surge of banks interest under the
significant accumulation of capital and so on. The founders of the
neoclassical approach assumed that theoretical constructions that
tie the abstract notions will enable them to go out of the realms
of the ``latest news'' and provide both prediction and control,
see \cite{Me}.

In particular, C.~Menger\index{Menger C.} saw the base of
theoretical models in economics in the separation of empirical
forms of economic activity
--- the types devoid of individual features~--- and in establishing
logical relations between the types.

In this way, the theoretical economics was alienated both from the
``economic politics'', that is the technology of decision-making,
and from studies on the history of economics. These studies
invariably, due to the logic of the historical science,
concentrated their attention on the individual peculiarities of
the situations studied.

Theoretical economics became, therefore, the domain of functioning of
interpretational models, i.e., the models that enable one to understand
the general properties of the reality studied, the principles of its
organization and functioning.

Still the considerable difference between the interpretational
models of theoretical economics and the type of theoretical
knowledge which in the natural sciences was used to be called
``laws of Nature'' was unbridgeable. The neoclassical models
helped to understand certain principal peculiarities of
functioning of the economic institutes, for example, the study of
market  equilibrium\index{equilibrium, market}  under the
conditions of ideal competition but could not offer anything
similar to the laws of mechanics that enable us to predict the
future state of the system on the base of a~small amount of
general characteristics of the system and initial data.

Interpretational models of neoclassical economics are rather
analogues of existence theorems in mathematics, i.e., logical
means that help to discuss the existence of certain objects and
their properties but tell nothing on the dynamics of these
objects. In order to understand this fundamental distinction one
has to make a~short digression in the domain of general principles
of methodology of natural sciences formed in the beginning of the
Modern Age and compare these principles with methodological
principles of theoretical economics.

The base of the modern science is, after Tycho Brahe,\index{Brahe,
Tycho} the experiment, more exactly, a~methodology that assumes
that there are means to verify the applicability of theoretical
constructions and models, that is to compare the results obtained
with the help of these models with an experiment.

Here I would like to make a~very essential remark, which is usually
missed not only by various laymen from the general public, but also by
highly qualified experts.

The point is that any ``comparison with an experiment'' is hardly
a comparison with natural situations encountered in Nature.
Strictly speaking, by means of comparison with ``natural
situations'' one cannot either confirm or disprove a~scientific
theory.

Indeed, the experiment deals not just with a~mere reality. It
deals with a~{\it specially prepared} reality in which all the
parameters considered to be given (and essential) in the
theoretical model are under control. Actually, any experiment has
no direct relation to Nature as such. This is a~construction which
is a~corollary of theoretical abstraction (see a~remarkable
analysis of this problem in \cite{Bib}).

For example in the ``theoretical dispute'' of
Aristotle\index{Aristotle} and Galileo\index{Galileo} on the
nature of motion in physics the practice is on Aristotle's side:
moving bodies, devoid of a~source of energy, have a~tendency to
stop and assume a~certain position. The idea of Galileo about the
motion by inertia is a~theoretical abstraction that ignores
friction and resistance of the media. Nevertheless it was
precisely this abstraction that Galileo realized constructing his
experiments.

Actually, a~thought experiment preceded the real experiment, i.e.,
the construction of a~device that develops certain properties of
the object under investigation.

Such a~device creates a~sort of microcosm that purifies the object
and deprives it of the properties inessential for the experiment.

I intentionally stress here the role of the  thought experiment as
a necessary predecessor of a~real experiment as a~theoretical
construction, which, being completely realized only in the ideal
world, possesses, nevertheless, an intrinsic value and ability to
prove. (M.~ Wertheimer suggested a~very interesting analysis of
the structure of Galileo's  thought experiments, see \cite{We}.)

The extent to which the conceptual pattern of an experiment can be
realized is always questionable. How can we be sure that the real
device does not possess certain factors unknown or not under our
control but with decisive influence on the results of the
experiment? To be so sure, we should believe in the theory
underlying the conceptual scheme of the experiment or more
precisely in the logic behind this conceptual scheme because
neglecting certain possible influencing factors during the
development of the conceptual scheme is often based not on
theoretical considerations but on the ``natural logic'' which is
often implicit.

Let us make one more step. What if the experimental device spoken
about in the conceptual scheme of the experiment is conceivable in
principle but so complicated that it cannot be realized for
technical reasons? Does it mean that we may not argue
appealing to the conceptual scheme of the experiment?

Similar situations are encountered all the time.

It turned out that the conceptual scheme of an experiment can be a~means
of the proof even without being realized in hardware. It
suffices to recall discussions of Bohr\index{Bohr N.} and
Einstein,\index{Einstein A.} see \cite{BE}\footnote{This volume
comprises Bohr's papers on relativistic quantum theory and his
philosophical texts regarding complementarity in physics from the
years 1933 to 1958, together with two introductory essays, some
unpublished manuscripts, and four relevant papers of other
authors. It is supplemented by selected correspondence mainly with
Heisenberg and Pauli.

The part on relativistic quantum theory documents Bohr's views on
field quantization, measurability, and early particle physics that
do not so much hinge on lengthy calculations and consistent
schemes as on detailed reflections on the principal problems.

The part on complementarity as a~``bedrock of the quantal
description'' presents Bohr's\index{Bohr N.} reaction to the
Einstein-Podolsky-Rosen\index{Einstein} paper (also reprinted),
that had a~remarkable effect on him, quite in contrast to, e.g.,
Heisenberg and Pauli,\index{Pauli W.} who considered it as no
news. Further papers deal with causality in atomic physics and
general epistemological problems.} on the basics of quantum
mechanics, ``Schr\"odinger's cat'',\index{Schr\"odinger's
cat}\index{Maxwell's Demon} ``Maxwell's Demon''\footnote{For an
analysis of Maxwell's thought experiment important for problems
considered in our paper, see \cite{Bri}.} and other examples of
thought experiments. A thought experiment verifies the logic of
the theory under the hypothesis of a~potential realizability of
however complicated technical devices whose principle is
understood.

Consider, for example, ``Maxwell's Demon'',\index{Maxwell's Demon}
a thought experiment that will be needed in what follows. Working
with problems of statistical physics, Maxwell suggested an idea of
a technical device whose performance was intended to disprove the
{\it second law of thermodynamics},\index{second law of
thermodynamics} the law of growth of the entropy.\index{entropy}
According to the second law of thermodynamics, {\bf heat cannot
naturally  pass from a~cooler body to a~hotter one}. Maxwell
suggested to consider two volumes of gas separated by a~door
managed by a~robot of a~kind or as he called it, ``Demon''. We
assume that the Demon is able to determine the speed of the
particles of gas approaching the door and either open it or not,
depending on this speed, and this way regulate the flow of the
particles from one volume into another without affecting their
natural motion. It seems that the Demon is able, without spending
any energy, make so that the faster particles will go from a~cooler
volume to the hotter one, and therefore raise the
temperature of the hotter volume still further, contrary to the
second law of thermodynamics. This is a~paradox from the
thermodynamic point of view.

This paradox was repeatedly studied afterwards\footnote{See the
paper by Szilard\index{Szilard L.} \cite{Sci}, seminal for the
development of\index{information theory} information theory.} and
it became one of the cornerstones of modern information theory.
L.~Szilard used this example to show that the performance of such
a device is impossible without exchange of information between the
Demon and the gas that the Demon controls, and that this
information may be defined as a~negative entropy, i.e., by
diminishing the entropy in the system, the Demon will invariably
increase its own entropy which proves the impossibility of this
type of devices.

Numerous examples of other thought experiments directed, as a~rule,
to elucidate fundamental questions of new theories show that
a thought experiment is a~powerful tool of analysis on the
borderline between theory and experiment that helps the
theoretical thought to penetrate the worlds inaccessible to
technical control, the domains of reality where the experimental
situation is not possible to be prepared and controlled
artificially.\footnote{For semantics of such ``twilight zone''
worlds, see, e.g., \cite{SPU}.}

Let us now return to the question formulated at the beginning of this
chapter.

{\bf What is economics as science?} Is it possible to experiment in
economics and is it possible to construct theoretical economics
along the classical pattern well known from the practice of physical
studies: theory~--- experimental verification~--- theory?

At first glance nothing contradicts such an approach. Theoretical
economics describes real processes and from the point of view of a~``naive
methodologist'' a~provable comparison of the theory with
``practice'' is possible. In other words, the ``experimental
verification'' of a~theory is possible in real life. The above
arguments, however, make it manifest that the experiment, as it
was understood during the last three centuries thanks to the
efforts of the classics of the European scientific tradition, is
hardly possible nowadays in economic investigations. In real life,
there are no artificially prepared situations specifically
designed for verification of theoretical ideas. And even if in
certain cases (for reasons unknown, the history of Russia is
exceedingly rich with such cases) ``economic experiments'' are
being performed, the conceptual schemes of these experiments are
very far from the complete description of potential influencing
factors on the artificially prepared experimental situation.

We say nothing about the price of such ``experiments'' and how
sound is the theoretical base of such ``experimentation'' if
evaluated on the level of theoretical models in natural sciences.

The impossibility, at least at present time, of a~real scientific
experiment in economics does not, however, mean that the idea of
an experiment as a~test of theoretical models in economics failed.
The rich possibilities of thought experimenting still remain. The
huge help of thought experiments, for example, in the construction
of quantum mechanics, leaves some hope. In such a~setting of the
problem we do not speak of course about verification of the
quantitative predictions of the theory. But it seems quite
possible to verify the theoretical models themselves and their
inner logic.

Having said so, we immediately encounter a~very serious question
whose scale is comparable with the logical problems of quantum
mechanics and which, to my mind, can be solved by analogous means.
This question is {\bf where are the limits of rationality of
economic behavior?}

{\it To what extent the economic situation formed by the choices,
preferences, plans and behavior of a~multitude of people can be
controlled rationally? Can there be certain principal constraints,
logical in their nature, on the rational control similar to the
second law of thermodynamics\index{statistical
thermodynamics}\index{second law of thermodynamics}
or\index{principle, complementarity, Bohr's}\index{Bohr N.} Bohr's
complementarity principle?}

Nothing contradicts the assumption that thought experiments might
suffice to solve such problems if the theoretical models will be
formulated clearly and logically.

Conceptual models possess their own logic and, as the example of
the analysis of ``Maxwell's Demon''\index{Maxwell's Demon} shows,
the rigorous adherence to this logic may reveal the inner
self-contradicting nature of the conceptual constructions, which
seemingly satisfy the principles of the ``natural, naive'' logic.

It is precisely such analysis which we think is of primary
importance in economic studies. But serious and rigorous study of
the inner logic of theoretical models is only possible if these
models are formulated in a~sufficiently lucid language subject to
logical analysis and possessing a~certain inner completeness that
does not let the theoretician get lost in the multitude of weak
and sometimes not explicitly formulated basic assumptions. In
other words, to be verifiable by a~thought experiment a~theoretical
model should be a~logical construction and not a~metaphor.\index{metaphor}

As we will see in what follows, this is precisely this requirement
that the current theoretical models of market economy fail to
satisfy essentially.

Theoretical models of market started with the metaphor of the
``invisible hand''\index{invisible hand} of the market suggested
by Adam Smith.\index{Smith A.} Surely, the metaphor is a~most
powerful weapon of creative thinking. But theoretical analysis
must go beyond metaphor. It should make explicit the inner logical
structure of the metaphor turning the metaphor into a~theoretical
model and, to an extent, destructing the metaphor.\index{metaphor}

The existing economic theories of market are metaphorical to the
marrow. They constantly discuss an
``equilibrium'',\index{equilibrium, market} the notion
metaphorically borrowed from mechanics and therefore our first
problem will be the analysis of the history of the conception of
this basic metaphor of economic theory. In order to turn a~metaphor into a~model
this metaphor\index{metaphor} should have
been subjected to ``deconstruction'',\index{deconstruction} say in
the same sense one deconstructs theoretical notions by
M.~Foucault\index{Foucault M.} and W.~Derrida.\index{Derrida W.}
Still, deconstruction alone is insufficient to construct a~theoretical model.
The deconstructed metaphor should be composed
again using the ``logical constructor'' and only after that the
logically constructed model may be fit for thought experimenting.
This is precisely what we intend to do with the notion of
equilibrium in market economy.

\chapter{General Equilibrium Theory and the ideas of
A.~Smith}

In economic theory, there seems to be no notion whose importance
can be compared with the notion of equilibrium.

The General Equilibrium Theory (GET)\index{GET, General Equilibrium
Theory} became the cornerstone of modern economic science and
subjected not only theoretical investigations but even the economic
politics. Strange as it may seem, but General Equilibrium
Theory\index{equilibrium, market} gained a~great deal of currency
and popularity in the countries with transitional economies whereas,
as is manifest, the transitional economy is constantly being in
a~state quite distant from equilibrium, at least in its traditional
reading.

Our aim here is to consider the ontological basis of GET. This
will enable us to offer in the subsequent chapters a~generalization
of the notion of equilibrium in economics. This
generalization will allow us, at least partly, to consider
non-equilibrium processes with approximately the same degree of
generality as the non-equilibrium statistical
thermodynamics\index{statistical thermodynamics}\index{statistical
thermodynamics, non-equilibrium} has in the framework of physics.

The creators of the General Equilibrium Theory were well aware
that this theory only describes economic situations in a~sufficiently
small neighborhood of an equilibrium state and that
the logical base for General Equilibrium Theory should be sought
outside its realms, since the mechanisms that force the system to
attain an equilibrium have nothing in common with the description
of characteristics of the equilibrium itself.

\footnotesize
One of the known American experts in mathematical
economics wrote not so long ago, see \cite{Fi}:
\begin{thetext}
{``{\sl For macro-economists ... the microeconomic theory is
primarily about positions of equilibrium.\index{equilibrium,
market} The plans of agents\index{agent} (usually derived from the
solution of individual optimization problems) are taken together,
and certain variables
--- usually prices~--- are assumed to take on values that make those plans
mutually consistent... In all this very little is said about the
dynamics of the process that leads an equilibrium to be
established... Attention is centered on the equilibria
themselves... and points of non-equilibrium are only discussed by
showing that the system cannot remain in such points.} ''}
\end{thetext}

\normalsize This fundamental position is often forgotten and those
who try to apply this theory in practice, assuming that
``liberating'' market forces and rigorous sustaining certain
parameters (such as, for example, the volume of the currency or
the level of the deficit of the budget) in certain limits
automatically leads economy to the state of equilibrium. Such
economists or, more exactly, politicians, since the one who
implements an economic theory in practice becomes a~politician
whether one wishes that or not, should recall the warning on
practical application of economic models expressed in one of the
most known books in the world literature on economic theory
(\cite{NM}):
\begin{thetext}
{``{\sl The sound procedure is to obtain first utmost precision
and mastery in a~limited field, and then proceed to another
somewhat wider one, and so on. This will do away with the
unhealthy practice of applying so-called theo\-ries to economic or
social reforms, where they are in no way useful}''.}
\end{thetext}
Thus it is very important to establish the limits of the basic
notions used in economic theories. This applies first of all to
the notion of economic equilibrium.\index{equilibrium, market}

There is a~number of key questions concerning General Equilibrium
Theory which are very hard to answer.

Is General Equilibrium Theory a~scientific theory or is it just a~``paradigm'',
an ontological base in the frame of which an
economic theory should be developed? Is it possible to falsify
General Equilibrium Theory? Is it possible to use it as a~foundation
for ``normative economics'', i.e., for practical
economic decisions? These questions are a~subject of heated and
incessant debates, see, e.g., \cite{Bl}, Ch.8.

The very existence of such questions during discussions testifies
that the status of General Equilibrium Theory as a~theory is
unsettled. It is hardly possible that such questions may arise
nowadays for example, concerning classical mechanics. Therefore,
instead of considering General Equilibrium Theory as a~base for
practical solutions, we should at least try to analyze the logical
foundations of General Equilibrium Theory and see what
mathematical models are adequate to these foundations. Let me make
several remarks that at first glance may seem rather trivial.

Any description of a~real situation is a~reduction. Certain
parameters of the situations are being taken into account, the
other ones are being ignored. For the scientific theory the
problem is to find out the degree to which the inevitable
reduction of the description is the subject of a~conscious
control. Since in the frameworks of the majority of
``non-scientific'' descriptions the reduction is as a~rule
implicit, the critics ``from inside'' become impossible. The
description becomes the subject of faith. Precisely this type of
transformation happens usually during the process of application
of scientific theories. Scientific descriptions differ from
``non-scientific'' ones in the very fact that the reduction
becomes the object of reflection. Methodological criteria clarify
what ``models of reality'' are the base of the description. And
what is very important, such a~practice (i.e., reflection) needs
its own ``metalanguage'' which is generally speaking, different
from the one used for the description of reality.

To what extent are these trivial arguments used during the
construction of economic models? To a~large extent this is related
with the understanding of the role of mathematical models in the
description of reality.

When mathematical models are being used, the level of requirements
to the understanding of implicit hypotheses of the model is much
higher than for the usage of conceptual models. We think that,
under the passage from the conceptual models of Adam
Smith\index{Smith A.} and the founders of the neoclassical theory
to the mathematical models of economics in 1930s--50s, the
understanding of the hypothesis did not improve in the majority of
cases considered in the economic theory, but instead, diminished.
Therefore we deem it highly important to analyze the genesis of
conceptions on market equilibrium\index{equilibrium, market} and
reveal the hidden hypothesis that used to lie in the foundation of
the molding representations of the classical and neoclassical
economic theory.

The first thing to do is to try to understand the inner
motivation and implicit suggestions lying in the foundation of the
notions used by the founder of the classical economic theory, Adam
Smith.

Observe, first of all, that A. Smith was not, generally speaking, an
economist. At least his main interests used to lie outside the
domain of economics. Smith was a~moral philosopher. He was
interested in the problems and principles explaining people's
behavior and correlation of the characteristics of this behavior
with the common well-being. Essentially, his economic theory was
precisely the theory purported to explain the character of the
relations between human behavior\index{human behavior} and common
well-being.\footnote{Before A.~ Smith wrote his famous book
\cite{Sm1}, he was for a~long-term plunged into studies of moral
philosophy. At the age of 29, A. Smith was elected the Chair of
Moral Philosophy at Glasgow University and at the age of 36 he
published {\em Theory of Moral Sentiments}. Only in 1767, being 44
years old, after he had traveled to Paris and acquainted with
Turgeau, A.~Smith began to study theory of economics.}

We think that the economic representations of Smith are impossible
to understand without first understanding the ``model of the human
being'' which he put into the foundation of his economic theory
(for a~shrewd analysis of this aspect of A. Smith's ideas, see
\cite{Wh}).

For Smith, a~man is, first of all, an emotional being whose behavior
is determined not by reason but by feelings and emotions.

According to Smith,\index{Smith A.} the ``moral'' expresses the
communal feelings and a~moral behavior is a~behavior concordant
with these feelings. Without difficulty Smith reveals
contradictions between the individual utility\index{utility} and
communal appraisal: he

\begin{thetext}
{\sl remained, however, under the opinion that the first and prime
reason of our approval or disapproval does not follow from the
understanding what can be useful or evil to us... To the feeling
of approval we constantly add our understandings on what is
natural and legal which have nothing to do with the understandings
on utility\index{utility}...}
\end{thetext}

He believed that human abilities in pure sciences are being
discovered in a~most wide and illustrious way especially in the
so-called higher mathematics. The utility\index{utility} of these
sciences, however, is known to precious few and, to prove it, one
would require explanations which far from all can grasp. Therefore
not the utility provided by these sciences is the reason of the
universal respect to them. There was no discussion about this
utility until one had to retort the reprimands of the people who,
having no inclination to these higher sciences, tried to diminish
their value accusing them of uselessness, see \cite{Sm2}, part IV,
Ch. II.

Actually, A.~Smith\index{Smith A.} spent his life seeking for the
principles and social institutions which could harmonize human
emotions and the common prosperity. Having found nothing
satisfactory in the domain of moral and political philosophies he
turned to the study of economic behavior where he found, at last,
the dominating emotion that governs behavior~--- the individual's
interest\index{interest} and a~regulative principle that matches
this emotion with common prosperity~--- the market. Observe,
however, that for A.~Smith the interrelation between the
individual's interest and the common prosperity was never a~postulate.
This was his deduction obtained from an economic
analysis. Smith needed the ``invisible hand''\index{invisible
hand} of the market not to describe the phenomenon of equilibrium.
He needed the ``invisible hand'' to find out and explain the
circumstances of a~rapid economic growth. He is most interested in
the questions of interrelation between well-being and growth. He
has found out, for example, that the price of a~labor force is
highest not in the regions with the higher well-being but where
the production growth is the highest.

A.~Smith formulates a~dynamical theory of the market in which the
interests of its agents\index{agent} under the conditions of free
competition\index{free competition} are the driving forces that
establish an equilibrium and turn themselves into this ``invisible
hand''. He describes the influence of the ``invisible hand'' as
follows:

\begin{thetext}
{{\sl All the people that use the land, force or capital to
provide the market with goods are interested that the quantity of
the goods would not exceed the actual demand whereas all the
others are interested in maintaining its quantity on the level not
lower this demand... If on the contrary, at some moment the amount
of goods delivered to the market would be lower than the actual
demand then certain constituents of its price should raise higher
its natural norm. If this is a~land rent, then the
interests\index{interest} of all the other landlords will
naturally prompt them to adjust more land for unearthing this
product; if this is a~wage or a~profit then the interests of all
the remaining workers and business people will soon force them to
invest more labor and capital into production and delivering this
item of goods to the market}, see \cite{Sm1}, book 1, Ch. 7.}
\end{thetext}

\noindent
A.~Smith\index{Smith A.} was not interested in the market
equilibrium\index{equilibrium, market} as such. He was interested
in the conditions which establish the fastest growth of communal
wealth. The problem of an ``invisible hand''\index{invisible hand}
and the study of equilibrium were only his means to investigate
the problem of harmonization of individual
interests\index{interest} (utility)\index{utility} and the common
prosperity (i.e., the rapid growth of the total wealth of the
people). This is a~principal difference between A.~Smith and his
neoclassical followers.

It is interesting to observe that the method Smith uses to prove
his statements is essentially a~thought experiment. Using an ideal
logical model~--- a~free competition\index{free competition} and
the lack of constraints on the nature of economic activity Smith
proves the existence of a~market equilibrium based on a~logical
deduction. Introduction of any constraint into his model (say
restrictions on the available resources of labor or capital) will
change the results.

Such a~model does not, of course, allow to make quantitative
predictions. The thought experiment of A.~Smith\index{Smith A.}
can be easily formulated as a~mathematical model but only on
a~``metaphorical level''.\index{metaphor}

The equilibrium of a~dynamical system\index{dynamical system} is
precisely a~metaphor which can be applied to the description of
a~market ``equilibrium'' starting from Smith's thought experiments.
A deviation of the system from the equilibrium state calls the
``forces'', i.e., interests\index{interest} that struggle to
return the system into the equilibrium\index{equilibrium, market}
state.

Of course, Smith himself did not consider market equilibrium as
a~singular point of a~system of differential equations but this
metaphorical model can be easily extracted from his arguments.

Smith's ideas were developed in the first half of the XIX century
by economists of the ``classical school'': Ricardo,\index{Ricardo
D.} J.~H.~Mill,\index{Mill J. H.} and so on. But this development
was not qualitative and did not considerably augment our
understanding of economic processes. These theories never gave any
``computational'' recipes for behavior nor models of economic
decision making.

The situation completely changed with appearance of works by
C.~Menger,\index{Menger C.} L.~Walras\index{Walras L.} and
W.~Jevons who changed the intellectual landscape of research in
economics.

As we mentioned above, the purpose of these studies was to
construct rigorous models in whose framework by logical
(C.~Menger)\index{Menger C.} and mathematical (W.~Jevons and
L.~Walras)\index{Walras L.} methods one could obtain significant
results. To do so, the neoclassics had to essentially restrict the
range of application of their analysis confining themselves to the
study of market equilibrium as such. In other words, the market
equilibrium became (especially in works of W.~Jevons and
L.~Walras)\index{Walras L.} not an institutional mechanism to be
analyzed, but a~rigorously defined situation of the market,
namely, the absence of a~surplus demand.

In the middle of the XX century, the development of the ideas of
Jevons and Walras\index{Walras L.} by A.~Wald\index{Wald A.}
(\cite{W}), K.~Arrow\index{Arrow K.} and G.~Debreu\index{Debreu
G.} (\cite{D}, \cite{AD}) led to creation of mathematical models
of market equilibrium with developed research tools; these models
digressed very far from the initial ideas of neoclassics.

To understand these highly sophisticated mathematical
constructions, one has to sort out the ontological shift that took
place in the passage from the classical theory to a~neoclassical
one. One of the central novelties here became the notion of {\it
utility}.\index{utility} In order to get a~possibility to
construct mathematical models of the market, one had to create
a~universal model of behavior of its agents.\index{agent} Whereas
the basis of such a~model for the sellers was considered
sufficiently obvious after works of A.~Smith\index{Smith A.} (as
maximization of the profit realizing the private\index{interest}
interest), it was not that easy to introduce a~similar universal
measure for the buyers. On the philosophical level, the notion of
utility introduced and justified by I.~Bentham\index{Bentham I.}
started to play this role.

Bentham thought that human behavior\index{human behavior} in
general is determined by a~utility\index{utility} derived from
decisions and actions, and the common welfare can be computed as
the sum of gains and losses in totality obtained by the members of
the society as a~result of their decisions, CR. \cite{Ben}.
Bentham suggested an essential change in understanding of human
behavior:\index{human behavior} for A.~Smith, a~private
interest\index{interest} was only one of emotions that determine
human behavior, and the role of this emotion in the general
picture of behavior was a~subject of the study; for Bentham, the
question was solved: utility became the only driving force that
determines the whole totality of behavioral acts.

This point of view was adopted by the founders of the neoclassical
school and turned into an instrument of mathematical analysis of
the market. With the help of the notion of utility\index{utility}
it became possible to construct a~model of rational decision
making by the buyer. Thus, the hypothesis that
\begin{thetext}
{\it the buyer possesses clearly described goals and preferences
and is capable to determine and compare utility of different
goods}
\end{thetext}
became the ontological postulate of the neoclassical
orthodoxy to the same extent as the hypothesis that
\begin{thetext}
{\it the utility\index{utility} the buyer obtains and the profit
the seller gets exhaust the conception of common welfare obtained
as a~result of an economic activity.}
\end{thetext}

Sufficiently simple hypotheses on the dependence of
utility\index{utility} on the quantity of the goods purchased (its
monotonic decay) together with the principle of maximization of
utility  led to the ``marginalist revolution''\index{marginalist
revolution} in the study of market economy namely to the
understanding that the utility of the last portion of goods
purchased determines the market situation.

Indeed, it is not difficult to see that the ratio of utility of this last
portion to its price should be the same for all goods, otherwise it would
have been possible to enlarge the common utility of goods to buy by changing
the assortment of purchases.

The discovery of this fact immensely influenced the whole style of
economic studies. New possibilities to study properties of the
state of market equilibrium\index{equilibrium, market} opened up.
The visible ontological change discernible already in works of
Bentham \index{Bentham I.}--- placement of utility\index{utility}
function and its properties in the center of analysis of human
behavior\index{human behavior}~--- became fossilized. The theory
of ``rational choice''\index{rational choice} was formed. The
analysis of decisions the buyer makes became the prime object in
the study of market equilibrium. Later it was dubbed the ``buyer's
sovereignty'',\index{sovereignty, buyer's} that is independence of
the buyer's decisions of anything except for the utility function.

The already mentioned basic metaphor\index{metaphor} of
mathematical economics emerged, namely, the problems of
mathematical economy became now understood as problems on extrema
of certain functions subject to given constraints (usually
constraints on resources). The problem of existence and properties
of the market equilibrium came to the center of research in
economics.

Such a~change of main goals of the study entailed the appearance
of a~specific mathematical apparatus: in 1950s, for the proof of
existence of the point of market equilibrium, i.e., a~certain
vector of positive prices under a~non-positive extra demand, the
researchers started to use topological methods such as the fixed
point theorems\footnote{A theorem stating that a~continuous
selfmap of a~compact set has a~fixed point is due to
Brouwer.\index{Brouwer L. E. J.} Later, Kakutani\index{Kakutani
S.} formulated a~fixed point theorem for semi\Defis continuous above
maps of sets, \cite{Ka}. Arrow\index{Arrow K.} and
Debreu\index{Debreu G.} used this theorem to prove existence of an
equilibrium in the dynamics of competition. Their idea is to
construct two maps of sets: a~map of the set of vectors of excess
demand to the set of vectors of prices and a~map of the set of
vectors of prices to the set of vectors of excess demand and to
apply Kakutani's theorem to the selfmap induced by the above maps
on the Cartesian product of these two sets. Under certain
conditions (whose fulfilment for concrete economic models is to be
verified) there exists a~vector of prices for which the excess
demand is non-positive, cf. \cite{ABB}.}.

Of course, the conclusions of the economists of the neoclassical
school, as well as the conclusions of the classics, were based on
thought experiments since it is hardly possible to actually
compute the utility\index{utility} function of the market
agents.\index{agent} By means of a~thought experiment it is
possible to prove the existence of a~point of market equilibrium,
but it is hardly possible to determine what would this equilibrium
be for given conditions on production and demand.

The most important happening in the framework of the marginalist
revolution\index{marginalist revolution} was the change in the
understanding of the nature of market equilibrium.

For A.~Smith,\index{Smith A.} a~{\it market
equilibrium}\index{equilibrium, market} was a~stable point of a~dynamical
system\index{dynamical system} with actually acting
``forces'' (interests\index{interest} of the market
agents)\index{agent} that react on deviations from the equilibrium
state. In other words, the metaphor\index{metaphor} of equilibrium
itself turned out to be justified by comparison with the behavior
of mechanical systems. In works of neoclassics and especially in
mathematical interpretations of neoclassical conceptual
constructions, even the relation with the etymology of the word
{\it equilibrium} on the metaphorical level becomes lost.

It is unclear why the vector of prices which is a~fixed point of a~certain
map and corresponds to a~non-positive extra demand should be called an
equilibrium point of whatever.

The model of the human in neoclassical economics becomes reduced
to triviality. A behavior is called ``rational'' if it maximizes
utility\index{utility} and utility is actually defined as
a~function which determines a~``rational behavior''.

This is a~logical vicious circle. The market is considered to be
a~priori free, i.e., the conditions for the ideal competition are
justified as if ``taught by themselves'' whereas A.~Smith, for
example, clearly understood that
\begin{thetext}
{\sl a~free competition\index{free competition} is a~social
institute which should be maintained and the justification of
conditions for free competition is a~function of political power.}
\end{thetext}
Implicitly, the neoclassics started to consider {\it the free
competition as a~self-sustaining system} which is a~very doubtful
assumption (at least it should be seriously justified).

The General Equilibrium Theory\index{equilibrium, market} formed
on the base of neoclassics' opinions became lately not even as
much scientific paradigm as the leading ideology of conservative
political movements and parties in many countries despite its very
doubtful performance in their economic effectiveness. One should
observe that in the post-war time the highest rate of economic
growth was achieved in the countries (Japan, South Korea, Taiwan
and recently China and Vietnam) where the general orientation to
market economy was combined with an economic policy very distinct
from the ideology of neoclassic.

The faults of the economic policy inspired by the principles of
the general theory of market equilibrium\index{equilibrium,
market} (as well as recent economic difficulties of the East
European countries) can, nevertheless, hardly be arguments in the
dispute on scientific consistency of this theory for reasons
discussed in Chapter 1. We need an inner logical analysis of the
thought experiments on which General Equilibrium Theory is based.

\chapter[Challenge to
utilitarianism]{C. Menger, F. Hayek and D. Hume: Challenge to
utilitarianism}

Observe that the basic for the classical and the neoclassical
economic theory idea of the market as a~means to directly
harmonize a~private interest\index{interest} and common welfare
was subject to a~serious criticism by economists who strived to
create a~rigorous economic theory; first of all, by
representatives of the Austrian school.\index{Austrian school}
Carl Menger,\index{Menger C.} one of the initiators of the
marginalist revolution,\index{marginalist revolution} insisted,
nevertheless, that a~private interest cannot be considered as
a~means for automatic achievement of common welfare and the idea
on an interrelation between a~private interest and common welfare
is not a~suitable foundation for a~rigorous economic theory.

According to Menger, certain economists consider the ``dogma of
private interest''\index{dogma of private interest} the founding
principle that pursuing a~private interest on the level of an
economic individuum without any influence of the political and
economic measures of the government should result in the highest
level of common welfare which can be achieved by the society in
the given spatial and time constraints. We will not nevertheless
deal here with this approach which is faulty at least in its most
general form, see \cite{Me}, p. 83. Even if, in their economic
activity, people would always and everywhere be guided exclusively
by their private interests, it would be, nevertheless, impossible
to assume that this feature determines economic phenomena, since
from experience we know that in uncountable many cases people have
faulty views on their economic interests or have no idea about the
situation in economics at all. \ldots The presumption of such
definition of the regulating force in economic phenomena, and
therefore within theoretical economics, however multivalued this
word might be, is not just a~dogma on a~private interest. It is
also a~dogma on ``faultlessness'' and ``omniscience'' of the
people concerning the situation in economics, \cite{Me}, p. 84.

Menger insists that the economic theory should disregard
inessential aspects in the human behavior\index{human behavior}
(in particular, possible errors) giving an ideal scheme. The
economic theory cannot provide with understanding of the human
phenomena in all its totality and concreteness. But it may provide
with understanding of one of the most important facets of human
life, \cite{Me}, p. 87.

Menger denies the economic theory in its claim to understand real
situations in which exterior factors influencing human behavior
(in particular, faults, delusions and prejudices) are impossible
to remove but believes that the reside after removal of these
occasional factors is of considerable interest\index{interest} for
understanding economic behavior of people.

Such a~position is, of course, both vulnerable and
self-contradicting because it is absolutely not obvious that,
having removed essential factors that affect a~real situation, the
residue will be of any interest.

This is impossible to postulate. This must be proved.

In particular, one should show that eliminating such a~factor as
a~faulty understanding of the economic situation by many market
agents\index{agent} does not destroy the scheme of thought
experiment which we used to obtain a~new notion on the nature of
economic life. To compare, observe that we know now that, in the
physical theory of microcosm, the experimental mistakes cannot be
eliminated from the theory.\footnote{This is one of the
fundamental results of quantum mechanics, see, e.g., \cite{JvN2}.
There exist quantities which cannot be simultaneously measured
with arbitrary precision, namely such are the physical quantities
whose corresponding quantum mechanical operators do not commute.
The measurement problem in modern physical theory does not reduce,
however, to impossibility of simultaneous precise measurements:
\begin{thetext}
{``... {\sl it goes without saying that any measurement or related
process of subjective perception is related to the external
physical world a~new entity that does not reduce to it. Indeed,
such a~process leads us out of the outer world or, more correctly,
leads into a~non-controllable situation since in each control test
the inner life of the individual is already supposed to be
known}.'' (\cite{JvN2}, Ch. VI) }
\end{thetext}
} We see, therefore, that while criticizing ontological
idealization that lie in the foundation of the neoclassical theory
Menger\index{Menger C.} remained on the ``classical'' positions in
quite another sense. In the sense of possibility to eliminate from
the theory the so-called ``inessential'' factors. Nevertheless,
the principal difference of his point of view from that of
neoclassical orthodoxy is manifest. And therefore it is not
accidental that the representatives of the neoclassical orthodoxy
are rather restrained in their appraisal of Menger\index{Menger
C.} and the Austrian economic\index{Austrian school} school as
a~whole.\footnote{S.~Littlechild\index{Littlechild S.} \cite{Li}
characterizes the Austrian school\index{Austrian school} as
follows:
\begin{thetext}
{``{\sl Austrian economists are subjectivists; they emphasize the
purposefulness of human actions, they are unhappy with
constructions that emphasize equilibrium\index{equilibrium,
market} to the exclusion of market processes; they are deeply
suspicious of attempts to apply measurement procedures to
economics, they are skeptical of empirical ``proofs'' of economic
theorem and consequently have serious reservations about the
validity and importance of a~good deal of the empirical work being
carried on in the economics profession today}.''}
\end{thetext}

Earlier, P. Samuelson\index{Samuelson P.} (\cite{Sam}, p. 761)
formulated his relation to certain critics of the neoclassical
orthodoxy as follows:
\begin{thetext}
{``{\sl In connection with the exaggerated claims that used to be
made in economics for the power of deduction and a~priori
reasoning by Carl Menger,\index{Menger C.} by Ludwig von
Mises...\index{von Mises L.} I tremble for the reputation of my
subject. Fortunately, we have left this behind us}.''.}
\end{thetext}}

Much later the eminent representative of the Austrian
school,\index{Austrian school} F.~Hayek,\index{Hayek F.} made the
next step in understanding of the deeper principles of market
functioning suggesting an interpretation of these principles
totally different from the traditional one. He started to consider
competition at the discovery procedure\index{discovery procedure}
(in particular, circumstances and by particular people) of the
facts and data on the market state which give preference to those
who know these facts and can use them.

Hayek\index{Hayek F.} accepted the principal incompleteness of
knowledge on the actual situation in economics the main
circumstance of any economic activity. According to him:

It is difficult not to concede with the accusation that during the
approximately 40--50 years the economists' discussions on
competition based on assumptions which had the actually mirror of
the reality would have made competition totally meaningless and
useless. If somebody could indeed know everything that economic
theory calls ``data'', then competition would be quite a~wasteful
method of adjustment to these ``data''... Having this in mind it
is useful to recall that each time when competition can be
rationally justified it turns out that the basics for it was the
lack of facts given beforehand that determine the rivals'
actions.\footnote{The first, it seems, to come to the idea to
apply evolutionary theory\index{evolutionary theory} to economics
was A.~Alchian\index{Alchian A.} \cite{Alc}. D.~North\index{North
D.} also used evolutionary approach in the study of evolutions of
social institutes and first of all the inheritance law to
a~considerable extent, see \cite{NoI, NoS}. }

Hayek\index{Hayek F.} poses a~principally comparative problem:
\begin{thetext}
{\sl what institutional organization in economics becomes more
effective, i.e., capable to survive, and under which conditions?}
\end{thetext}

Such a~formulation of the question makes him to turn to the
evolutionary theory for justification of deductions made. He
cannot any longer consider just empirical facts, or even thought
experiments, to justify his opinion:
\begin{thetext}
{... {\sl We come to an inevitable conclusion that in cases where
the competition is meaningful the validity of the theory is
definitely impossible to verify empirically. We can verify it on
abstract models and conjecturally in the artificially reproduced
real situations where the facts on which discovery the competition
is aimed are already known to the observer. But in such situations
the theory has no practical value... At best we may hope to be
able to establish that the communities relying on competition will
eventually achieve their goals with more success than the other
ones. This is the deduction which I think is remarkably confirmed
by the whole history of civilization}.}
\end{thetext}

\noindent This excerpt lucidly shows what determines
Hayek's\index{Hayek F.} skepticism concerning mathematical models
of market economy. Hayek's turned to evolutionary
theory\index{evolutionary theory} for justification of the
advantages of market economy as compared with the methods of
centralized planning let to reconsider ontological assumptions
concerning the nature of society as a~whole. Hayek started to
consider evolution\index{evolution} of market structures as
a~``third world'' of a~kind as compared with two other worlds~---
the world of ``nature with its laws and the world of rational
human activity'' (\cite{H2}, Ch.~1).

Competitive structures (i.e., not only economics but also
biological species) circumnavigate he thinks the limits of human
rationality, thus stating the problem of creating a~new type of
arguments~--- the evolutionary epistemology\index{evolutionary
ontology} (I would have said: evolutionary ontology).

Hayek\index{Hayek F.} is well aware that approaching the study of
market economy from the position of availability of knowledge he
destroys the conception on harmony of private
interests\index{interest} and the global goals of the society. He
is forced to refute the notion of economics as a~system of
rational ``management'' thus destroying the suggested by
A.~Smith\index{Smith A.} interpretation of interest as the inner
regulating force, i.e., Hayek goes much further than
C.~Menger\index{Menger C.} who refused to accept the idea on
a~correspondence of a~private and common interest in market
conditions as a~dogma (i.e., without discussion) but it seems
would have nothing against accepting it as a~deduction (in
accordance with the general methodology of A.~Smith). Hayek
writes:
\begin{thetext}
{``{\sl The direct meaning of the word ``management'' is an
organization or a~social structure where somebody consciously
places resources in accordance with a~unique scale of goals. In
the spontaneous order\index{spontaneous order} created by market
none of this exists and it functions principally differently than
the above ``management''. In particular, it differs in that it
does not guarantee necessarily to satisfy first more important due
to general opinion needs and then less important ones}...''
(\cite{H2}, Ch.~1).}
\end{thetext}

Hayek\index{Hayek F.} denied the notion of ``management'' the
prime role and instead introduced the idea of ``order''.

This conceptual change requires a~particular attention. The role
of idea of ``order'' in statistical physics where ``order'' is an
antonym of ``chaos''\index{chaos} is well known. In statistical
physics same as in the conceptual model of economics suggested by
Hayek the order establishes itself under certain circumstances
since such a~state becomes more probable than the chaos (for
example, this is the way crystallization of the fluids under low
temperatures occurs). This happens not because each molecule knows
its place. The movement of molecules is only determined by local
information on the position of its neighbors. The order is
established because the number of possible ordered states with
given energy turns out to be very high, greater than the number of
chaotic states.

We can therefore try to transform Hayek's arguments based on the
just mentioned metaphor\index{metaphor} of order in the
statistical theory of economic activity which uses in principle
the same notions as the statistical physics.

I think that Hayek himself came rather close to this idea:

\smallskip

``It goes without saying that it is worthwhile to try to create
conditions under which the chances of any randomly selected
individual for the most effective realization of his goals would
have been rather high, even if it would have been impossible to
predict beforehand for which particular goals these conditions
will be favorable and for which unfavorable'', \cite{H2}, Ch.1.

\smallskip

Therefore Hayek\index{Hayek F.} assumes that under the conditions
when ``general prosperity'' would be considered simply as the sum
of satisfied interests\index{interest} of the participants of the
economic activity the market as an economic institute would ensure
the attaining the ``common prosperity'' simply due to the high
probability of satisfaction of interests of individuals. In such
conditions to predict how would the ``common prosperity'' look
like is impossible.

Such an approach to the study of market can hardly be called the
study of equilibrium\index{equilibrium, market} in the sense in
which this metaphor\index{metaphor} was used by
A.~Smith.\index{Smith A.} Here we see no ``balance'', no
``forces'' that return the system to a~certain natural state.
Hayek is very sensitive to this. He writes:
\begin{thetext}
{``{\sl Economists usually call the order created by competition
{\it an equilibrium}.\index{equilibrium, market} This term is not
quite adequate since such an equilibrium assumes that all defects
are already discovered (compare with A.~Smith's thoughts that the
prices inform us on ``hidden'' parameters such as profit or
interest rate. {\it V.~S.}) And the competition therefore is
stopped. I prefer the notion of ``order'' to that of equilibrium
--- at least during the discussion of problems of economic
politics}'' (\cite{H2}, Ch.~12).}
\end{thetext}

Hayek \index{Hayek F.}further makes a~number of particularly
interesting remarks which hint that the ideas of a~fundamental
relation of physical statistics with modeling of economic processes
were not alien to him and his skepticism towards modeling in
economics was occasioned by perhaps the fact that both conceptual
constructions and mathematical methods widely used in the modern
mathematical economics were in Hayek's view inadequate. It seems
that Hayek was just not acquainted with the main ideas of
statistical physics and therefore related even purely statistical
ideas with cybernetic ideas on positive feedback, i.e., with
a~mechanical metaphor.\index{metaphor} (For the sake of justice
observe that in N.~Wiener's\index{Wiener N.} papers applications of
the idea of the feedback to the study of behavior of complicated
systems were largely inspired by his interest to the problems of
statistical physics.) Studying the problem of interaction of
economic agents\index{agent} of the market Hayek \index{Hayek
F.}writes:
\begin{thetext}
{``{\sl This mutual adjustment of individual plans is being
performed along the principle that we following the pattern of
natural sciences that also turned to the study of {\it spontaneous
orders}\index{spontaneous order} (my italics, V.~S.) or
self-organizing systems became known the ``negative feedback''}\
'' (\cite{H2}, Ch.~1).}
\end{thetext}

Feeling that the market works rather spontaneously as
a~statistically organized system in which the order appears not
because the parts have information on the whole and they, the
parts of the system, make a~choice in favor of the ``order'' but
just because the order is more probable in certain conditions.
Hayek still cannot completely get rid of the mechanical
metaphor\index{metaphor} in the study of principles of market
functioning. Actually from the above arguments Hayek\index{Hayek
F.} deduces a~principle of universal applicability of market
economy refusing to distinguish the conditions and the social
problems in solution of which the principles of market economy are
indeed effective and the conditions and problems for which their
effectiveness is doubtful. We think that such a~mixture is
a~result of coexistence in Hayek's perceptions of two different
metaphors\index{metaphor} that explain functioning of market
economy~--- the mechanical metaphor of A.~Smith\index{Smith A.}
and the statistical metaphor developed by Hayek himself without
due logical separation. Speaking about statistical metaphor in the
excerpt cited above Hayek mentions self-organization. Indeed,
there were multitude attempts to prove a~possibility for
self-organization starting from statistical
principles.\footnote{The study of such problems started, it seems,
from the book by E.~ Schr\"odinger\index{Schr\"odinger E.}
\cite{Sch}. At the end of his life, J. von Neumann\index{von
Neumann J.} studied selforganization of automata, see \cite{JvN}.}

Strictly speaking, a~self-organization means diminishing of the
entropy\index{entropy} of the system. In the closed system this
cannot happen, for self-organization we need an influx of energy
from outside. In other words, if we speak about self-organization
in economics, such a~process should assume existence not of
equilibrium\index{equilibrium, market} but metastable
states.\footnote{Life as a~metastable state was discussed as early
as in \cite{W}. In relation with inevitability of eventual
equilibrium of Maxwell's demon\index{Maxwell's Demon} and the
environment, Wiener\index{Wiener N.} wrote: ``Nevertheless, before
the Demon becomes muddled a~considerable laps of time may pass and
this period may turn out to be so prolonged that we have right to
say that Demon's active phase is metastable. There are no reasons
to believe that metastable Demons do not exist in reality,
contrariwise, it is quite probable that enzymes are such
metastable Maxwell's Demons that diminish entropy if not by
separation the fast particle from the slow ones but by some other
method. We can very well consider the live organisms, and the Man
himself, in this light.''}

We believe that one should be very careful applying perceptions on
self-organization to the evolutionary theory\index{evolutionary
theory} (cf. Chapter 8).

Discussing the problem of self-organization Hayek says that the
rules of behavior of market's agents\index{agent} are the result
of a~long cultural process. They are not natural and often are
being followed contrary to the interests\index{interest} of the
economic agents.\index{agent} The market as a~whole exists as
a~system not because following the rules of behavior the market
agents\index{agent} gain an immediate profit but because the
system of rules of behavior that determines market institutes wins
in the process of competition with other systems of rules of
behavior. In other words, there exists a~competition between the
types of economic institutes and in the larger time scales that is
in ``macrotime'' the market institutes are more effective.

This is a~very radical transformation of the viewpoint on the
nature of market. Placing the economic and generally speaking
social institutes of the society in the evolutionary line similar
to the process of evolution\index{evolution} of species, Hayek
changes the ontological level of his analysis from the analysis of
separate behavioral acts to the analysis of the culturally
determined systems of behavioral\index{principle, behavioral}
principles. It is not people who are being dragged into the
process of competition but rather cultures and social institutes.
In such an approach there is nothing left of
metaphor\index{metaphor} of mechanical
equilibrium.\index{equilibrium, market} It is the dynamical
systems\index{dynamical system} of different types that compete.
Similar systems were considered in order to describe biological
mechanisms of evolution\index{evolution} on the molecular level
where the chemically reacting flows of biological quantities are
the dynamical systems and the metastable states win as a~result of
selection, see now widely know works \cite{E}; similar problems
were earlier discussed from different position in \cite{Wdd, Th,
Kas}.

We cannot consider here biological theories in any detail but we
may make one principal deduction: if Hayek\index{Hayek F.} is
right in his opinion above the nature of market competition and
evolution\index{evolution} of social institutes then the
mathematical metaphors used in order to construct models
describing such processes should be cardinally modified. A natural
language for description of such systems is the information
theory\index{information theory} and equilibrium and
non-equilibrium statistical thermodynamics.\index{statistical
thermodynamics} Such modifications in the language of description
should naturally completely modify the theory. The models of
market equilibrium should completely change both the meaning and
the apparatus.

The metaphors\index{metaphor} of mechanical equilibrium, dynamical
system,\index{dynamical system} feedback will hardly be
applicable. It is interesting to investigate if the history of
economic thought contain some attempts on alternative
conceptualization of economic equilibrium\index{equilibrium,
market} in the early period of the development of economic theory
when the neoclassical orthodoxy was not fossilized?

Indeed, such an attempt is known, it is due to David
Hume.\index{Hume D.} Even before the fundamental work of
A.~Smith,\index{Smith A.} Hume \cite{Hu} wrote several essays on
economic problems. In one of them, named ``On Market Balance'', he
suggested an interesting metaphor of an economic equilibrium.
Discussing the problem of relation between the amount of currency
in the country and the prices Hume writes (\cite{Hu}, pp.
112--113):

``Let us suggest, for example, that during one night the amount of
currency in Great Britain will be multiplied five-fold... Will not
this raise the prices on labor and goods until neither of the
neighboring countries will be able to buy anything from us while
contrariwise their goods will become comparatively cheap so much
so that despite possible preventive laws they will saturate our
market and our money will flow out of the country until we become
equal with our neighbors in relation of money and will not lose
this access well which placed us in such an unfavorable position.
Water, wherever it penetrates, always stands at one level. Ask
physicists the cause of this phenomenon and they will answer that
if water raises at one place then the raising weight of the water
in this place without being in equilibrium makes the water lower
its level until the equilibrium is attained.''

Apparently we have here a~``mechanical'' equilibrium metaphor. But
essentially this is the same metaphor (a liquid in joint jars),
which was the initial point for the construction of thermodynamic
theory.\index{statistical thermodynamics} It is approximately in
this way that the temperature of touching bodies becomes equal.

The meaning of Hume's\index{Hume D.} metaphor\index{metaphor} is
not mechanical. He draws our attention to the equalizing of the
value of the essential parameter in two systems coming into an
interaction. The fact that the system comes to an equilibrium is
not so important. What is important is the values of the essential
parameter become equal, as soon as the systems come into
a~contact.

Let us draw a~particular attention to Hume's remark concerning
possible preventive measures which nevertheless cannot forbid the
attaining of an equilibrium. Hume assumes that the means of
attaining equilibrium (in this case the ways of foreign goods to
infiltrate the country) are so numerous that it is not worth even
discussing this question: what is important is that such means
will always be found. This attitude cardinally distinguishes
Hume's model of equilibrium from the model of equilibrium of A.
Smith\index{Smith A.} who explicitly indicates the mechanism for
gaining the equilibrium.

Hume's\index{Hume D.} idea is close to that of F.~Hayek. It is
exactly the same distinction that appears between the mechanical
metaphor of equilibrium and the thermodynamic metaphor in which
the ways of attaining equilibrium are not known and are not
important. The only important thing is that the final state is the
most probable one.

Hume's\index{Hume D.} ideas on the economic equilibrium did not
make an influence on the economic theory comparable with the
influence of A. Smith's ideas. But we see that a~certain germ of
a~new sense, namely pregnant with possibilities theoretical
metaphor contained in Hume's essay can be in principle unfolded
into a~theory that uses completely different language and has
a~different ontology, in a~theory which leads in certain cases to
principally different conclusions than the neoclassical orthodoxy.
Hayek's approach to the market as a~system in which the means of
actions are impossible to calculate and predict whereas the
``equilibrium'' or ``order'' are attained as the most probable
state are essentially quite concordant with Hume's metaphorical
conceptualization. Let us try to make one more step and construct
an economic theory that would systematically use the language of
information theory\index{information theory} and statistical
thermodynamics.\index{statistical thermodynamics}

For this, we first have to ``destruct'' the mechanical
metaphor\index{metaphor} of equilibrium.

\chapter{Metaphors of equilibrium}

The theory of economics is an example of a~non-trivial case in
which ontological assumptions that underly mathematical models are
mathematical models themselves or, to say more precisely, are
mathematical metaphors.\index{metaphor} A basic mathematical
metaphor for the classical model of the market is that of a~{\it
mechanical equilibrium}.\index{equilibrium, market}

The basic idea of such a~model is that small deviations of the
system from the point of equilibrium produce ``forces'' which try
to return the system to the equilibrium state.\footnote{Clearly,
economists have in mind the stable equilibrium. A nonstable
equilibrium is no less interesting, especially after
\index{Kapitsa P.}Kapitsa explained how to stabilize an inverted
or even slanted stick \cite{BP}. The recent results on cyclic
nature of (stock) markets indicate that this remark is, perhaps,
deeper than one might think. {\it D.L.}}

In some, very important, sense ``the invisible
hand''\index{invisible hand} of the market in this model is
equivalent to a~mechanical force. The economics is considered as
a~dynamical system.\index{dynamical system} {\it Time} stands as
a~key notion, and the mathematical structure of the economic models
is represented by a~system of differential equations.

In the mechanical models, the equilibrium is considered as a~state
at which the forces applied to the system counterbalance each
other and the potential energy attains its extremum.\footnote{It
is worth remembering, among other things, that the subject of the
modern research in mathematical economics is, as a~rule, the state
of equilibrium in itself. The dynamics of the system is only
considered as a~metaphor,\index{metaphor} pointing out the way
this equilibrium state may be attained, see \cite{Fi} and the
works of the classics: \cite{W}, \cite{D}, \cite{AD}. In general,
further evolution\index{evolution} of the theory produced such
a~state of affairs, in which the system's dynamics in the vicinity
of the equilibrium point was completely neglected by the
researchers.} Consequently, to apply the mechanical metaphor of
equilibrium in the economics, some analogs of the mechanical
notions are needed. Such a~conceptualization is not ``harmless''
at all: it implies that the system, having slightly digressed from
the state of equilibrium, will return to this very state being
left alone. As far as the dynamics of the market is concerned, the
neo-classical economics inherited the classical approach. Here lie
the roots of ideas how to revitalize the economics by means of
financial stabilization,\index{financial stabilization} the
essence of monetarist approach to vitalization of economics.
According to it, it suffices to release the prices while
preserving the volume of money for the system to immediately come
to an equilibrium.

The practice of ``shock therapies''\index{shock therapies}
illustrates that this is not always the case. Still, practice may
lead us astray. To unearth the reasons why the economic systems
refuse to come to an equilibrium\index{equilibrium, market} as
predicted, we have to deeply analyze, first of all, the
``mathematical metaphor''\index{metaphor} used. The question is
{\bf are the dynamical systems\index{dynamical system} adequate
and sufficient metaphors for description of the equilibrium of
economic systems?}

In physics, it is well-known, there are other, distinct,
approaches to the conceptualization of the intuitive notion of
equilibrium.\index{equilibrium} Our construction is based on the
thermodynamic notion of equilibrium. According to this concept,
the system gets in the state of equilibrium not because it is
being affected by ``forces'', but simply because this is the most
probable state of the system, consisting of numerous parts, each
of which is characterized by its independent dynamics.

This approach may refer as well to mechanical systems obeying the laws
of mechanics. But, if the system is very complex, its general
behavior is determined by absolutely different principles, very unlike
those of mechanics.

This distinction in the mathematical description of how the system
changes its state is fundamental. In terms of thermodynamic
approach to equilibrium, the system, instead of evolving in time,
simply changes its position in the space of macroscopic
parameters, remaining on certain surface, the {\it surface of
state},\index{surface of state} singled out by the ``equation of
state''.\index{equation of state}

Time is not included into a~set of parameters important for the
description of the system's equilibrium.\index{equilibrium, market}
The equation of state is given by (linear) dependencies between the
differentials of macroscopic parameters, in other words, by a~system
of {\it Pfaff equations}\index{Pfaff equation}\index{equation,
Pfaff}.\footnote{A {\it Pfaff equation} is a~particular linear
equation for vector fields $X$; namely, any equation of the type
$\alpha(X)=0$, where $\alpha$ is a~differential 1\defis form. For
a~modern exposition of the theory of Pfaff equations, see
\cite{BCG}. For another interpretation of the term, see footnote in
Editor's preface and \cite{BCG}. {\it D.L.}} By changing one or
several macro-parameters of the system we simply moves the system
along the surface of state.

In essence, from the mathematical point of view, the investigation
of equilibrium in such a~model is a~problem of differential
topology\index{differential topology} of the surface, described by
the equation of state \cite{B2}.

Such a~metaphor\index{metaphor} of equilibrium\index{equilibrium,
market} essentially differs from the mechanical one. Time does not
occur here as an internal parameter of the system, the parameter
that determines its dynamics, but as an external one.

Dependencies between the differentials of changeable macro-parameters
are determined by the internal structure of the system, for example,
by the energy values of subsystems.

In the beginning of the XX century,
C.~Carath\'eodory\index{Carath\'eodory C.} proved \cite{Cy} that
it is possible to logically develop the thermodynamic
theory,\index{statistical thermodynamics} drawing, exclusively, on
the assumption that ``the equation of state'', i.e., the surface
in space of thermodynamic variables, corresponding to a~system of
Pfaff equations, exists. In this case, the second law
of\index{second law of thermodynamics} thermodynamics is
formulated as existence, in an infinitesimal neighborhood of each
state of the system, of such states that cannot be reached without
the change of entropy.\index{entropy}

In other words, a~simple assumption that the differentials of
``generalized positions'' are constrained (and this constraint is
nonintegrable, in Hertz's words, {\it nonholonomic}), appears to
be sufficient to develop the system of thermodynamic equations.
Accordingly, the idea of equilibrium will look completely
differently. This thermodynamic equilibrium, unlike that from
a~mechanical metaphor, would mean not the existence of a~singular
point of a~solution of a~system of differential equations or an
extremum of a~potential function, but movement along the surface
of state.

In what follows I will show that there are most serious grounds to
believe that the Pfaff equations (and, consequently, thermodynamic
metaphor of equilibrium) are often more adequate for the
description of various economic phenomena than the mechanical
metaphor.

Under some additional assumptions on the thermodynamic system
described by the Pfaff equations\index{Pfaff
equation}\index{equation, Pfaff} the {\it Le Chatelieu
principle}\index{principle, Le Chatelieu} \cite{LL} is applicable.
Namely, the system demonstrates a~behavior obstructing influences
exercised on it in the result of changes of the external
macro-parameters.

The systems in economics also demonstrate such a~behavior under
certain conditions. For example, the increase of prices can be an
incentive for the production in order to support the consumption
level; attempts to impose total control over levels of production or
consumption trigger the process of corruption of executive bodies,
which diminish the effect of such control, and so on.

An important discovery of the past was that the phenomena of such
homeostasis\index{homeostasis} in physics, sometimes appearing as
almost reasonable behavior, can be explained on the basis of the
thermodynamic metaphor,\index{metaphor} proceeding from the very
simple assumption, namely, that
\begin{thetext}
{\sl for the most time the system is staying in the most probable
state.}
\end{thetext}

In case of economics, it may look as if the system is directed,
citing A.~Smith,\index{Smith A.} by an ``invisible
hand''.\index{invisible hand} At the time when A.~Smith was
working on his book, the principles of
thermodynamics\index{thermodynamics} were not known yet, so the
``invisible hand'' was interpreted in terms at hand, those of
a~mechanical metaphor. A.~Smith's conceptual model that identifies
human interests\index{interest} with ``forces'' of the market
gives really serious grounds for such an interpretation.

Let us consider a~thought experiment of A.~Smith a~bit closer.
Smith assumes \cite{Sm1} that an increase of commodity prices
brings about an increase in at least one of the price-making
components~--- the rent, workers' salaries or profit
--- giving a~signal to actors, or rather forcing them, to change
their behavior (to exploit more facilities, increase the offer of
jobs or to expand manufacturing), and this change of behavior
finally leads to the reduction of price.

There is, however, one very weak spot in the \lq\lq thought
experiment'' of A.~Smith.
\begin{equation}
\label{eq1}
\boxed{\begin{array}{l} \text{{\it The only immediately accessible
information for}}\\
\text{{\it  the market actors is the price}.}\end{array} }
\end{equation}
Whatever changes it~--- up or down~---
the reasons for this change are not immediately revealed to the
observer. It is most often not clear to the buyer at all if the
changes in the price were due to change-making factors (say, raise or
fall of profit) or the reason was in increase of demand? Usually,
such information is the seller's most guarded secret.

In the studies of researchers from the Austrian
school\index{Austrian school} we find serious arguments in favor
of the hypothesis that
\begin{equation}
\label{eq2}
\boxed{\begin{array}{l} \text{{\it no complete
information on the hidden components}}\\
\text{{\it  of the price is available for the market actors}.
}\end{array} }
\end{equation}
This, consequently, means that they simply cannot behave in the
way, described in the A.~Smith's\index{Smith A.} ``thought
experiment''. In blunt terms, this ``thought experiment'' was
based on false assumptions. This necessitates to put under doubt
the mathematical metaphor\index{metaphor} underlying the classical
concepts of the market dynamics, i.e., the mechanical metaphor.

In other words, the ontology of the classical model of economics is,
indeed, not indisputable.

F.~Hayek\index{Hayek F.} \cite{H} despite of his vigorous
diatribes of socialist ideas of regulation of economics and
adherence to market principles, negated liberal viewpoints on the
role of selfishness in market economy. Hayek directly asserts that
the market economy is based on the observance of moral principles,
ensuring survival of the community in competition with the other
communities, and that these principles not infrequently directly
contradict mercantile egoistic interests.\index{interest}

M.~Weber,\index{Weber M.} in his famous work about the role of
protestant ethics in genesis of capitalist economy, also developed
similar views. Weber shoed that the protestant
ethics\index{protestant ethics}\index{ethics, protestant} (the
scrupulous honesty and workaholism in the frames of the capitalism
formed in the Northern Europe) are necessary conditions for
richness growth.
\begin{thetext}
{\sl The interests of the members of society are being fulfilled
{\bf as a~result} of abiding the moral laws in the society as
a~whole.}
\end{thetext}

This directly contradicts the idea of A.~Smith\index{Smith A.}
that
\begin{thetext}
{\it the common well-being is a~resulting corollary of individuals
attempting to fulfill their egoistic interests.}
\end{thetext}

In neo-classical models of the economic
equilibrium\index{equilibrium, market} constructed later, such,
for example, as the ones due to Arrow-Debreu-McKenzie,\index{Arrow
K.}\index{Debreu G.}\index{McKenzie L.} in order to prove the
existence of equilibrium, one assumes that the subjects of
economic activity maximize their utility\index{utility} functions.
Thus, the arguments on in-built incompleteness of the information
available for the market actors were ignored.

This postulate~--- about unavoidable incompleteness of the information
--- seems to be one of the basic reasons why the economists of the
Austrian school\index{Austrian school} rejected applicability of
mathematical methods in economics. It looks as though these
scientists were not so much dissatisfied with the mathematics
itself, but rather with the mechanical metaphor\index{metaphor}
used to construct models of equilibrium~--- the metaphor in which,
in the end, the utility played the same role as the potential in
the classical dynamical systems.\index{dynamical system}

The dissatisfaction with such models of equilibrium was, however,
discernible not only on the part of the opponents of mathematical
economics, but also among its champions. The latter were anxious
with the absence, within the framework of the mechanical metaphor,
of satisfactory stability theory of the market equilibrium. Most
profound disappointments were connected with impossibility to
consider, in all cases, the excess demand as continuous function
of the price. Controversial examples are well known, see, for
example, the work by B.~Arthur\index{Arthur B.} \cite{AAP}.

One can, of course, try to ``improve'' the theory, remaining in
the confines of the mechanical metaphor and liberal dogma. We
believe, however, that although the arguments of the Austrian
School\index{Austrian school} of economics are insufficient to
completely reject possible applicability of mathematics in
economics, they suffice to be the reason to change the
mathematical metaphor of the equilibrium.

F.~Hayek's\index{Hayek F.} concept of market, as the process of
discovery, emphasizes the key role of {\it information} in the
market economy (as opposed to the priority of unobservable
utility\index{utility} functions in neo-classical models, based on
the mechanical metaphor\index{metaphor} of an
equilibrium).\index{equilibrium, market} In what follows we
develop a~model of market equilibrium proceeding from the
information theory:\index{information theory}
Brillouin\index{Brillouin L.} showed \cite{Bri} that the
mathematical information theory is, in essence, identical to
thermodynamics, if we identify {\it information} with {\it
entropy}.\index{entropy}

The quantity of information, obtained at the instance of
interaction of the subject with the system, is measured in the
information theory by the logarithm of the {\it relative reduction
of opportunities for choice enjoyed by the subject before the
information had been obtained}. It is easy to understand that such
an interpretation of the information theory directly links it to
the behaviorial description of the market actors, thus making
entropy the most important parameter of the market interactions.

This idea is not new. At the end of 1960s, A.~ Wilson\index{Wilson
A. G.} suggested that the entropy analysis should be used for the
examination of transport flows \cite{WE}. Wilson mainly used the
method of the maximization of entropy to tackle transportation
problems.

In his three papers published in the Russian journal {\it
Avtomatika i Telemekhanika} in 1973\footnote{Cover-to-cover
translated starting 1978 as {\it Automat. Remote Control.}} (nos.
5, 6, 8), L.~Rozonoer\index{Rozonoer L.} formulated very general
principles of \lq\lq resource dynamics". Unfortunately, I was
unaware of these papers while writing this book and became
acquainted with them only a~year after the publication of the
Russian version. The basic idea of Rozonoer was to suggest a~theory
that could unite on most general level the economic and
thermodynamic theory.\index{statistical thermodynamics}
On a~deeper level, his aim was to formulate a~theory in which the
exchange of resources and the idea of equilibrium can be expressed
and to extract the most general consequences from such a~theory in
the manner of the Second Law of Thermodynamics and Le Chatelieu
principle. \index{principle, Le Chatelieu} For such a~purpose,
Rozonoer introduced two specific functions: the \lq\lq function of
effect" and a~\lq\lq structural function", the first being an
analog of entropy,\index{entropy} the second an analog of
equilibrium entropy.\index{equilibrium, market} The definitions
for these functions were suggested on the most abstract level, and
the state of equilibrium was defined as a~condition when the
structural function reaches its maxima for subsystems of a~general
system, other conditions being the additive property of resources,
conservation principle and the possibility of measuring resources
in terms of non-negative numbers. In all his three papers,
Rozonoer argued that economics with traditional concept of
utility\index{utility} and thermodynamics with classical
definition of entropy (in Carath\'eodory style) are two basic
examples of resource dynamics. Rozonoer assumed no specification
for any concrete economic or physical system. Such an approach is
reasonable because Rozonoer wanted to formulate a~theory with
maximum abstractness and the most wide range of application.

Rozonoer succeeded in formulating a~number of rather general
theorems, implementing the idea of \lq\lq basic exchange resource"
(an analog of gold or money in economics or energy in
physical\index{statistical thermodynamics} thermodynamics).
Rozonoer showed an analogy between pressure divided by temperature
and \lq\lq value of resource" divided by a~\lq\lq basic relation
of effect", Rozonoer's term for an analog of temperature in his
theory. Rozonoer obtained an impressive long list of analogies
between economic theory and thermodynamics (Avtomatika i
telemekhanika, 1973, no.~6, pp. 75-76).

What is interesting for our subject is to try understand why in an
excellent series of papers devoted to \lq\lq resource dynamics"
(Rozonoer's term) Rozonoer did not suggest any methods to analyze
concrete economic systems along his lines of thought. It is
interesting to wonder why there was no response to his ideas from
the economics community in general.

To my mind, Rozonoer considered similarities between economics and
thermodynamics not as direct analogies allowing one to construct
a~new theory of economic equilibrium, but as a~\lq\lq
metaphor"\index{metaphor} or, even less stringently, as an
analogy, because he was not ready to adopt the idea that a~\lq\lq
narrow", statistical definition of entropy \index{entropy}could be
used instead of the vague concept of \lq\lq utility". As a~result
his {\it resource dynamics} was too \lq\lq abstract", and this
made it practically useless for the description of economic
systems. Rozonoer incorporated into his theory only the theorems
which basically were known by scholars in economics, and could be
formulated within the language elaborated on the basis of General
Equilibrium Theory,\index{General Equilibrium Theory} possibly in
a more elegant if not more complicated way.

What was really important~--- insisting that {\it utility} must be
considered as entropy in models formulated using a~specific
conceptual language~--- was not, unfortunately, accomplished.
Using the language of {\it resource dynamics} it was possible to
make abstract statements about economic systems, but it was
difficult to derive an equation of state for a~given economic
system, or to analyze for such a~given system the role of
institutional constraints and these tasks were not performed.

 The use of entropy\index{entropy} approach is
also well-known in the economic studies in relation to the
estimation of uncomplete data on the basis of the ideas by E.~T.~
Jaynes\index{Jaynes E. T.} \cite{J}. For details, see \cite{Le},
\cite{GJM}. These works do not include, however, analysis of
economic equilibrium.\index{equilibrium, market} On the precedents
of application of other methods from statistical physics for
socio-economic studies, see, for instance, \cite{Dur1, Dur2}.

Nevertheless, until now, the idea to apply the notion of
entropy\index{entropy} to the study of economics was only realized
for particular purposes (mainly, in relation to the transportation
problems, and was not used with the aim to build up the whole
theory of economic equilibrium).

The entropy alone is not sufficient to create the thermodynamic
theory\index{statistical thermodynamics} of economic
equilibrium.\index{equilibrium, market} To this end, we need the
whole spectrum of thermodynamic variables, such as temperature,
pressure, chemical potential, free energy, and so on.

The basic idea of thermodynamic approach to the analysis of
economic equilibrium is as follows. If the system is described on
two ontological levels~--- a~``macroscopic'' and a~``microscopic''
ones~--- and one macroscopic state is characterized by a~multitude
of microscopic states (their number is called the {\it statistical
weight} of the macroscopic state), and the system will, generally,
remain in the most probable state, i.e., in the state with the
greatest statistical weight.

The conditions of applicability of the thermodynamic approach can
be formulated in very general terms. It becomes clear thereby,
that the applicability of thermodynamics goes beyond the realm of
physical systems.

One can imagine The Large System that can be decomposed, or rather
consists of, a~huge number of Small Systems, each with its independent
dynamics. We assume that the state of the Large System, and its
relatively large parts, is described by a~certain number of
macro-parameters, which are additive, i.e., {\it the total values of
the macro-parameters appear to be the sums of values of the same
macro-parameters of parts}.

In physics, {\it energy} is an example of such a~parameter if we
neglect superficial interaction between the parts of the Large
System. This is a~basis for application of thermodynamic methods
in physics. Let further each of the systems considered be
additionally characterized by a~set of micro-parameters which can
have distinct values at the same value of the fixed
macro-parameter. Their values are determined by the system's
dynamics and, generally, can be of interest in relation with our
task in one aspect only. Namely, having fixed them, an exact state
of the system becomes known and one can answer, how many various
micro-states correspond to one macro-state of the system.

At this stage, we can introduce the notion of statistical weight,
as a~number of micro-states corresponding to one macro-state, and
the notion of entropy,\index{entropy} as the measure of
uncertainty of the macro-state of the system, which is a~function
of the number of micro-states.

If the system is such that the micro-states of the parts of the Large
System are statistically independent, it is possible to compute the
statistical weight of the Large System as a~whole, given the
statistical weights of all Small systems. For that purpose, it
suffices to multiply the statistical weights of Small Parts.

If we wish the entropy be additive, we may regard it as the logarithm
of the statistical weight. Since $\ln(ab)=\ln a+ \ln b$, it follows
that the uncertainty, or entropy, is additive.

Assuming that nothing is known about the dynamics of the system,
except that it is very complex, the natural assumption for the
probability value of any macro-parameter is that it is proportional to
the number of the appropriate micro-states, i.e., to the statistical
weight. Since the logarithm is a~monotonous function, the most
probable state, i.e., the state with the greatest statistical weight,
is, at the same time, the state with the greatest entropy, i.e., with
the greatest extent of uncertainty.

This is the core of the {\it second law of
thermodynamics}\index{second law of thermodynamics}~--- {\it the
entropy tends to increase}\index{entropy}~--- as every part of the
Large System goes in the most probable state during interactions
with the other parts.

If we ``isolate'' some part of the Large System in order to
observe the distribution of the probable states, we have to
consider the rest of the Large System as a~{\it thermostat}, that
is a~reservoir that ensures equilibrium\index{equilibrium, market}
(in a~thermodynamic sense) of the distribution of states within
the subsystem isolated.

To find out the form of this distribution, it is necessary to
introduce the notion of ``temperature'', equal to the inverse of
the derivative of entropy of the Large System with respect to the
macro-parameter for which there exist a~conservation law.

The introduction of the parameter of equilibrium~--- temperature~---
is simply a~result of the condition that {\it there are no flows of
the conserved macro-parameter between the parts of the system}. If
there are several conserved macro-parameters, then there are as many
parameters of equilibrium as there are conservation laws.

Observe that we said nothing related to either physics, or physical
laws and observables. All the arguments are applicable to Large
Systems of any nature, subject to the above hypotheses. These
arguments look rather natural, in relation to the large economic
systems, if we regard the total income, the total value of products or
the total value of consumption of goods as macro-parameters, whereas
distribution of income and products or consumption of goods between
the subjects of economic activity are viewed as micro-parameters.

In this approach, to the study of economic equilibrium, we avoid
the necessity to explore the subsystem's dynamics, so the
knowledge of institutional restrictions\index{institutional
restrictions} on the goods production and distribution suffices.
Thus,
\begin{thetext}
{\sl we offer not only a~new approach, appropriate to describe the
economic equilibrium, but, also, the instruments to investigate
the impact of institutional restrictions on the state of
equilibrium. }
\end{thetext}

Namely, we get an opportunity to build up a~mathematical apparatus
for analysis of the transaction costs theory, to suggest
quantitative methods of the study of the impact of the
informational asymmetry on the behavior of the market's
agents\index{agent} and to tackle many other problems as well.

Observe that, for more than a~century, the endeavors to prove
thermodynamic predictions within the framework of theoretical
physics by analyzing equations of motion were not successful.
Born\index{Born M.} \cite{B1} used to remark a~radical distinction
of the methods and mathematical techniques of thermodynamics from
those of other branches of theoretical physics: ``In the classical
physics the logical processing of a~branch of science is
considered finalized when it is reduced to one of the chapters of
the ``normal mathematics''. There is one astounding exception~---
the\index{statistical thermodynamics} thermodynamics. The methods
usually applied in this discipline to deduce the main postulates
markedly differ from those accepted in other domains of physics.''

Nobody doubted that the principles of thermodynamics work
regardless of possible reductionist interpretations. The two
levels of description pose the problem: how to single out certain
parameters, perhaps, completely ``inconspicuous'' or not obvious,
that govern the conditions for equilibrium, intuitively understood
as the absence of significant flows between the parts of the
system.

If the macro-parameters are functionally dependent, and the
surface of state is differentiable, then the differentials of the
macro-parameters produce are related by a~system of Pfaff
equations.\index{Pfaff equation}\index{equation, Pfaff}

Thus, the idea of thermodynamic equilibrium\index{statistical
thermodynamics} is quite appropriate for description of economic
systems. They have observable flows of money, goods and people,
and, like physical systems, have two levels of description.

Consequently, the description of the economic system should be
equivalent to the description of physical systems in
thermodynamics, though the parameters of equilibrium~---
``temperature'', ``pressure'', ``chemical potential''
--- will certainly have quite different interpretations, the ones
that mirror the peculiarities of economic systems.

Observe that thermodynamic terms had often been already used, by
folk, not scientists, in relation to the economic systems in
a~``naive'' manner, as metaphors\index{metaphor} of description: the
stock exchange is ``overheated'', the national economic problems
``boiled over'' or ``cooled down'', stock exchange indices are
associated with temperature degrees of a~thermometer,\index{thermometer} and so on.

One of the aims of this work is, besides all, to show that, not
rarely, there is more sense in the ``naive'' metaphors of such
sort, than in the complex mathematical models based on mechanical
metaphor of equilibrium. Proceeding from the thermodynamic
metaphor, a~thermodynamic theory\index{statistical thermodynamics}
of economics can be developed. It not only catches hold on certain
realities of markets by no means less than the one built on the
mechanical metaphor, but also accounts for the role of
institutional restrictions for the establishment of economic
equilibrium~--- the task unfeasible to the theories based on the
mechanical metaphor.

\chapter{Entropy and temperature in models of economics}

\section{Entropy and temperature}

We begin our description of a~thermodynamic
model\index{statistical thermodynamics} of economics with the
simplest example. Let an economic system consist of $N$
agents\index{agent}, among whom the income, constant for the
system as a~whole, is distributed. We assume that there are many
ways to distribute income, and we are incapable to foresee all
possible alternatives. This assumption fully corresponds to
Hayek's\index{Hayek F.} concept of market in which the market is
described as an arena of discoveries of new procedures and
operations. For ultimate simplicity, we assume that the income is
quantum, i.e., presented in integers (which is natural, as the
smallest unit of currency is operational in economics).

Now, for each value of the total income $E$ it is possible to find
the quantity of modes of income distribution between the
agents,\index{agent} as a~characteristic $n(E, N)$ of this value,
called the {\it statistical weight of the state}\index{statistical
weight} with income $E$.

At this stage we can introduce the concept of equilibrium. The
idea is that two systems (under the above hypothesis) are in {\it
equilibrium},\index{equilibrium, market} if the distribution
function of income does not change when they enter in a~contact,
hence, there is no income flow between the systems. By a~``contact''
we understand here an ``open list'' of possible modes
of redistribution. It is remarkable that it is possible to
calculate the statistical weight\index{statistical weight}
regardless of uncountable variety of various institutional
limitations imposed on the agents'\index{agent} incomes, so
functions $n(E, N)$ may be different for different systems.

The given model is rather simple: at this phase of reasoning we do not
really turn our face to the market. The restrictions on income may be
sustained coercively, but the actual means are irrelevant for our
investigation.

If two systems interact, one with the total income $E_1$ and the
number of agents $N_1$, and another one, with $E_2$ and $N_2$,
respectively, then the total system is characterized by the total
income $E_1 + E_2$ and the number of agents equal to $N_1 + N_2$. How
to find conditions necessary or sufficient for the equilibrium, i.e.,
for the state with no income flow between the systems?

In order to build up an appropriate theory, we have to adopt one more,
extremely important, hypothesis on the nature of the systems under
investigation. Namely, we assume that {\it all elementary states of
income distribution have the same probability}.

The main ground for such an assumption is symmetry of states. As in
the probability theory and in statistics, we assume equal probability
of elementary events just because there are no grounds to prefer one
event to another. Thus, it is of utmost importance for the theory, to
include all probable states of distribution. The change of function
$n(E, N)$ will, certainly, change the results obtained in this model.

In statistical physics, the literature devoted to justification
the principle on equal probability of elementary states is
uncountable. Still, for the majority of models used,
\begin{thetext}
{\sl this principle remains a~subject of faith, the principle that
nobody was able to prove}
\end{thetext}
\noindent and we rely on it because it is justified by the
remarkable effectiveness of the statistical theory and brilliant
agreement of the theory with the practice.

To find the state of equilibrium,\index{equilibrium, market} is to
determine conditions under which the income is not to be
redistributed between the interacting systems. For this purpose,
consider a~redistribution of income during an interaction. Let
a~certain part of income, $\Delta E$, pass from system 1 to system
2. Then, the states of the systems change and their statistical
weights become equal to $n_1 (E_1 - \Delta E, N_1)$ and $n_2 (E_2
+ \Delta E, N_2)$, respectively.

The principle of equal probability\index{principle, of equal
probability} implies that the most probable state of the
integrated system is the one with the greatest statistical
weight.\index{statistical weight} So we should seek the maximum of
function $n_{tot}(E_1, E_2, N_1, N_2)$, with the proviso that the
total income $E_1 + E_2$ is a~constant. If no transfer of
agents\index{agent} from one system to another is possible, then
the statistical weight of the integrated system is equal to:
\begin{equation}
\label{eq3}
n_{tot} (E_1, E_2, N_1, N_2) = n_1 (E_1, N_1) n_2 (E_2, N_2).
\end{equation}
Since $E_1 + E_2=\text{const}$, it follows that $\Delta E_1 =-\Delta E_2$.

Instead of seeking the maximum of $n_{tot}$, we can seek the maximum
of $\ln n_{tot}$, because the logarithm is a~monotonous function.
>From $\ln n_{tot} = \ln n_1 + \ln n_2$ we derive the condition on the
maximum. It is very simple:
\begin{equation}
\label{eq4}
\pderf{\ln
 n_{1}(E_1, N_1)}{E_1}=-\pderf{\ln n_{2}(E-E_1, N_2)}{E_2}
\end{equation}
or, as $dE_1 = - dE_2$, we have
\begin{equation}
\label{eq5}
\pderf{\ln n_{1}(E_1, N_1)}{E_1}=\pderf{\ln n_{2}(E_2, N_2)}{E_2}
\end{equation}
So, two systems are in equilibrium, if they are characterized by the
same value of parameter $\pderf{\ln n(E, N)}{E}$.

In thermodynamics,\index{statistical thermodynamics} the logarithm
of the statistical weight\index{statistical weight}is called the
{\it entropy}\index{entropy} (of the system), and its derivative
on energy is the inverse temperature,
\begin{equation}
\label{eq6}
\pderf{\ln n(E, N)}{E}=\frac1T.
\end{equation}
{\it In order to reach the state of equilibrium, the interacting
systems should be at the same temperature. }

The italicized statement above, together with its deduction, can be
found in any textbook on statistical thermodynamics \cite{Ki}. Let us
analyze here the adequacy of such an approach to the economic systems,
at least, under the above assumptions. The economic system is in the
state of equilibrium if it is rather homogeneous and there are no
income flows from one of its part to another. We suppose, certainly,
that the homogeneity is kept only unless there is no division into
parts so tiny that significant income flows are observed.

The same postulates are available in statistical physics.
Subdivision of the system into too small parts results in
significant fluctuations. In physics, the question of an
equilibrium\index{equilibrium, market} of small parts of the
system is solved by assuming (and this assumption is equivalent to
a postulate of equal probability of elementary states), that if we
observe a~small part of system for ``sufficiently long'' time,
then we will be able to adequately describe the distribution of
probabilities of its state. It is one of the formulations of the
so-called {\it ergodic hypothesis}, see \cite{To}.

A similar hypothesis can be made for the economic systems. We see
that in our thermodynamic\index{statistical thermodynamics} model,
there are two extremely important characteristics~--- the
entropy\index{entropy} and temperature. If we do not know these
parameters, we can not correctly infer the conditions for the
system's state of equilibrium: as the system is in the state of
equilibrium only when its subsystems have an identical
temperature, and the temperature cannot be calculated without
knowledge of entropy.

In physics, to measure temperature, one uses
thermometers.\index{thermometer} These are special devices whose
equations of state are known and calibrated; so by introducing the
thermometer into a~contact with a~body we may find the temperature
of the body by a~change of the state of the thermometer. As we
will see\footnote{See the book \cite{SLR}.}, (stock) markets may,
to an extent, be considered as thermometers in economics.

The basic possible objection against the thermodynamic approach to
economy is that the number of ``particles'' involved (in the given
example~--- the number of the market agents)\index{agent} is much
less as compared with numbers of particles in the usually
considered physical systems. In the physical systems, the number
of particles is comparable, as a~rule, with the Avogadro number,
whereas in the economic systems it is usually is $\sim 10^3 -
10^8$.

In statistics, the order of dispersion is equal to
$\frac{1}{\sqrt{N}}$, where $N$ is the number of particles in the
system. Thus, in physical problems, if the statistical errors related
with the fact that the number of particles is finite are completely
insignificant due to the largeness of their number, in economic
problems we should expect much larger errors, $\sim 3$ \%, or less.
Such errors do not look too large ones, actually, taking into account
the extreme roughness of economic models.

One of the tendencies in contemporary physics is to apply
thermodynamic\index{statistical thermodynamics} approach to
systems with a~rather small number of particles $\sim 10^3 - 10^8$
(nuclear physics, cluster physics, and so on\footnote{This idea
was suggested by J.~Frenkel, see \cite{F}}.), and the results
appear to be quite valid not only qualitatively, but also
quantitatively.

It seems that both models of interaction and the data on
interactions are far from being accurate, and to strive for a~better
accuracy is meaningless: there are in-built limits of
accuracy, like the uncertainty principle.\index{principle,
uncertainty} The situation with applicability of thermodynamic
approach to economics seems to be similar.

An important remark. The statistical models in physics show that
the energy of the system has a~peculiar quality responsible for a~success
of the thermodynamic theory: if the system has some
parameter of inhomogeneity, then, provided there are sufficiently
many particles, the system has a~very sharp maximum of
entropy\index{entropy} attained in the limit as the parameter of
inhomogeneity tends to zero. Thus, not only the maximally
homogeneous state is most probable, but even small deviations from
it are {\it most improbable}.

To illustrate this thesis, let us analyze the entropy of one very
simple system. Suppose that the income distribution between the
subjects of economic activity is arranged as follows: each subject has
either zero income, or a~fixed income, $A$ assuming that the system is
organized in such a~way that all deviations are annihilated through
special institutional redistribution mechanism.

Such an example is not so much unreal, if we recall the economic
experience of some countries aspiring to implement various
``leveling principles'' in distribution of income, stipulating that
a~part of population is totally excluded from economic activities,
being allowed to have a~very low level of income (subsidized by
a~social security or by Nature). If the total number of economic
subjects is $N$, the number of the subjects with income $A$ is $L$
(hence, the size of the total income is $E_{tot}=LA$), then,
cleanly, number of probable states of this system would be equal to
$\binom{N}{L}$.

By comparing the statistical weights\index{statistical weight} of
the systems with different levels of the total income, we conclude
that the biggest statistical weight is the characteristic of the
state with $E = LA$, where $L = \displaystyle\frac{N}{2}$ (to
simplify arguments, let $N$ be even). Introducing a~parameter of
inhomogeneity, $m = \left|L - \displaystyle\frac{N}{2}\right|$, we
apply the Stirling formula for the factorial. We obtain the
following simple approximation to the dependence of the
statistical weight on the parameter of inhomogeneity:
\begin{equation}
\label{eq7}
n(N, m)\simeq 2^N\sqrt{\frac{2}{\pi
N}}\exp\left(-\frac{2m^2}{N}\right).
\end{equation}
Formula (\ref{eq7}) shows that the statistical weight (and,
consequently, the entropy)\index{entropy} has an extremely sharp
maximum depending on the parameter of inhomogeneity, with width
$\frac{1}{\sqrt{N}}$.

The principle of equal probability\index{principle, of equal
probability} of elementary states immediately implies that the
entropy\index{entropy} of interacting systems tends to increase.
The most probable state is the state with the greatest statistical
weight, i.e., the state with maximum entropy. As the number of
probable system's states sharply decreases with the increase of
the parameter of inhomogeneity, it is hardly possible to find the
system in a~state for which the parameter of inhomogeneity exceeds
a certain value determined by the number of particles in the
system (in the example above, this value is $\frac{1}{\sqrt{N}}$).

\section{Thermostat and function of income
distribution}\index{thermostat}

Now we will investigate how the income is distributed among the
subjects of economic activity. Within the framework of
thermodynamic\index{statistical thermodynamics} model, it appears
that there exists a~universal function of distribution, whose
configuration, if number of the subjects is a~constant, depends
only on the temperature. This situation is well known in
statistical thermodynamics, where such distribution is called as
the {\it Boltzmann distribution}.\index{Boltzmann
distribution}\index{distribution, Boltzmann}

To examine this distribution, imagine that the economic system $X$ is
very large, and, out of it, a~small part, $Y$, is separated.
Naturally, the system $X$ is considered as a~reservoir. If we
consider a~certain state of the small subsystem with income $E_1$, the
probability for the small subsystem to have income $E_1$ is
proportional to the number of possible states of the reservoir with
the range of income $E - E_1$, under the hypothesis that the total
income of the system is a~constant. Then the ratio of probability
$P(E_1)$ for the subsystem to have income $E_1$, to probability to
have income $E_2$, is equal to the corresponding probability ratio for
the reservoir:
\begin{equation}
\label{eq8}
\frac{P(E_1)}{P(E_2)}=\frac{n(E-E_1, N)}{n(E-E_2, N)}.
\end{equation}
Since $n = e^{S(E, N)}$, where $S(E, N)$ denotes the
entropy\index{entropy} of the system, we can express (3.1) as:
\begin{equation}
\label{eq9}
\frac{P(E_1)}{P(E_2)}= e^{S(E-E_1, N)-S(E-E_2, N)}=e^{\Delta S}.
\end{equation}
If the reservoir is far larger than the subsystem in question, we can
make expansion of $\Delta S$ into Taylor series in $\Delta E$, thus obtaining
in the first order $\Delta S\simeq\frac{E_2-E_1}{T}$, where $T$ is the
temperature of the reservoir, see (\ref{eq6})

It follows that
\begin{equation}
\label{eq10}
\frac{P(E_1)}{P(E_2)}= e^{-\frac{E_1-E_2}{T}}.
\end{equation}
This is an approximate formula for the probability distribution for a~small system.

If the number of probable states of subsystem $Y$ with incomes $E_1$
and $E_2$ is equal to $n_Y (E_1)$ and $n_Y (E_2)$, respectively, then
the distribution formula takes the form
\begin{equation}
\label{eq11}
\frac{W_Y (E_1)}{W_Y (E_2)}=\frac{n_Y (E_1)}{n_Y
(E_2)}e^{-\frac{E_1-E_2}{T}}.
\end{equation}
In this formula we proceed from the probability of the state, $P$, to
the probability of the level of income, $W$. Observe that no
assumptions whatsoever were made, except the two: the system is
homogeneous, so it is possible to subdivide it into interacting parts
without producing income flows from one part to another, and the total
system's income is a~constant.

The arguments laid out above, are, actually, a~replica of the
traditional deduction of Boltzmann distribution in statistical
thermodynamics.\index{statistical thermodynamics}

Hereby we got something more than just ``thought experiment'' for
the verification of the logic of reasoning. When the appropriate
data are at hand, it is possible to verify this thesis with real
economies, comparing, for example, the income distribution
function of various sectors of economics.

The Boltzmann distribution function is realizable, with respect to
the income, in any system which interacts with the reservoir kept
at certain temperature. In the isolated system, on the contrary,
the temperature depends on both income and entropy.\index{entropy}

The Boltzmann distribution of income allows one to understand the
relationship between the average income of the subjects inside the
system, and the system's temperature. Suppose that there are no
restrictions on the agent's\index{agent} income and all levels of
income are allowed. Such a~situation is commonly associated with
the market economy. Then, with the help of the Boltzmann
distribution function, we easily calculate the average income of
the agent:\index{agent}
\begin{equation}
\label{eq12}
\bar E= \frac{\int\limits_0^{\infty}
Ee^{-E/T}dE}{\int\limits_0^{\infty}e^{-E/T}dE}=T.
\end{equation}
We see that in absence of any restrictions on the agent's income,
the mean income is equal to the temperature. The upper limit of
integration here is equal to $\infty$, despite the fact that even
the total income of the real system is bounded. However, since the
exponent steeply decreases, this does not matter, provided the
total income greatly surpasses the average income of a~single
agent;\index{agent} in the real systems this is true, of course.

As we will see, under restrictions on income the relationship between
the temperature and the average income may look quite different.

In conclusion of this section, observe that, like in statistical
thermodynamics,\index{statistical thermodynamics} in the
thermodynamic model of economics, the parameter called
``statistical sum'',\index{statistical sum}
\begin{equation}
\label{eq13}
Z_0=\mathop{\sum}\limits_E n(E, N)e^{-E/T}
\end{equation}
is extremely useful. The sum here runs over all possible values of
$E$. The probability of the system interacting with
thermostat\index{thermostat} with temperature $T$ can be expressed
as
\begin{equation}
\label{eq14}
p(E)=\frac{e^{-E/T}}{Z_0}.
\end{equation}
With a~fixed number of agents,\index{agent} the average income in
the system at temperature $T$ is given by the formula:
\begin{equation}
\label{eq15}
\bar E=\frac{\sum\limits_E En(E, N)e^{-E/T}}{Z_0}=T^2\pderf{\ln Z}{T}.
\end{equation}
This means that if we know how the statistical
sum\index{statistical sum} depends on the temperature, it is
possible to obtain the value of the average income by
differentiation. Again, this is fully agreeable with the standard
technique of statistical thermodynamics.

Consider now one paradoxical example. It demonstrates that
application of the thermodynamic approach to economic systems
helps to deduce unexpected, though true, conclusions.

\section[Systems with and without
restrictions on income]{On interaction of systems with and without
restrictions on income}

In this section the ``spin model''\index{spin model}\index{model,
spin, see spin model} with two possible values of income, 0 and
$A$, already addressed above, will be examined in more detail.
Despite its seemingly too abstract nature, this model is rather
useful, as it catches some important features of systems with
restrictions on income.

What the results would be once such a~spin system enters into
interaction with a~``free'' market system, the one without any
restrictions on income? In the free market system, the
entropy\index{entropy} increases together with the increase of
energy, and the temperature is always positive. This is not always
the case for the system with restrictions on income. Actually, if
the number of agents\index{agent} $L$ with non-zero income
surpasses the half of a~total number of agents, $N$, the number
$\binom{N}{L}$ of probable states of the system, starts to
diminish, and the inverse temperature given by formula (\ref{eq6})
becomes negative.

\medskip

What is the meaning of the {\it negative temperature}?

\medskip

This phenomenon was investigated quite well in the laser theory
\cite{Kl}. The negative temperature implies the existence of
``inverted population'', i.e., a~situation, in which the levels
with higher energy are more densely populated than the ones with
lower energy.\footnote{This means that the system with negative
temperature is ``overheated". Such a~system is ready to throw out
a portion of its energy at any contact with a~``normal" system in
equilibrium.} In such a~state, the system is imminently ready to
release its energy at the contact with any system with positive
temperature.

How can this be interpreted in terms of the income distribution
described above? If two systems with positive temperature are in
contact, the redistribution of income goes from the system with
higher temperature to the one with the lower one. As we saw, for
the free market, the temperature is equal to the average income
per agent.\index{agent} The redistribution of income from the
``richer'' system to the ``poorer'' one takes place in accordance
with our intuitive understanding.

Contrariwise, if a~free market system interacts with a~system with
restrictions on income, then, under certain conditions, a~counter-intuitive
process is observed, when redistribution the
income goes from the ``poorer'' system to the ``richer'' one.

Let $X$ be a~free market system, with the temperature $T_X$ equal
to the mean income per market agent. Let $Y$ be a~spin model of a~market
system\index{spin model} with income bounded from above by
$A$ and with the total number of its agents $N$. Let $T_X>NA$ and
let the number of $Y$'s subjects with non-zero income exceed the
half of their total number, so the temperature of $Y$ is,
evidently, negative.

What is going to happen when these systems start to interact? The
entropy\index{entropy} of the joint system will increase, and the
entropy of system $Y$ {\bf will also increase} simultaneously,
thereby precipitating disorder of system $Y$. Therefore, $Y$ {\bf
must} transfer a~part of its income to $X$ despite the fact that
average income of the system $X$ is already higher than the
average income in $Y$.

As a~result, system $Y$, with restrictions on income, will get still
poorer, while system $X$, without any restrictions on income, will get
richer and richer.

This redistribution of income will not be an effect of coercion or
looting, but just a~consequence of the fact that the combined system
tends to acquire the most probable state. As for coercion, it may
play a~certain role, but not in the form of $X$ raping $Y$, but
coercion inside system $Y$, aimed at obstructing any rise of income
above certain level.

Applying the above most simple model to real economic situations, we
can draw two important conclusions:

\medskip

1) the policy of restriction of income is dangerous if the contact
with a~free market surroundings is unavoidable;

2) if such a~policy was already embarked on, and the economy in
which it was implemented is isolated from the free market, then
the effect of the ``market reforms'' will depend on the sequence
of two main steps, the release of incomes and the establishment of
contact with the free market surroundings.

\medskip

For the country with a~``non-market'' economy, it would be
necessary first to release the income from restrictions and then
to ``open'' the economy to the market only {\bf after the
equilibrium\index{equilibrium, market} had been attained}.
Otherwise, the resources of the country will be ``sucked out''
outward before the equilibrium is established.

Of course, our arguments are based on a~very simple idealized model.
Still, our reasoning gives an explanation of different results
of market reforms observed in the Eastern Europe, on
the one hand, and in China and Vietnam, on the other.

In the Eastern Europe, the reforms were conducted by means of
a~``shock therapy'': the national economies were ``opened up'' for
outside activities without preliminary creation of market
institutions inside the country.

The results were catastrophic (flight of capital and collapse of
production).

In China and Vietnam the reverse order was implemented: first, the
income was released from restrictions inside the country and then
the market was gradually opened up for outsiders, in accordance,
more or less, with the process of creation of a~domestic market.
Such a~policy resulted in an astoundingly rapid economic growth.

Consider another case when a~system with a~restriction on income
interacts with a~free market system.

Let the ``spin system'',\index{spin model} a~community of $N$
workers, with fixed individual income $A$, have $L$ vacancies.
Obviously, for $A$ fixed, the system's temperature is negative for
$L>\frac{N}{2}$.

This means that

\begin{thetext}
{{\bf in the state of equilibrium,\index{equilibrium, market} the
free market would prefer that half of the workers should not get
salary},}
\end{thetext}
i.e., should be unemployed,\index{unemployment} because this is a~far
more probable state than the other ones with a~lower rate of
employment. So, this model includes unemployment as an inherent
characteristic of the system in the state of equilibrium.

For the neo-classical school of economics, the following famous
paradox stood as a~stumbling block:
\begin{thetext}
{\it why the market equilibrium\index{equilibrium, market} is not
realized when the free labor force market is in
operation\footnote{The problem of unemployment\index{unemployment}
is of utmost importance for the discourse on economic equilibrium,
as it represents by itself a~counter-example for neo-classical
theory of equilibrium, stating that no excess demand is possible.
This served as one of greatest incentives for the emergence of
institutional approach to the economy. See \cite{Wi},
\cite{Af2}.}?}
\end{thetext}

In the above model this paradox is dismissed.

A negative temperature may also emerge in the systems with
restrictions on the salary's {\it from above}. Indeed, to increase
the total value of wages in the situation of a~salary-restricted
economy with the value of salary bounded from above by $A$, {\it
means} that as the total value of wages $E$ equals $NA$, the
entropy\index{entropy} vanishes. This implies that, at least for
some interval $E_0 < E < NA$, the temperature,
$T=\frac{1}{\pderf{S}{E}}$, is negative, i.e., at any contact with
a free market the value of total wages will fall, at least to
$E_0$.

If, moreover, there is a~{\it minimal wage value}, $B$, the number of
gainfully employed will be bounded from above:
\begin{equation}
\label{eq16}
L<\frac{E_0}{B}.
\end{equation}
This restriction will keep salary from dropping down, and
necessitate a~certain rate of unemployment,\index{unemployment}
which will be $\geq \alpha=\frac{N-L}{N}$. In reality, the level
of wages is always limited ``from below'' by the level of
biological survival (at least, in towns, where Nature's resources
are unattainable). Therefore, if salary is bounded, the
opportunity for unemployment is created.

The analysis given above shows that {\bf unemployment emerges in
the state of equilibrium}\index{equilibrium, market} at the
positive temperature if the wages are bounded simultaneously from
above and below.

The above can be considered as a~proof of Keynesian\index{Keynes
J. M.} arguments on the reasons of unemployment. In real
situations, the administration aspires to put a~limit on the
salary ``from above'', while the trade unions (are supposed to) do
it ``from below''. This combination {\bf must}, according to our
model, result in the emergence of unemployment.

\section{Migration potential}\index{migration potential}

In the previous sections we showed some applications of the
thermodynamic\index{statistical thermodynamics} approach to
economics and interpreted in terms of economics several
thermodynamic parameters.

One more very important parameter is known in statistical physics
under the name the {\it chemical potential}. Here it will be
referred to as {\it migration potential}. It describes the state
of equilibrium in the systems when the number of
agents\index{agent} is not fixed.

Take two such systems, assuming that the total number of their
agents\index{agent} is a~constant, $N_1 + N_2 = N$, but the agents
can migrate from one system to another. Consider the problem: {\it
under what conditions these two systems will get into an
equilibrium?}

Having applied the standard technique, it is possible to find out
when the entropy\index{entropy} of two interacting systems,
characterized by restrictions on the total amount of income and
the total number of agents, will be maximal. The total entropy is
the sum of entropies of the two systems. As a~result, we get two
conditions (as a~corollary of $dN_1 =-dN_2$ and $dE_1 =-dE_2$
implied by $dN=0$ and $dE=0$):
\begin{equation}
\label{eq17}
\pderf{S_1(N_1, E_1)}{N_1}\Big |_{E_1}=\pderf{S_2(N_2, E_2)}{N_2}\Big
|_{E_2};
\end{equation}
\begin{equation}
\label{eq18}
\pderf{S_1(N_1, E_1)}{E_1}\Big |_{N_1}=\pderf{S_2(N_2, E_2)}{E_2}\Big
|_{N_2}.
\end{equation}

The factor $\mu=-T \pderf {S(N, E)}{N}\Big |_{E}$ will be called
the {\it migration potential}.\index{migration potential} When the
temperatures are equal, the two systems in diffusion contact,
i.e., when migration of agents\index{agent} from one system to
another is allowed, are in equilibrium (i.e., in the most probable
state), if their migration potentials are equal.

As $\delta N_1$ agents\index{agent} from the second system migrate
to the first one, the change of entropy\index{entropy} will be as
follows
\begin{equation}
\label{eq19}
\delta S=\delta S_1+\delta S_2=\pderf{S_1}{N_1}\Big |_{\delta
N_1}+\pderf{S_2}{N_2}\Big |_{\delta N_2}= \left ( \frac{\mu _2}{T}-
\frac{\mu _1}{T} \right ) \delta N_1
\end{equation}
(since $\delta N_2=-\delta N_1$ and $\delta N_1>0$).

This makes clear what happens with the systems, if their migration
potentials are not equal. If $\mu_2<\mu_1$, the change of entropy
with the increase of the number of agents in the first system is
positive, hence, it is a~change of state towards the maximal
probability.

In other words, the agent\index{agent} flow proceeds from the
system with a~larger migration potential to the system with
a~smaller migration potential.

Now, we can find the relative probability of various states for
the system interacting with a~thermostat\index{thermostat} and
exchanging with it either income, or agents, or both. In the same
way as before, we obtain
\begin{equation}
\label{eq20}
 \frac{P_Y(E_1, N_1)} {P_Y(E_2, N_2)}=\exp \left (S(E_1-E_2, N-N_1) -
 S(E-E_2, N-N_2) \right),
\end{equation}
where $(E_1, N_1)$ and $(E_2, N_2)$ are different states of subsystem
$Y$, interacting with the thermostat, while $E$ and $N$ are the total
income and the total number of agents, respectively, in the total
system (sub-system $Y$ plus the thermostat).

On expansion into Taylor series and simplifications, we obtain:
\begin{equation}
\label{eq21}
 \frac{P_Y(E_1, N_1)} {P_Y(E_2, N_2)}=\frac { \exp \left
 (\displaystyle\frac{N_1\mu
 -E_1}{T} \right) }{\exp \left
 (\displaystyle\frac{N_2\mu -E_2}{T}\right)}
\end{equation}
This means that the probability for subsystem $Y$ to exchange
income and agents\index{agent} with thermostat is proportional to
the Gibbs's factor
\begin{equation}
\label{eq22}
\exp \left (\displaystyle\frac{N\mu -E}{T} \right).
\end{equation}
As it was mentioned earlier, the thermostat is just the remaining
part of the large system, out of which the subsystem $Y$ was
isolated. As other thermodynamic parameters, we obtained the
Gibbs's factor in a~way usual for statistical physics.

A question arises: {\it are the hypotheses on the total system's
income and the number of agents too binding and non-realistic?}

It looks that hypotheses of this kind are always needed in the
study of the properties of ideal systems. We observe very similar
problems in statistical physics as well. There, certainly, no
absolutely closed system can be found, still, nevertheless, the
theory ``works''.

In real economic systems, the number of agents is, of course, much
smaller than that in physical systems, but, on the other hand, the
expected accuracy of prediction is not so high, either.

The Gibbs's factor paves the way for introducing a~very useful
parameter, called in statistical\index{statistical thermodynamics}
thermodynamics the {\it large statistical sum}:\index{statistical
sum}
\begin{equation}
\label{eq23}
 Z= \sum _{N=0}^{\infty } \sum _k \exp \left( \displaystyle\frac{N\mu
 -E_k(N)}{T} \right).
\end{equation}
The large statistical sum\index{statistical sum} allows one to
compute, effortlessly, some important parameters of the system.
For instance, the mean number of the system's agents can be found
by differentiation with respect to $Z$.

Set $\lambda=e^{\mu/T}$. Then the large statistical sum takes the
form:
\begin{equation}
\label{eq24}
 Z= \sum _N \sum _k \lambda ^Ne^{-\frac{E_k}{T}}.
\end{equation}
Now it is easy to deduce (see \cite{LL}) that the mean number of
agents\index{agent} is equal to
\begin{equation}
\label{eq25}
\bar N= \lambda \pderf {}{\lambda } \ln Z.
\end{equation}
This relation is very important. It provides with a~way to determine
$\lambda$ in the investigated systems by equating the number of agents
in the system to $\langle N\rangle$.

Here is an example from the economics of migration, aimed to
demonstrate the usage of the migration potential.\index{migration
potential} Consider two systems: $A$, with $N_A$ vacancies, and
$B$, with $N_B$ vacancies, with the total of $n$ agents capable to
freely migrate between the two systems. Let the per capita income
in $A$ be $E_A$, that in $B$ be $E_B$. Besides, we assume that
both systems are immersed in a~much greater system, a~thermostat\index{thermostat}
with temperature $T$.

How do probabilities of filling vacancies in systems $A$ and $B$
depend on the parameters of the model? The
equilibrium,\index{equilibrium, market} considered here as the
most probable state of the joint system, will be determined by the
system's temperature and migration potential. Using the large
statistical sum, one easily obtains the function of distribution
of agents for $A$ and $B$, see \cite{Ki}.

Let one vacancy in system $A$ be isolated out as a~subsystem.
Assuming that this subsystem is in equilibrium with the remaining
part of the system, it is possible to determine the large
statistical sum\index{statistical sum} of the subsystem. As this
subsystem can only be in one of the two possible states, with
income 0 (the empty state) and with income $E_A$, the large
statistical sum is equal to:
\begin{equation}
\label{eq26}
 Z_A= 1+ \lambda e^{-\frac{E_A}{T}}, \text{ where
 $\lambda=e^{\mu/T}$}.
\end{equation}
Similarly, for system $B$:
\begin{equation}
\label{eq27}
Z_B=1+ \lambda e^{-\frac{E_B}{T}}
\end{equation}
As systems $A$ and $B$ are supposed to be in equilibrium, and,
therefore, their migration potentials are equal and $\lambda$ is the
same for both systems. Hence, the probability of one vacancy in
system $A$ to be occupied is equal to:
\begin{equation}
\label{eq28}
\varphi_A=\frac {\lambda e^{-\frac{E_A}{T}}}{1+ \lambda
e^{-\frac{E_A}{T}}}.
\end{equation}
Similarly, for system $B$:
\begin{equation}
\label{eq29}
\varphi_B=\frac{\lambda e^{-\frac{E_B}{T}}}{1+ \lambda
e^{-\frac{E_B}{T}}} .
\end{equation}
So we have:
\begin{equation}
\label{eq30}
\frac{\varphi_A}{\varphi_B}=\frac { \lambda +e^{\frac{E_B}{T}}}{\lambda
+e^{-\frac{E_A}{T}}}.
\end{equation}
Since the total number of the system's agents\index{agent} is
equal to $n$, we have
\begin{equation}
\label{eq31}
n= N_A\varphi_A + N_B\varphi_B.
\end{equation}
We determine the migration potential\index{migration potential}
from this equation; we substitute it into the expressions for
$\varphi_A$ and $\varphi_B$ thus finding the probabilities for
agents to belong to $A$ and $B$, respectively, in the
equilibrium\index{equilibrium, market} state. It is clear that as
the temperature of the thermostat\index{thermostat} changes, the
migration potential also changes, because the relative probability
for the joint system $A+B$ to have total income $E$ depends on the
Boltzmann's factor $ e^{-\frac{E_A}{T}}$.

Now, consider the case when the numbers of vacancies in the
systems in question are equal and both of them are almost totally
occupied, i.e., both $\varphi_A$ and $\varphi_B$ are close to 1.
This means that we can expand $\varphi_A$ and $\varphi_B$ with
respect to $\lambda^{-1}e^{-\frac{E_A}{T}}$ and
$\lambda^{-1}e^{-\frac{E_B}{T}}$. We will confine ourselves to the
first order terms. Simple calculations result in:
\begin{equation}
\label{eq32}
\frac{\varphi_A}{\varphi_B}= \frac{x\left (1 - \frac{\Delta} {N} \right )
+ 1}{x+\left (1 - \frac{\Delta}{N} \right )},
\end{equation}
where $N=N_A= N_B$ is number of vacancies, $\Delta=2N-n$ is total
number of vacancies in the joint system, and $x= e^{\frac{E_A -
E_B}{T}}$.

Clearly, if $E_A - E_B$ is far greater than the temperature, then the
ratio of probabilities becomes independent of the income, and tends to
$1-\frac{\Delta} {N}$. If $E_A - E_B$ is small, we have
\begin{equation}
\label{eq33}
e^{ \frac {E_A - E_B}{T} }\sim 1+ \frac {E_A - E_B}{T} \text{ and }
\frac{\varphi_A}{\varphi_B} \sim 1- \frac { (E_A - E_B) \Delta }{2TN}.
\end{equation}
Thus, not the values of $E_A$ and $E_B$ are essential, but the
ratio $\frac{E_A - E_B}{T}$, the parameter determined by the
thermostat,\index{thermostat} that is by the environment.

Assuming that the thermostat is a~free market system, and, therefore,
the temperature in it is equal to the mean income, it becomes manifest
that to find the relative probability of employment in two systems,
the ratio of the difference of wages to the mean wage in the
environment is essential.

So far we had considered very simple model problems of employment, in
order to demonstrate a~possible way of solving this kind of problems
in general. Clearly, one can consider more realistic assumptions,
e.g, the investigate migration processes in interacting systems
characterized by distinct restrictions on income.

Such models are of particular importance in the study of regional
economies.

\chapter{The thermodynamics of prices}

\section{A setting of the problem}

Let us try now to extend the thermodynamic\index{statistical
thermodynamics} approach to the study of markets and prices. For
this purpose, consider a~simplified model situation. It is
necessary, nevertheless, to simplify with caution, so as not to
lose the most essential situational characteristics.

Our first model is intended to analyze how the size of the flow of
goods and money influences the market prices.

In the above sections we showed that if the money flow is
constant, it is possible to introduce the concept of equilibrium
in the income distribution. To implement this, we have to know the
entropy\index{entropy} of the system, hence its temperature. Using
temperature we are able to find out the conditions under which the
system stays in equilibrium with the environment.

Now, we add to this model a~flow of goods. It is still possible to
calculate the new entropy and introduce an additional parameter of
equilibrium. In what follows we will discuss how this additional
parameter can be interpreted in terms of the price of the goods.
Moreover, it is also possible to deduce the {\it equation of the
market state}, i.e., to find dependence between the flow of goods, the
price, the number of buyers and the temperature.

First, consider an intuitive concept of the market and the market
equilibrium, and then try to formalize it.

The market is characterized by the presence of goods which are
sold, and money spent at purchases. Let $V (t)$ be the amount of
``units of goods'' sold per unit of time during which the buyers
spend $E (t)$ of units of money. We say that the market is {\it
stationary} if $V (t)$ and $E (t)$ are constants independent of
time and all goods are bought, i.e., there is no accumulation of
goods in the hands of the sellers. While the market functions, the
deals are made, i.e., agreements on exchange of some of the goods
for some money.

Bargaining regulations are typical for the market. Some kinds of
deals may be ruled out, for instance, the ones bidding price too high
or too low. What is important for us is that the regulations on deals
do not vary with time.

Observe that even if the price regulations are not officially
fixed, some rules exist anyway, e.g., the ones that ensure
contracts' fulfillment. Thus, the market is a~social institution
due to existence of rules of dealing. No doubt, the rules imply
that a~control apparatus should be present, its purpose to enforce
the rules. It has to be entitled with certain enforcement powers,
in order to punish violators and guarantee reliability of
contracts. In this sense, the market is certainly not the arena
for totally spontaneous activity of its agents,\index{agent} but
an organized social institution.

Consider a~model with several co-existing markets able to
interact: exchange goods and money resources. The intuitive
concept of equilibrium\index{equilibrium, market} of these markets
is that the situation in each of these markets remains, in certain
essential aspects, the same even after the interaction. Clearly,
this cannot take place for any values of the amounts of goods and
money $V_1$, $E_1$ and $V_2$, $E_2$, in each system, respectively.
Besides, the interaction can vary, i.e., include an exchange only
of money resources, only of goods, or both.

Consider the simplest case, when the values of the flows of money and goods are
discrete, as is the case in real life:
\begin{equation}
\label{eq34}
 E_n = nE_0 , \quad V_m = mV_0 .
\end{equation}
In addition, let the values of $n$ and $m$ be sufficiently large,
to make it possible to treat small changes of flows as
insignificant and differentiate. Let $N$ be the number of buyers
in our model. (Generally speaking, $N$ should be the number of
bargains, but it is more convenient to consider $N$ as the number
of buyers.) The market is assumed to be stationary in the above
sense.

The set of probable states of the market is the set of all contracts,
allowed by rules, i.e., the ways of distribution of the goods between
buyers. Let us examine the market with a~single seller assuming that
the seller is not capable to influence the flow of arriving goods.

For this model, introduce the entropy\index{entropy} in the same
way as earlier, namely, as the logarithm of the number of probable
states. In this case, the entropy depends on both the total money
expenditure, $E$, and the total amount of purchased goods, $V$.
Various states of the market can be regarded as legalized by the
rules of distribution of the total supply of money and goods among
$N$ buyers. Observe that now we study the ``market of the
buyers'', so, we do not have to include in the entropy the
distribution of goods among sellers.

If, however, we do take them into consideration, we have to
incorporate in the model the distribution of the money among the
sellers. This is a~far more complex task.

To determine the price, it does not matter where the buyer purchased
the goods. What is important, is how many goods he or she has
received and how much he or she has paid.

So, for the analysis of the ``buyer's market'',\index{market,
buyer's} the entropy,\index{entropy} due to the presence of many
sellers at the market, is not considered.

To avoid all these complications, we could have counted the number
of contracts, instead of the numbers of sellers and buyers. Such
an assumption, however, requires to introduce the migration
potential\index{migration potential} corresponding to possible
changes of the number of contracts. This case will be considered
later.

By analogy with the pressure in statistical\index{statistical
thermodynamics} thermodynamics, the notion of {\it marginal
price}\index{marginal price} $P$ is introduced as follows:
\begin{equation}
\label{eq35}
P=T\pderf{S}{V} .
\end{equation}
A bit later we will justify the introduction of this parameter.
The marginal price is a~characteristic that points out whether or
not the two interacting systems are in
equilibrium.\index{equilibrium, market}

First, consider the simplest case, referred to as the ``free
market''. In this model there are no restrictions on goods and
money distribution among the buyers. This means that the buyer can
pay any price~--- either infinitesimally low or infinitely high.

The free market is rather easy to study, because the total number
of probable market states is the product of the number of possible
distributions of goods between the market agents\index{agent}
times the number of possible distributions of money between the
same agents. Thus, the statistical weights\index{statistical
weight} of the system are obtained by multiplying the statistical
weights determined by the flows of goods $(V)$ and money $(E)$.
Consequently, the entropy\index{entropy} of the system (the
logarithm of the statistical weight) consists of the two
components: one, is determined by the flow of goods, the other
one, by the flow of money:
\begin{equation}
\label{eq36}
 S (E, V) = S (E) + S (V).
\end{equation}
Let the money and goods flows be quantized, as in (\ref{eq34}). In what
follows we will derive the equation of state for such a~system.

First, compute the number of probable distributions of goods and money
flows among $N$ agents of the market. The statistical weight $g (E_n,
N)$ of the flow $E_n$ , distributed among $N$ agents is equal to the
number of non-negative integer solutions of the equation
\begin{equation}
\label{eq37}
 n = x_1+ \dots +x_N.
\end{equation}
The calculational technique for such equations is well-known. Namely,
add 1 to each $x_i$. Then the quantity to be determined is the number
of positive integer solutions of the equation
\begin{equation}
\label{eq38}
n + N = y_1 + ... +y_N .
\end{equation}
Now, let us subdivide the integer segment of length $n + N$ into
$N$ integer subsegments. Thus, the statistical
weight\index{statistical weight} is equal to
\begin{equation}
\label{eq39}
g (E_n, N) = \binom{N-1+n}{N-1} = \frac{(N-1+n)!}{n!(N-1)!}.
\end{equation}
For sufficiently large $n$ and $N$, the Stirling formula gives
\begin{equation}
\label{eq40}
g (E_n, N)\approx \frac{1}{\sqrt{2\pi }} \frac {(N+n-1)^{N+n- \frac12}}
{{N- \frac12}^(N-1){n+\frac12}^{n}} .
\end{equation}
Hence,
\begin{equation}
\label{eq41}
\renewcommand{\arraystretch}{1.4}
\begin{array}{l}
S (E_n)= \ln g (E_n, N)\approx \\
(N+n-1)\ln (N+n- \frac12) - (N-1)\ln (N-\frac12) n\ln (n+\frac12)
-\frac12 \ln 2\pi . \end{array}
\end{equation}
The temperature $T$ is calculated in terms of $\pderf{S}{E}$:
\begin{equation}
\label{eq42}
\frac1T=\frac{1}{\eps}\pderf{S(E_n)}{n}\approx\frac{1}{\eps}
\ln\frac{N+n- \frac12}{n+\frac12}.
\end{equation}
For $n\gg N$, which is a~rather natural condition for the free market,
the expression for the inverse temperature is:
\begin{equation}
\label{eq43}
\frac1T=\pderf{S(E_n)}{n}=\frac{1}{\eps}\frac{N}{n}=\frac{N}{E}.
\end{equation}
This means that the temperature $T=\frac{E}{N}$ is equal to the mean
value of income per capita.

To compute $\pderf{S}{V}$ is a~completely analogous matter (since
the entropy\index{entropy} $S$ can be described as the sum of two
terms, one of which only depends on $E$, the other one, on $V$,
see (\ref{eq36})):
\begin{equation}
\label{eq44}
\frac{P}{T}=\pderf{S(V_m)}{V_m}=\frac{1}{W}\frac{N}{m}=\frac{N}{V}.
\end{equation}
Thus, a~relation between the marginal price,\index{marginal price}
the goods flow and the ``temperature'' of the free market is of
the form
\begin{equation}
\label{eq45}
P=T\frac{N}{V}.
\end{equation}
As $TN=E$, it is easy to show that for the free market {\it the
marginal price is equal to a~median price}. This, actually, allows to
define the {\it price} as $T\pderf{S}{V}$.

In presence of restrictions on price, the marginal price may deviate
from the median one.

Manifestly, the equation of the free market (\ref{eq45}) is totally
analogous to the equation of the ideal gas.

In our arguments, {\bf the marginal price has the same
thermodynamic origin as the temperature or the chemical
potential}, i.e., the theory is completely derived from
thermodynamic principles. The marginal price\index{marginal price}
is the parameter of equilibrium,\index{equilibrium, market}
identical in meaning to the temperature. It indicates whether or
not the interacting systems are in the equilibrium.

Indeed, the same pattern of reasoning as at the introduction of
the temperature, can be repeated and we similarly prove that two
interacting systems for which the exchange of the goods is
possible are capable to be in the state of equilibrium only when
their marginal prices\index{marginal price} for the goods are
equal.

Indeed, let the statistical weights\index{statistical weight} of
the interacting systems depend on the flows of money and goods,
$E$ and $V$, while the total values of both parameters are
constants (speaking about a~flow, we mean the expenditure capacity
per unit time):
\begin{equation}
\label{eq46}
E_1+E_2=E, \qquad V_1+V_2=V,
\end{equation}
then the statistical weight of the combined system is
\begin{equation}
\label{eq47}
g_1(E_1, V_1)\cdot g_2(E-E_1, V-V_1).
\end{equation}
The probability of the state becomes maximal when the function
(\ref{eq47}) attains its maximum. Introducing
entropy\index{entropy} as $\ln g$, the conditions for the entropy
extremum become
\begin{equation}
\label{eq48}
\left(\pderf{S_1}{E_1}dE_1+\pderf{S_1}{V_1}dV_1\right)+
\left(\pderf{S_2}{E_2}dE_2+\pderf{S_2}{V_2}dV_2\right)=0.
\end{equation}
Taking into account that
\begin{equation}
\label{eq49}
dE_1=-dE_2, \qquad dV_1=-dV_2,
\end{equation}
we express the conditions of equilibrium\index{equilibrium,
market} as follows:
\begin{equation}
\label{eq50}
\pderf{S_1}{E_1}\Big|_{V_1}=\pderf{S_2}{E_2}\Big|_{V_2}, \qquad
\pderf{S_1}{V_1}\Big|_{E_1}=\pderf{S_2}{V_2}\Big|_{E_2}
\end{equation}
The first expression, as we already discussed, means the
coincidence of ``temperatures''. The second one means coincidence
of values of the marginal prices.\index{marginal price}

Thus, the coincidence of values of the marginal prices is a~necessary
condition for the interacting markets to be in the state of
equilibrium.

Now, proceed to the case that the market is not ``free''; let some
restrictions be imposed on its operation.

Let us investigate, first, what is going to happen if something
similar to the ``allocation system'' is adopted, i.e., the
possibilities to buy goods are forcefully restricted. Recall also
our assumption that the market is {\it stationary}, which means
that the flows of goods and money are completely distributed among
the market participants.

Suppose that none of the market agents\index{agent} is allowed to
get more goods than a~certain amount, $K$. In all the other
respects the distribution remains free, and~--- this is of utter
importance~--- {\it does not depend on the money distribution},
i.e., there are no explicit restrictions on the prices. This
means, that the statistical weight\index{statistical weight} of
the state with $V$ goods, and $E$ money is still the product of
statistical weights:
\begin{equation}
\label{eq51}
g_N(E, V)=g_N(E)g_N(V),
\end{equation}
while the entropy\index{entropy} includes two components, one of
which depends on $E$, and the other one on $V$:
\begin{equation}
\label{eq52}
S_N(E, V)=S_N(E)+S_N(V).
\end{equation}
To determine the marginal price,\index{marginal price} we should
known $\pderf{S}{V}\Big|_{E, N}$, i.e., we should compute
$S_N(V)$. This appears to be a~difficult combinatorial problem:
{\it find the number of non-negative integer solutions of the
equation}:
\begin{equation}
\label{eq53}
m=x_1+x_2+\dots+x_N, \text{ where $ x_i\leq K$ for each $i$}.
\end{equation}
The solution of this problem is distinguished by a~special feature,
which renders meaningful conclusions.

It is clear that if $m>KN$ this problem has no solutions at all.
The logarithm of the number of solutions is the system's entropy
$S^K_N(V_m)$. This means that when $m$ (i.e., the flow of goods)
approaches $KN$ from below, then at some point the
entropy\index{entropy} begins to decrease. But this, in turn,
means that $\pderf{S(V_m)}{V}$ becomes negative. Taking into
account that the distribution of money flow is free, which implies
positive temperature, the result would be that, under
restrictions, the system's marginal price becomes negative when
the volume of the flow of goods becomes sufficiently large.

This means that during interaction with other systems with
positive marginal prices,\index{marginal price} our ``allocation
system'' with imposed restriction on consumption of the goods will
begin to eject goods at dumping prices: any more-than-zero price
of the exported goods will contribute to destabilization of the
equilibrium.\index{equilibrium, market}

Of course, it is possible to say that when the flow of goods is close
to the value of $KN$, there is no sense to restrict the distribution.
The fact, however, is that the marginal price can become negative {\it
long before} the flow of goods approaches the critical value, $KN$.

For instance, for $K=1$, this will happen for $m>\frac{N}{2}$.

Since the introduction of restrictions such as in our example is
a~political decision, the question should be addressed: {\bf how the
persons responsible to make such decision could learn that the
marginal price has already became negative}?

Indeed, the price of market transactions is positive, and the median
price is also positive. In order to know how far the system that
interacts with other systems is from the state of the equilibrium, it
is necessary to calculate certain, not directly observable, parameters
--- the temperature and the derivative of the entropy\index{entropy} with respect to
the flow of goods.

As we have just shown, this problem is rather tough, even for a~very
simple model.

We see, nevertheless, that the restrictions imposed on the market
can create, in the range of the system's states, a~zone of
``latent instability''.\index{latent instability}

Let us consider now whether the zones of latent instability would
emerge if the price is bounded from below. This model is very
important indeed because real market systems are hardly free: it is
not realistic to sell the goods, on the large scale, at the price
below the production cost.

If in our model the price of the bargain is bounded from below,
then the entropy\index{entropy} of the system cannot be described
any more as the sum of two components, each depending on only one
variable, one being the entropy of the flow of goods, the other
one the entropy of the money flow.

The mathematical problem of computing the entropy is to determine $\ln
g_N(E, V_m, \lambda)$, where $\lambda$ is a~parameter bounding the
prices from below, and $g_N=\sum g_N(y_1, \dots, y_n)$, where
$g_N(y_1, \dots, y_n)$ is the number of positive integer solutions of
the equation
\begin{equation}
\label{eq54}
m=x_1(y_1)+x_2(y_2)+\dots+x_n(y_n),
\end{equation}
In (\ref{eq54}) $y_n$ is an arbitrary partition of the number $n$ into $N$
non-negative integers, and
\begin{equation}
\label{eq55}
x_i(y_i)=0 \text{ for $y_i=0$ and } x_i(y_i)<\frac{y_i}{\lambda}
\text{ for $y_i>0$}.
\end{equation}
(Here $x$ is the allowed number of goods in the deal while $y$ is the
corresponding amount of money.)

This is a~still more difficult calculational problem. But, as in
the previous cases, some qualitative dependencies can be
discovered relatively easy. If we increase the goods flow but the
money flow does not increase simultaneously, the problem has no
solutions. Indeed, any possible price remains below the allowed
level once $V_{k, m}$ is great enough for limitations on $x_i$,
where the $y_i$ are $\leq n$ and $n$ is such that $ne=E$), i.e.,
the entropy\index{entropy} $S_N^{\lambda}(E, V)$ vanishes.

Like in the previously given model, the marginal
price\index{marginal price} becomes negative earlier than that
because the distribution of the money flow is free, the
temperature is positive, but the derivative $\pderf{S_N^\lambda(E,
V)}{V}$ at values of $V$ smaller than but close to $V_K$ becomes
negative.

So, we observe the same effect as under the terms of the
restrictions on goods, namely, the latent zone of
instability\index{latent instability} with a~negative marginal
price, and the same consequences~--- an outward dumping of goods
by the system striving to reach an equilibrium.

Apparently, this explains phenomena of mass destruction of the
goods for the sake of maintaining the level of prices during the
crises of overproduction. It is not the ``malicious will'' of
owners of the commodities, but simply the shortest way to the most
probable state of the system: getting rid of the goods the system
assumes a~statistical equilibrium, i.e., a~most probable state.

As a~byproduct of our arguments, we see that {\bf introducing
restrictions in the system we provoke occurrences of zones of
latent instability}.\index{latent instability} The borders of
these zones are extremely scarcely discernible even for the
simplest models.

Such parameters, as the derivatives of the entropy,\index{entropy}
e.g., the temperature and the marginal price, are crucial for the
description of the behavior of systems with restrictions after
they became engaged in interactions with similar systems or with
free-market systems.

The marginal price\index{marginal price} becomes a~very important
parameter, a~major parameter of the market equilibrium. It
coincides with the median price only for the free market systems.
In the systems with restrictions, it may be negative. If this
happens, the system is unstable.

Thus, restrictions on economic activity can, by no means, be held
as ``harmless''. First of all, {\bf no restriction is harmless
because the range of its influence is unclear, the range within
which its influence renders the system {\it {\protect {\bf
unstable}}}}.

Observe that the so-called ``economic'' reasons for various
restrictions on deals (e.g., the minimal price determined by the
cost of production) turn out, at deeper scrutiny, purely political
reasons: the cost of production often can not be lowered
``thanks'' to a~monopoly, i.e., a~political control of the market.
In other words, the restrictions imposed on the freedom of market
activity, are capable to generate uncertainties in the system,
instead of, as politicians use to think and preach during the
election campaigns, making it more predictable.

The above study of the market with restrictions on prices, is
related to the problem which lately aroused considerable interest
in theoretical economics. In 1970, G.~Akerloff\index{Akerloff G.}
published an article \cite{Af1} that soon enjoyed much popularity.
The article dealt with the markets with ``asymmetric''
information, i.e., markets, where the seller and buyer have
different opportunities to estimate the quality of the item on
sale.

Akerloff\index{Akerloff G.} showed, further developing his
contention in later publications \cite{Af2}, that, for asymmetric
markets, {\bf the presence of low quality goods and dishonest
sellers can result not only in sweeping away of high quality goods
from the market, but may also result in a~collapse of the market
as such}.

Briefly, Akerloff's idea was as follows. Assume that the article
of goods (say, a~used car) may appear to be of low quality with
certain probability, $q$ (and, accordingly, of high quality with
probability $1-q$). Assume further that the buyer ``knows'' to an
extent these {\it a~priori} probability because $q$ may reflect
the ordinary index of production rejected by the factory.

But the seller in question knows about the article he or she is
selling far better.

The buyer just follows the general opinions concerning the item. By
this reason, the seller of a~{\it good} second-hand car is seldom able
to get the real price for it: the buyer wants to insure himself or
herself against the dishonest seller.

The seller of the {\it bad} car, contrariwise, enjoys all the
chances to get for it more money than it is really worth. As a~consequence,
good cars are ``washed away'' from the market, and
the bad ones dominate. Using the standard technique of the
utility\index{utility} functions, Akerloff\index{Akerloff G.} has
shown that under certain conditions the
equilibrium\index{equilibrium, market} may be unattainable, e.g.,
when the graph representing dependence of supply on prices does
not overlap with the graph representing dependence of demand on
prices. Akerloff interprets the problem of asymmetric markets
linking it to the {\it Grasham law}\index{Grasham law} which
states that the ``bad'' money oust the ``good'' money out of
circulation: people tuck away the ``good'' money. Akerloff
conjectures that such an approach gives an opportunity to estimate
the losses of the market from crooked dealings.

Akerloff\index{Akerloff G.} concludes: the price of dishonest
behavior amounts not just in the losses of the buyer, but also in
the undermining consequences for honest businesses \cite{Af2}.

The model with restrictions on prices from below discussed above,
shed, in our view, some new light on the problem of asymmetric
markets. We believe that Akerloff has proved not so much the
possibility of collapse of the asymmetric market, but rather the
impossibility to apply the standard equilibrium techniques used in
neo-classical analysis. Indeed, even in an asymmetric case, the
flow of deals will last anyway, but for the study of the market
state, the definition of equilibrium, based on the notions of
statistical thermodynamics,\index{statistical thermodynamics}
looks to be more appropriate.

Indeed, consider two markets in interaction, one, with
restrictions on prices from below, and another one, without such
restrictions (say, a~market of used cars). Consider a~model with
an ``asymmetric information'', we see that a~market with
restrictions on price from below is a~market of ``good'' cars,
while the market without restrictions is a~market with defective
cars (Akerloff\index{Akerloff G.} uses a~slang term:
``lemons'')\index{lemon} in circulation. As we have shown above,
at a~certain value of parameters of the flow of goods and
``temperature'' of the market (namely, at low temperature), the
marginal price\index{marginal price} in the market with
restrictions (here: in the market of ``good'' cars) becomes
negative. This means that this market is collapsing. But {\bf the
actual collapse will only happen at a~rather low temperature}. At
higher temperature the market of high quality cars is quite
viable. Thus, it is possible to make some amendments to Akerloff's
statements on the market of ``lemons'', and on the opportunities
to operate honest business in the developing countries, where
``standards of honesty'' are low.

The market of ``good'' cars will not rarely be on the brink of
collapsing but will not actually and totally collapse. According
to our thermodynamic approach, the crucial factor lies not in the
standards of honesty, but in the ``low temperature'' of the
market. If the temperature is not low any more, the ``honest
market'' (the one with restrictions on prices from below) is quite
capable to coexist with the market of ``lemons''. The marginal
price\index{marginal price} formally corresponds to the pressure
and the only thing needed to render the market ``alive'' is that
the ``pressure'' in the market of ``good'' cars were not below a~certain minimal value.

Next, consider in more detail the thermodynamic model of the
market with prices bounded from below. For this purpose, we apply
the technique of large statistical sum.\index{statistical sum}
But, first of all, let us make some remarks.

If we forget sellers and buyers, and only consider the amount of
contracts, $N$, then the temperature of the market, according to
our previous consideration, can be defined as:
\begin{equation}
\label{eq56}
T=\frac{E}{N} ,
\end{equation}
where $E$ is the amount of money.

Introduce the concept of a~``potential contract'' which means the
possibility of striking a~deal, with the goods flow $V_l=lV_0$, and
money flow $E_k=kE_0$, where $V_0$ and $E_0$ are the least values of
the respective flows. Then some further simplification would be
expedient. Introducing discrete time labeled by moments $T_k$, we
assume that the goods and money flows only occur at discrete time
moments. Such a~model of market resembles the market of von
Mises\index{von Mises L.} \cite{Mis}. It allows us to treat every
deal as a~single-time event.

Bearing in mind the statistical aspect of the model, it looks quite
natural to consider the market as the set of simultaneously coexisting
systems covering all admissible deals, and to carry out averaging over
the ensemble instead of averaging over time.

The potential contract $C(E_k, V_l)$ with parameters $V_l$, $E_k$
may be ``filled in'' with a~real content, but also may remain
``empty''. The notion of potential contract corresponds to the
notion of an orbital in quantum mechanics. We are able now to
refer to the set of potential contracts, $C=(E_k, V_l)$ subject to
some restriction, for instance, the contracts with
$\frac{E_k}{V_l}<Q$ may be ruled out (which, actually, mirrors the
restriction on prices from below, $Q$ being the least allowed
price).

Now, apply the standard technique of large statistical
sum.\index{statistical sum} If $N\ll nm$, where $n$, $m$ stand for
the total amount of goods and money in the market, respectively,
the market discussed is a~``classical one'': the ``population
density'' of every contract is very low ($\sim\frac{N}{nm}$), so,
we can neglect the probability that one ``potential contract'' may
be ``filled in'' with two real contracts with equal values of
$V_l$ and $E_k$. This means that, in the large statistical sum, we
can neglect the terms of degree $\geq 2$ with respect to
$\lambda$. Hence, the probability for
 the potential contract to
be ``filled in'' is given by the formula
\begin{equation}
\label{eq57}
W(k, l)=\lambda e^{-\frac{E_n}{T}}=\lambda W_0(k, l).
\end{equation}
If the number of actual contracts is equal to $N$, it is possible
to find $\lambda$ because $N$ can be found by summing over all
probability values $W(k, l)$ for all potential contracts:
\begin{equation}
\label{eq58}
N=\lambda\mathop{\sum}\limits_{l, k}W_0(k, l).
\end{equation}
Therefore
\begin{equation}
\label{eq59}
\lambda=\frac{N}{\mathop{\sum}\limits_{l, k}W_0(k, l)}.
\end{equation}
Since $\lambda=e^{\frac{\mu}{T}}$, where $\mu$ is the migration
potential,\index{migration potential} we can express $\mu$ as
follows:
\begin{equation}
\label{eq60}
\frac{\mu}{T}=\ln\frac{N}{\mathop{\sum}\limits_{l, k} W_0(k, l)}.
\end{equation}
So, the entropy\index{entropy} of the system can be derived from
the formula
\begin{equation}
\label{eq61}
\pderf{S}{N}\Big|_{E, V}=-\frac{\mu}{T}.
\end{equation}
Substituting this value for $\frac{\mu}{T}$, we find that
\begin{equation}
\label{eq62}
S=-\int_0^N\frac{\mu}{T}dN=\int_0^N\ln N dN+ N\ln\mathop{\sum}\limits_{k, l}
W(k, l),
\end{equation}
or
\begin{equation}
\label{eq63}
S=-N\ln N+N+N\ln\mathop{\sum}\limits_{k, l}W(k, l).
\end{equation}
The sum $\varphi(T, V):=\sum W(k, l)$ depends, generally, on $T$
and $V$.

Now, differentiating the entropy \index{entropy}with respect to
$V$ we derive the equation of state (\ref{eq35})
\begin{equation}
\label{eq64}
\frac{P}{T}=\pderf{S}{V}\Big|_{E, N}.
\end{equation}
It requires to calculate $\varphi(T, V)$.

First, consider the already discussed case, the one without
restrictions on prices:
\begin{equation}
\label{eq65}
\varphi(T, V)=\mathop{\sum}\limits_{k=1}^{\infty}C(k)\cdot e
^{-\frac{kE_0}{T}},
\end{equation}
where $C(k)$ is the number of contracts with $E_k$. Since a~certain definite
number $m=\frac{V}{W}$ of potential contracts
which can be ``filled'' exists for each value of $E_k$, it follows
that $C(k) = m$. In order to simplify the calculations, we assume
that $n$ is very large ($E=nE_0$), and the upper limit of
summation is equal to $\infty$. We thus obtain
\begin{equation}
\label{eq66}
\varphi(T, V)=\frac{V_0}{V}\frac{e^{-\frac{E_0}{T}}}{1-e^{-\frac{E_0}{T}}}=
\frac{V_0}{V}\frac{1}{e^{-\frac{E_0}{T}}-1}.
\end{equation}
Hence, the already known result:
\begin{equation}
\label{eq67}
\frac{P}{T}=\pderf{S}{V}\Big|_{E, N}=N\pder{V}\ln\varphi(V, T)=\frac{N}{V}.
\end{equation}

Now, let us calculate $\varphi(T, V)$ for the case of prices
bounded from below. The number of potential contracts
$C_k(E_k, Q)$ now depends on $E_k$ and the minimal price, $Q$:
\begin{equation}
\label{eq68}
C_k(E_k,
Q)=\mathop{\sum}\limits_{n=1}^{\frac{kE_0}{QW}}1=\frac{kE_0}{QV_0}.
\end{equation}

In this case $\varphi(T, V)$ splits into the two components.

One of them is obtained by summation up to $k=\frac{mQW}{E_0}$,
hence, $C_k(E_k, Q)=\frac{kE_0}{QV_0}$. For these values of $k$,
the number of potential contracts depends on $k$. For very large
$k$ such dependence vanishes since we assume that the amount of
money used in any potential contract suffices to ensure any of
$m=\frac VW$ possible values of goods purchased:
\begin{equation}
\label{eq69}
\begin{split}
\varphi(T, V)=&\sum\limits_{k=1}^{\frac{QV}{E_0}}
C_k(E_k, Q)e^{-\frac{kE_0}{T}}+\sum\limits_{k=\frac{QV}{E_0}}^\infty
me^{-\frac{kE_0}{T}}=\\
&\sum\limits_{k=1}^{\frac{QV}{E_0}}
\frac{kE_0}{QV_0}e^{-\frac{kE_0}{T}}+\sum\limits_{k=\frac{QV}{E_0}}^\infty
\frac{V}{V_0}e^{-\frac{kE_0}{T}}.
\end{split}
\end{equation}
After simplifications we have
\begin{equation}
\label{eq70}
\varphi(T, V)=\varphi_1(T, V)\left(
\frac{E_0}{QV_0}-\frac{E_0}{QV_0}e^{\frac{-VQ}{E_0}}\right),
\end{equation}
where
\begin{equation}
\label{eq71}
\varphi_1(T)=\mathop{\sum}\limits_{k=1}^\infty
ke^\frac{kE_0}{T}=\frac{e^{\frac{E_0}{T}}}{\left(
e^\frac{E_0}{T}-1\right)^{\!2}}.
\end{equation}

Now, substitute $\varphi(T, V)$ in the expression for the
entropy,\index{entropy} so the equation of state takes the form:
\begin{equation}
\label{eq72}
\frac{P}{T}=\pderf{S}{V}\Big|_{E, N}=N\pder{V}\ln
\varphi(T, V)
\end{equation}
or
\begin{equation}
\label{eq73}
\frac{P}{T}=\frac{NQ}{E_0}\frac{1}{e^\frac{QV}{E_0}-1} .
\end{equation}
The equation of state (\ref{eq73}) is very interesting. When
$Q\ll\frac{T}{V}$, equation (\ref{eq73}) turns into the familiar
equation of the ideal gas:
\begin{equation}
\label{eq74}
\frac{P}{T}=\frac{N}{V}.
\end{equation}
But, as it turns out, even small restriction on price from
below completely transforms the equation of state.

In the range of larger $V$, the marginal price\index{marginal
price} $P$ (i.e., the equilibrium price)\index{equilibrium,
market}\index{equilibrium price} becomes smaller than the minimal
price of the contract, which means that the market collapses. Such
a situation is equivalent to the negative marginal price for the
free market. It is quite clear that if the least price is
implemented, the large volumes of goods cannot be sold in
equilibrium provided the temperature is bounded. This does not
mean that the absence of deals. This means that the equilibrium
cannot be attained.

If $T=\frac EN$, the minimal price is equal to $Q$, and
$V=\frac{E}{Q}=\frac{T N}{Q}$, then a~part of goods equivalent
to $V-\frac{T N}{Q}$ is impossible to realize.

A hasty observation may lead to the premature conclusion that
the critical mass of the goods when the price is bounded from
below will lie in a~neighborhood of
\begin{equation}
\label{eq75}
V_{crit}=\frac{T N}{Q}.
\end{equation}
But, as is clear from the above analysis, this is not the case:
in reality the critical mass is considerably smaller. Namely,
assuming that
\begin{equation}
\label{eq76}
P=\frac{T NQ}{E_0}\frac{1}{e^\frac{QV_{crit}}{T}-1}>Q
\end{equation}
or
\begin{equation}
\label{eq77}
\frac{T N}{E_0}>e^\frac{QV_{crit}}{T}-1,
\end{equation}
we obtain
\begin{equation}
\label{eq78}
V_{crit}=\frac{E_0}{Q}\ln\left(\frac{T N}{E_0}-1\right).
\end{equation}
As a~result, instead of $V_{crit}=\frac{T N}{Q}=\frac{nE_0}{Q}$,
deduced by the ``naive'' argument, $V_{crit}$ is close to $\ln V$,
i.e., is of several orders of magnitude smaller. Therefore, the
market collapses much earlier than it seems at the first glance.

Coming back to the market with ``lemons'', this means that the
ousting of good cars by the bad ones will happen much earlier than
the naive {\it a~priori} assumptions allow for.

Clearly, the asymmetric information narrows the market's ``zone of
stability'' to a~very small size. Thus, our analysis justifies
Hayek's\index{Hayek F.} conjecture that the market exists thanks
to the culture of ``honest business'', and survives in the process
of competition of business cultures.

The investigation above also points to another condition for
existence of markets with high quality goods. The market operating
with high quality goods has to be somehow separated from the
market of low quality goods. Otherwise, with the influx of high
quality commodities, the market becomes unstable.

The model outlined above shows that due to a~limited supply of
high quality goods, their market can coexist and interact with the
market of ``bad'' goods, but only as the ``market of elite'',
which means that outside its aegis the market is stunted by the
flaws in the ``business code of behavior''. An alternative is the
certification of goods.

We see that, in a~sense, our theory exorcises ``human component'':
the equilibrium\index{equilibrium, market} theory in economics
becomes identical to that of statistical physics.


It is also clear that hypothesis on equal probability of market states
--- the basis of any statistical theory~--- may be justified only if, following
Hayek,\index{Hayek F.} {\bf we consider the human activity a~principal factor
subject to incomplete
information\index{incomplete information} on what's on at the
market}. Otherwise the hypothesis on equal probability of market
states becomes absurd. It is as a~result of human ability to
discover new opportunities together with all possible errors, the
hypothesis of equal probability becomes natural.

Human component is essential in one more aspect. The market
requires abiding certain rules of contracts. In our theory we
assume this as a~given reality. This, however, is extremely
important. Even if contracts are not restricted by ``external''
rules, {\bf the mere necessity to abide the contracts makes the
market into a~social institute}. The behavior of this institute,
considered as a~machine, in particular, its
thermodynamic\index{statistical thermodynamics} properties are of
huge interest, but, first of all, such a~machine~--- such an
institute
--- must be created and being created they should be
maintained. 

\section{Two markets with two items of goods}

Consider now the problem of the market\index{equilibrium, market}
equilibrium for the two markets with two items of goods, capable
to replace each other. The market with replaceable goods poses one
of the most known problems of the mathematical economics. The
study of such markets prompted the marginalist
revolution.\index{marginalist revolution}

In terms of the neo-classical theory, to work out this task, it is
necessary to define the utility\index{utility} functions of the
goods, the function depending on the volume of consignments. When
the derivative of the utility function with respect to the volume
of consignment decreases as this volume grows, the techniques of
the classical analysis show that the summary utility function
reaches its maximum at such a~volume of the consignment of goods
that the increase of utility by one unit of the expenditure is
equal for all nomenclature of goods.

Intuitively, this is a~plausible statement. If, at the
expenditure of one unit of funds per one item of goods, it were
possible to increase the total utility of the purchased goods by
replacing one item by another, this replacement have been
implemented, until the utility had not dropped due to the
increase of the goods' volume.

We see that the dependence of utility\index{utility} on volume is
indeed very important in terms of maintenance of sustaining the
system's stability: otherwise all funds could have been invested
in the most useful goods only.

Let us now study the problem of replaceable goods in the framework
of thermodynamic approach.\index{statistical thermodynamics}

Clearly, it makes sense to speak about replacement only if there
is a~certain equivalence between goods. If there is no
equivalence, we can not proceed. So, let such an equivalence exist
in the model, i.e., $n$ units of item 1 are equivalent (never
mind, in what sense precisely) to $m$ units of item 2. The
equivalence relation can be used to find the maximum of
entropy,\index{entropy} in exactly the same manner, as it was done
before for only one item of goods.

It is important to observe that, in our hypothesis, the equivalence
relation does not depend on the flow of goods. If such dependence
takes place, the theory of statistical equilibrium is, all the same,
possible to deduce, but it is a~bit more involved.

The most probable or, what is the same here, equilibrium, state of
the system is attained as the entropy of the system attains its
maximum. To find the maximum of entropy, we use the equivalence
relation between two items of goods, in completely the same
fashion as earlier on these pages.

Thus, consider a~market with $N$ agents, with a~flow of money $E$
and flows of items of two goods, $V_A$ and $V_B$. To find the
entropy of system, we have to count the number of probable deals,
i.e., the number of ways to attribute to each agent\index{agent}
four numbers $(E_{A_{i}}, E_{B_{i}}, V_{A_{i}}, V_{B_{i}})$ that
show how much money was spent by an agent per unit time on
purchasing the goods $A$ and $B$, and how many goods had been
bought over the same period, respectively. Adding up over all
agents we obtain the values of macroscopic parameters of the
system.

Observe that in order to estimate the system's state, it does not
suffice to simply fix the amount of money spent by the $i$th agent
because some microscopic states may be ruled out, owing to, say,
a~restriction of the prices from below.

In the general case, the dependence of the entropy\index{entropy}
on macroscopic parameters cannot be represented as the sum of
components each depending on one or two parameters. For the free
market, however, this is so.

Consider the system with entropy $S(E_A, E_B, V_A, V_B)$. Replacing
$E_A$ by $E_B$ and using the fact that $E_A + E_B = E$ we
see that
\begin{equation}
\label{eq79}
\delta S=\pderf{S}{E_1}\delta E_1+\pderf{S}{E_2}\delta E_2
=\delta E_1\left(\pderf{S}{E_1}-\pderf{S}{E_2}\right).
\end{equation}
In the state of equilibrium,\index{equilibrium, market} the
derivatives $\pderf{S}{E_1}$ and $\pderf{S}{E_2}$ are equal and
$V_A$ and $V_B$ are fixed. If there is an equivalence between the
items $A$ and $B$, we may ``unite'' these items and consider the
equilibrium problem. Let $V_A=V_A^0 n_A$, $V_B=V_B^0 n_B$ where
$n_A$ and $n_B$ are the amounts of goods, purchased per unit time.

Having introduced $V_A$ and $V_B$ we imply that the replacement
relation is known, i.e., we have a~common unit of measurement. So,
again, we have:
\begin{equation}
\label{eq80}
\renewcommand{\arraystretch}{1.4}
\begin{array}{c}
V_A+V_B=V, \qquad \delta V_A=\delta V_B, \\
\delta S=\pderf{S}{A}\delta A+\pderf{S}{B1}\delta B =\delta
A\left(\pderf{S}{A}-\pderf{S}{B1}\right)
\end{array}
\end{equation}
and the condition for the equilibrium state can be expressed as:
\begin{equation}
\label{eq81}
\pderf{S}{V_A}=\pderf{S}{V_B}.
\end{equation}
Observe that we made no assumptions on utility,\index{utility} but
only on possibility to replace the goods by other goods.

Now consider the free market where we know how the
entropy\index{entropy} depends on macro-parameters. As we outlined
above, the equilibrium\index{equilibrium, market} conditions with
respect to the money flow are of the form:
\begin{equation}
\label{eq82}
\pderf{S}{E_A}=\frac{N}{E_A}, \qquad
\pderf{S}{E_B}=\frac{N}{E_B}.
\end{equation}
As $E_A + E_B = E$, the equality of $\pderf{S}{E_A}$ and
$\pderf{S}{E_B}$ result in $E_A = E_B =\frac{E}{2}$. Similar
calculational techniques for $\pderf{S}{V_A}$ and
$\pderf{S}{E_B}$ produce $\pderf{S}{V_A}=\frac{N}{V_A}$ and
$\pderf{S}{V_B}=\frac{N}{V_B}$
 and, with the constraint that $V_A + V_B
= V$, $V_A = V_B =\frac{V}{2}$ . Returning to the condition of
substitution, expressed by $V_A=V_A^0 n_A$, $V_B=V_B^0 n_B$, it
means:
\begin{equation}
\label{eq83}
\frac{n_A}{n_B}=\frac{V_B^0}{V_A^0},
\end{equation}
i.e., the volume of goods, purchased in the equilibrium state of
the free market, is inversely proportional to its ``replacement
capacity''.

Recall our earlier assumption that the equilibrium of the money
flow is attained. The conditions for the equilibrium of the goods
flow takes the form
\begin{equation}
\label{eq84}
T\pderf{S}{V_A}\Big|_{E_A, E_B}=
T\pderf{S}{V_B}\Big|_{E_A, E_B}.
\end{equation}
If the money flow is constant, $T dS$ can be interpreted as the
expenditure on the purchase of the last, ``marginal'' portion of
the goods. So, we have obtained the well-known marginalists's
formulation:
\begin{thetext}
{\sl In the state of equilibrium, the amount of money spent on the
last portion of goods per unit of replacement capacity is the same
for all items of goods represented at the market. }
\end{thetext}

\section{The market at constant temperature}

Above we have considered models with economic systems with a~fixed
flow of money. In the case of actual markets, it is of course very
difficult to determine how the actual flow of money looks like.
Moreover, the flow of money is not a~parameter that affects the
equilibrium of the system, i.e., even if we know what are the flows of
money in two distinctly organized systems, we cannot say if the systems will
come to an equilibrium after they come into a~contact.

Therefore it is much more interesting to consider economic models
at a~constant temperature since the temperature is a~parameter of
equilibrium.\index{equilibrium, market}

To pass to new variables, we can apply a~mathematical technique
called {\it Legendre transformation}.\index{Legendre
transformation} The geometric meaning of the Legendre
transformation is that, for the new variables in a~system given by
a Pfaff equation\index{Pfaff equation}\index{equation, Pfaff}
(i.e., a~linear constraint on the differentials of the initial
variables that single out a~surface), one uses the coordinates on
the tangent plane for this surface, see \cite{Cou}.

This technique enables one to obtain a~large number of new
relations. Under the passage to new independent variables the
extensive quantity whose conservation was used to derive the
parameter of equilibrium\index{equilibrium, market} is replaced by a
new one in which this parameter enters linearly. It turns out that
such a~reparameterization of basic quantities is very useful. In
statistical thermodynamics,\index{statistical thermodynamics} the
quantities obtained after Legendre transformation are usually called
{\it potentials}. They possess a~number of remarkable properties.

Let us look how to work with the Legendre transformation sending
independent variables $(E, V)$ into independent variables $(T,
V)$. If we start from the main  relation for our theory
\begin{equation}
\label{eq85}
\fbox{$dE=T dS+\mu dN-PdV$}\, ,
\end{equation}
then, assuming that the temperature $T$ and the number
of particles $N$ are constants, we immediately derive by dividing (\ref{eq85}) by $dV$
that
\begin{equation}
\label{eq86}
P=-\left.\pderf{E}{V}\right|_{T, N}+T\left.\pderf{S}{V}\right|_{T, N}
.
\end{equation}
If we introduce the function $F=E-T S$~--- the Legendre
transform,\index{Legendre transform} then
\begin{equation}
\label{eq87}
P=-\left.\pderf{F}{V}\right|_{T, N}
\end{equation}
Therefore given $F$, differentiating yields the value of the price
at the constant temperature and constant number of market
agents.\index{agent}

This (\ref{eq87}) is a~very essential relation. Let our system
interact with a~thermostat\index{thermostat} at temperature $T$.
Now we are not bounded by a~particular value of the flow of money:
the system exchanges its money with the thermostat and it suffices
to assume that the total system plus thermostat satisfy the law of
money conservation.

It is not difficult to see that if the flow of goods changes from
$V_1$ to $V_2$, then at constant temperature the corresponding
variation of the flow of money will be equal to
\begin{equation}
\label{eq88}
\int\limits_{V_1}^{V_2}PdV =
-\int\limits_{V_1}^{V_2} \left.\pderf{F}{V}\right|_{T, N}dV=
F(V_1)-F(V_2).
\end{equation}

Under the increase of the flow of goods the flow of money spent in order to
make such an increase possible becomes equal to the increment of the function
$F$.

In statistical thermodynamics, \index{statistical thermodynamics}
$F$ is called the {\it free energy}.\index{free energy} We will
call $F$ the {\it free flow of money}. It is easy to see that
$F=-\ln Z$ where $Z$ is the statistical sum\index{statistical sum}
of the system.

Indeed,
\begin{equation}
\label{eq89}
F=E-T S=\sum\limits_iE_iW_i-T\sum\limits_iW_i\ln W_i,
\end{equation}
where $W_i=\displaystyle \frac{e^{E_i/T}}{Z}$ is the probability of
the system to be in the state $i$. Substituting this $W_i$ in the
formula for $F$ and computing the difference we get
\begin{equation}
\label{eq90}
F=-\ln Z.
\end{equation}
Observe that $F$ is a~negative quantity. The relation (\ref{eq88}) shows that the
free flow of money are exactly the highest expenditures possible in the
system to increase the flow of goods.

Let us show now that a~free flow of money is extremal at the most
possible configuration and at a~constant temperature and constant
flow of goods. Indeed, in the system which interacts with
a~thermostat,\index{thermostat} the total flow of money is
preserved:
\begin{equation}
\label{eq91}
dE_C+dE_T=0,
\end{equation}
where $dE_C$ is the infinitesimal increment of the flow
of money in the system and $dE_T$ is the infinitesimal increment in the flow
of
money of the thermostat.

The total entropy\index{entropy} of the system $S_C+S_T$ is
maximal, and therefore $dS_C+dS_T=0$ and the temperature $T$ of
the system is determined by the thermostat\index{thermostat} and
\begin{equation}
\label{eq92}
dE_T=T dS_T.
\end{equation}
Therefore
\begin{equation}
\label{eq93}
dE_T=-dE_C=T dS_T=-T dS_C
\end{equation}
or
\begin{equation}
\label{eq94}
dE_C-T dS_C=0, \quad dF_C=0
\end{equation}
In other words, $F$ has an extremum at the point of
equilibrium\index{equilibrium, market} between the system and the
thermostat. This extremum is the minimum which follows from the
maximality of the entropy. For any change of the situation which
leads the system out of the equilibrium, the total sum of
increments of the entropies of the system and the thermostat is
\begin{equation}
\label{eq95}
\Delta S_C+\Delta S_T \le 0.
\end{equation}
But the energy can only sneak into the system due to the
diminishing of the thermostat's\index{thermostat}
entropy:\index{entropy}
\begin{equation}
\label{eq96}
\Delta E_C=-T\Delta S_T
\end{equation}
and since $\Delta S_T\le-\Delta S_C$, it follows that
\begin{equation}
\label{eq97}
\Delta F_C=\Delta F_C-T\Delta S_C\ge 0.
\end{equation}
This means that

\begin{thetext}
{{\sl the system is in equilibrium\index{equilibrium, market}at the
minimum of $F$ at constant $V$ and $T$}, see~\cite{Ki}.}
\end{thetext}\vskip 1mm

This is a~very important property. It means that we can find the
equilibrium point by looking for the minimum of a~function which
can be computed if we know the statistical sum. To find the
statistical sum,\index{statistical sum} it suffices to know the
statistical weights\index{statistical weight} of the states with
distinct admissible values of the money flow and the temperature.
All other parameters of the system (the flow of money and the
price) can be found from the simple formulas:
\begin{equation}
\label{eq98}
E=-T^2\left.\pder{T}\frac{F}{T}\right|_{V, N},
\end{equation}
\begin{equation}
\label{eq99}
P=-\left.\pderf{F}{V}\right|_{T, N}.
\end{equation}
It is easy to see that the Legendre transformations\index{Legendre
transformation} can be applied differently passing to other
independent variables, for example, $T$ and $P$. In this case in
statistical thermodynamics\index{statistical thermodynamics} one
uses the {\it thermodynamic potential}\index{thermodynamic
potential}
\begin{equation}
\label{eq100} \Phi:=E-T S+PV.
\end{equation}
We will retain this name in our case as well. It is not difficult
to show that the thermodynamic potential attains a~minimum in an
equilibrium at constant values of $T$ and $P$. For the
thermodynamic potential, we have the following relations:
\begin{equation}
\label{eq101} d\Phi=-SdT+VdP, \quad
S=-\left.\pderf{\Phi}{T}\right|_P, \quad
V=\left.\pderf{\Phi}{P}\right|_T.
\end{equation}
These relations imply (thanks to possibility to interchange the order of
differentiation) an important relation:
\begin{equation}
\label{eq102} -\left.\pderf{S}{P}\right|_T=
\pder{P}\left.\pderf{\Phi}{T}\right|_P=
\pder{T}\left.\pderf{\Phi}{P}\right|_T
\end{equation}
or
\begin{equation}
\label{eq103} -\left.\pderf{S}{P}\right|_T=
\left.\pderf{V}{T}\right|_P.
\end{equation}
Now we can construct a~thermometer\index{thermometer} to define
the absolute temperature of an arbitrary system.

One should, of course, observe that to apply such a~thermometer in
economics will be not as easy as in physics. In physics, a~thermometer
is an arbitrarily graded physical system which comes
into a~heat contact with the system to be measured and by the
change in the state of thermometer which is supposed to be in the
heat equilibrium\index{equilibrium, market} with the body to be
measured the ``conditional temperature'' is defined.

Further, if we know the equation of the state of the thermometer, the
conditional temperature expressed, say, in the volume of mercury, as
one does in the medical thermometers, one can derive the absolute
temperature.

This procedure, however, can be performed even without beforehand
knowledge of the equation of the state of the
thermometer\index{thermometer} by performing a~number of
measurements of thermodynamic quantities.

Let $T=T(T_{arb})$, where $T_{arb}$ is an arbitrarily graded scale
of a~thermometer. Relations $(\ref{eq103})$ imply
\begin{equation}
\label{eq104}
\left.\pderf{E}{P}\right|_T=
T\left.\pderf{S}{P}\right|_T=
-T\left.\pderf{V}{T}\right|_P.
\end{equation}
We can express $\left.\pderf{V}{T}\right|_P$ in the ``empirical
scale'' in terms of $T_{arb}$:
\begin{equation}
\label{eq105}
\left.\pderf{V}{T}\right|_P=\left.\pderf{V}{T_{arb}}\right|_P\pderf{T_{arb}}{T}
\end{equation}
This implies that
\begin{equation}
\label{eq106}
\left.\pderf{E}{P}\right|_{T_{arb}}=
-T\pderf{V}{T_{arb}}\pderf{T_{arb}}{T}
\end{equation}
or
\begin{equation}
\label{eq107}
\pderf{T_{arb}}{T}=
-\frac{\left.\pderf{V}{T_{arb}}\right|_P}{\left.\pderf{E}{P}\right|_{T_{arb}}}
\end{equation}
The right-hand side of this relation only involves the functions
that can be measured on the conditional scale (\cite{LL}). If we
can measure them we can, therefore, determine the law of
dependence of the conditional scale on the relative temperature.

For real economic systems, this would mean the necessity to have
a~certain model market which one could append to an arbitrary market
and measure the variation of the derivative of the flow of goods
with respect to the conditional temperature at the fixed price and
the derivative of the flow of money with respect to the price at
the fixed conditional changes.

To collect such statistical data for a~large number of
artificially created conditions is hardly possible in reality.
Observe, however, that such a~device is possible as a~thought
experiment. In other words, a~recovery of the scale of the
absolute temperature from a~collection of statistical data is
possible, in principle.

Indeed, the fact that there is an analogue of the gas
thermometer\index{thermometer} in economics~--- free market~--- is
a favorable circumstance. As we saw above, we can calculate the
absolute temperature of the free market in terms of the mean value
of the flow of money per market agent\index{agent}. Therefore we
have thermostats\index{thermostat} with different temperatures in
the form of free markets of considerable volume and, at favorable
circumstances, we might be able to use them as instruments to
study non-free markets.

Creation of such favorable circumstances, however, is rather
expensive and requires a~great number of large-scale economic
experiments whose price is difficult even to imagine. They can,
however, be replaced, to an extent, by the study of historical cases
where such experiments were performed for some reasons but it would
be hardly possible to collect a~sufficient number of cases for one
particular non-free market.

It seems that the only possible way to study the properties of
non-free markets is mathematical modeling or computer simulation,
the results of which can be compared with the results of analysis of
specially selected historical cases.

\chapter{Thermodynamic inequalities and Le Chatelieu's
principle}

\section{Thermodynamic inequalities}

In the framework of the theory developed in this book, we get an
interesting possibility to derive inequalities that relate
different variables in an economic system in precisely the same
way as similar inequalities are obtained in thermodynamics, namely
thanks to the technique of the change of variables.

The inequalities obtained are, as we suggest, precisely the
``economic laws'' Carl Menger\index{Menger C.} wrote about. These
inequalities are essentially corollaries of assumptions that under
any spontaneous changes the system only increases its
entropy,\index{entropy} i.e., tries to pass to the most probable
state. The technique used in the proof of inequalities is based on
Jacobians (\ref{eq108}). Recall that the Jacobian of two functions
of two variables (we will not need more complicated cases) is the
determinant of partial derivatives
\begin{equation}
\label{eq108}
\frac{\partial(u, v)}{\partial(x, y)}=\left|\begin{array}{cc}{\frac{\partial
u}{\partial x}}&{\frac{\partial u}{\partial y}}\\{\frac{\partial
v}{\partial x}}&{\frac{\partial v}{\partial
y}}\end{array}\right|=\frac{\partial u}{\partial x} \frac{\partial
v}{\partial y}-\frac{\partial u}{\partial y} \frac{\partial
v}{\partial x}.
\end{equation}

Obviously, under composition of $U$ and $V$ the Jacobian changes its
sign
\begin{equation}
\label{eq109}
\frac{\partial(u, v)}{\partial(x, y)}=-
\frac{\partial(v, u)}{\partial(x, y)}.
\end{equation}

Moreover,
\begin{equation}
\label{eq110}
\frac{\partial(u, y)}{\partial(x, y)}=\frac{\partial u}{\partial
x} \frac{\partial y}{\partial y}-\frac{\partial u}{\partial y}
\frac{\partial y}{\partial x}= \left.{\frac{\partial u}{\partial
x}} \right|_y.
\end{equation}

The multiplicativity of Jacobians reflects the chain rule:
\begin{equation}
\label{eq111}
\frac{\partial(u, v)}{\partial(x, y)}=\frac{\partial(u, v)}{\partial(q, p)}
\frac{\partial(q, p)}{\partial(x, y)}.
\end{equation}

By using the properties of Jacobians we may easily derive various
relations between derivatives of the parameters of the system
(\ref{eq109}). Actually, the basic inequality $\delta S\leq 0$ in
the equilibrium\index{equilibrium, market} state can be expressed
by a~multitude of ways that show what restrictions imposes this
inequality onto relations between various variables in the system
and their derivatives. Consider an important example.

The thermodynamic potential\index{thermodynamic potential} $\Phi$
given as $E-T_0 S+P_0V$, where $T_0$ and $P_0$ are fixed, attains
its minimum at the equilibrium point. Therefore
\begin{equation}
\label{eq112}
\delta \Phi = \delta E -T_0
\delta S+P_0 \delta V > 0
\end{equation}
under deviations from equilibrium. Let us decompose the flow of
money, as a~function of entropy\index{entropy} $S$ and the flow of
goods $V$, into the variations $\delta \Phi$ and $\delta V$ up to
the second order of magnitude:
\begin{equation}
\label{eq113}
\delta E=\frac{\partial E}{\partial S} \delta S+\frac{\partial
E}{\partial V}\delta V +\frac{1}{2}\left(\frac{\partial^2
E}{\partial S^2}(\delta S)^2+ 2\frac{\partial^2 E}{\partial S \partial
V}\delta S\delta V +
\frac{\partial^2 E}{\partial V^2}(\delta V)^2 \right).
\end{equation}

Further, since at the equilibrium point, where
\begin{equation}
\label{eq114}
\frac{\partial
E}{\partial S}=T\text{\quad and \quad}\frac{\partial
E}{\partial V}=-P,
\end{equation}
we have $T =T_0$ and $ P=P_0$, it follows that substituting $\delta
E$ into $\delta \Phi$ and canceling terms linear in $\delta S$ and
$\delta V$ we get
\begin{equation}
\label{eq115}
\frac{\partial^2 E}{\partial S^2}\delta S^2+ 2\frac{\partial^2
E}{\partial S \partial V}\delta S\delta V + \frac{\partial^2
E}{\partial V^2}\delta V^2 > 0.
\end{equation}

This inequality holds provided the following two conditions are
fulfilled:
\begin{equation}
\label{eq116}
\frac{\partial^2 E}{\partial S^2}>0, \qquad
\frac{\partial^2 E}{\partial S^2} \frac{\partial^2 E}{\partial
V^2} -\left(\frac{\partial^2 E}{\partial S
\partial V} \right)^{\!2} >0.
\end{equation}

The first of them means that,
under the constant flow of goods, to increase the flow of money, we
have to increase the temperature if it is positive or decrease
the temperature if it is negative. Indeed, by the definition of
the temperature,
\begin{equation}
\label{eq117}
\frac{\partial^2 E}{\partial S^2}=\left.
\frac{\partial T}{\partial
S}\right|_V=\frac{T}{\displaystyle
\left.T\cdot\frac{\partial S}{\partial T} \right|_V}.
\end{equation}

In statistical thermodynamics,\index{statistical thermodynamics}
the expression in the denominator is called the {\it heat capacity
under constant volume}. In our case this is the quantity by which
we have to augment the flow of money in the system in order to
heat it by a~unit of temperature. We will call it {\it the heat
capacity of the market under constant flow of goods}\index{heat
capacity of the market under the constant flow of goods} $C_V$. We
thus derive that
\begin{equation}
\label{eq118}
\frac{T}{C_V} > 0.
\end{equation}

This is the first of thermodynamic inequalities\index{inequality,
thermodynamic, first} we intend to obtain.

The second one\index{inequality,
thermodynamic, second} can be expressed in terms of Jacobians as
\begin{equation}
\label{eq119}
\frac{\displaystyle \partial \left( \left. \frac{\partial
E}{\partial S }\right|_V , \left. \frac{\partial E}{\partial V
}\right|_S\right)}{\partial (S, V)}>0.
\end{equation}
or, which is
the same, as
\begin{equation}
\label{eq120}
\frac{\partial (T , P)}{\partial (S, V) }<0,
\end{equation}
where $T$ is the temperature and $P$ is the price. Let
us replace variables $S$, $V$ by $T$, $P$. We get
\begin{equation}
\label{eq121}
\frac{\partial (T , P)}{\partial (S, V)}=\frac{\displaystyle
\frac{\partial (T , P)}{\partial (T , V)}}{\displaystyle
\frac{\partial (S, V)}{\partial (T , V)}}=\frac{\displaystyle
\left.\frac{\partial P}{\partial V}\right|_T }{\displaystyle
\left.\frac{\partial S}{\partial
T}\right|_V}=\left.\frac{T}{C_V}\frac{\partial P}{\partial
V}\right|_T <0.
\end{equation}

Having recalled the preceding inequality (\ref{eq118}) we deduce that
\begin{equation}
\label{eq122}
\left.\frac{\partial P}{\partial V}\right|_T <0,
\end{equation}
In other words,

\begin{thetext}
{\sl under the increase of the flow of goods at constant
temperature the prices plunge down.}
\end{thetext}

Observe that we have proved a~quite general and far from obvious
statement. Note also how this statement was derived. We have
started from the fact that $\delta \Phi >0$, and having expanded
$\delta E$ into a~series with respect to $\delta S$ and $\delta
V$, obtained a~new inequality. But this inequality means that
$\delta \Phi$ has a~strict minimum at the
equilibrium\index{equilibrium, market} point. The relation
obtained is nothing but the condition on the convexity of the
domain determined by the condition $\Phi > \Phi_0$ in a~neighborhood
of the equilibrium point. Under the passage to new
variables the transformations of Jacobians fix the same fact. The
thermodynamic inequalities obtained is just the same statement
expressed in other terms.

This is an extremely general method for obtaining thermodynamic
relations. Usually this circumstance is not being stressed in the
courses of thermodynamic but in the base of various thermodynamic
relations lie simple statements on convexity or concavity of
certain functions primarily the entropy\index{entropy} and also
the free energy\index{free energy} and thermodynamic
potential.\index{thermodynamic potential}

The fact that for a~twice differentiable function to have a~strict
maximum the second differential should be negative is sufficient to
make statements on convexity of this function in a~neighborhood of the
maximum. We have seen what inequalities are needed to express this
fact in terms of the derivatives. We can express these inequalities
in various coordinate systems; the property of the function to be
convex in a~neighborhood of the maximum is coordinate invariant. This
is exactly the fact that lies in the method of obtaining these
inequalities. This is the place to draw a~parallel with the role of
convex admissible domains in the standard problems of mathematical
economics.

In these problems the convexity is important to establish such
properties as the uniqueness of the maximum of the
utility\index{utility} function or the existence of a~fixed point
of the maps. We see that there exists something in common between
the standard approach of mathematical economics and the
thermodynamic approach developed in this book: the differential
and topological invariants are characters of importance. From a~very
abstract point of view any ``equilibrium theory'' under all
the distinctions between mechanical and thermodynamic approaches
mentioned above should be a~theory of critical points of maps,
i.e., a~branch of differential topology \cite{M}. Regrettably, as
we have already mentioned, this fact did not draw due attention.

Let us now define the expression for heat capacity of the market
under constant price as the change in the flow of money needed to
raise the temperature by one unit under the constant price in the
form
\begin{equation}
\label{eq123}
C_P=\left.T \frac{\partial S}{\partial T}\right|_P.
\end{equation}
Let us now try to compare $C_V$ and $C_P$~--- the heat capacities
of the market under constant price and constant value:
\footnotesize
\begin{equation}
\label{eq124}
\begin{split}
 C_V=T \displaystyle\frac{\partial (S, V)}{\partial
(T , V)}=T \frac{\displaystyle \frac{\partial (S, V)}{\partial
(T , P)}}{\displaystyle \frac{\partial (T , V)}{\partial (T
, P)}}
=\frac{\displaystyle T
\left(\left.\frac{\partial S}{\partial
T}\right|_P\left.\frac{\partial
V}{\partial P}\right|_T-\left.\frac{\partial S}{\partial
T}\right|_T\left.\frac{\partial
V}{\partial P}\right|_P\right)}{\displaystyle \left.\frac{\partial
V}{\partial P}\right|_T}
=T
\left. \displaystyle\frac{\partial S}{\partial
T}\right|_P-T\displaystyle \frac{\displaystyle \left.\frac{\partial S}{\partial
T}\right|_T\left.\frac{\partial
V}{\partial P}\right|_P}{\displaystyle \left.\frac{\partial
V}{\partial P}\right|_T}.
\end{split}
\end{equation}

\normalsize

If we substitute into this formula the earlier obtained relation
$\displaystyle \left.\frac{\partial S}{\partial P} \right|_T
=-\left.\frac{\partial V}{\partial T} \right|_P$ and recall that
$\displaystyle C_P=\left.T\frac{\partial S}{\partial T}\right|_P,
$ we get
\begin{equation}
\label{eq125}
C_P=C_V-T\frac{\displaystyle {\left( {\frac{\partial V}{\partial
T}}\right)_{\!P}^{\!2}}} {\displaystyle \left. {\frac{\partial V}{\partial
P}}\right|_T}.
\end{equation}

We have just established that $\displaystyle \left.{\frac{\partial
P}{\partial V}} \right|_T<0$ which means that, at positive
temperature, we have
\begin{equation}
\label{eq126}
C_P>C_V.
\end{equation}

In other words, in order to heat the market by a~unit of
temperature under the constant flow of goods we need lesser flow
of money than in order to perform the same under the constant
price, i.e.,
\begin{thetext}
{\sl under the increase of the flow of money the temperature of
the market under the constant flow of goods grows faster than
under the constant price.}
\end{thetext}

Translating this statement into the language of economic
politics, we may say that in the conditions of restriction onto
import the inflation in the system grows faster than under the
conditions of free market if the world prices are constant. (Under
the conditions of free market the price will be determined by the
prices of the world market.)

Of course as we have repeatedly underlined, statements of this
type are only valuable for the models. Under a~model we understand
here ideally typically abstraction in the sense of Weber,
abstraction from certain though essential perhaps but immaterial
for the {\sl given model factors}. Nevertheless, it was only with
this type of models that theoretical economics dealt with,
starting with Walras,\index{Walras L.} Javons, Menger\index{Menger
C.} through Arrow,\index{Arrow K.} Samuelson\index{Samuelson P.}
and Debreu.\index{Debreu G.}

\section{The Le Chatelieu principle}

The {\it Le Chatelieu principle}\index{principle, Le Chatelieu}
reflects the following remarkable fact:
\begin{thetext}
{\bf systems in their strife to the most probable state are
capable to exert a~certain resistance to the modification of
external conditions. Under such a~modification they produce
``forces'' that resist the modifications.}
\end{thetext}
In order to see this, let us consider a~system consisting of a~market
submerged into a~universum. Suppose there are two
parameters, $\alpha$ and $x$ that parameterize the market and $S$
is the entropy\index{entropy} of the system as a~whole. Let
further the parameter $\alpha$ characterize the inner
equilibrium\index{equilibrium, market} of the market, i.e., if $
\displaystyle \frac{\partial S}{\partial \alpha }=0 $ this means
that the market is already in the most probable state but not
necessarily in the equilibrium with the universum. If, moreover,
$ \displaystyle \frac{\partial S}{\partial x}=0 $ this means
that the equilibrium exists also between the universum and the
market. Under disequilibrium conditions the quantities $
\displaystyle A=-\frac{\partial S}{\partial \alpha }\: \text{ and
}\: X=-\frac{\partial S}{\partial x}$ will be nonzero. Since in
equilibrium the entropy\index{entropy} should be maximum this
imposes certain restrictions not only on A and X but also on their
derivatives with respect to $\alpha $ and $x$ in the precisely the
same way as in the preceding section where we have obtained
thermodynamic inequalities. Under equilibrium we should have
\begin{equation}
\label{eq127}
\begin{split}
 &A=0, \quad X=0, \\
 &\displaystyle \left.{\frac{\partial A}{\partial \alpha
}}\right|_\alpha
>0, \quad \left.{\frac{\partial X}{\partial x}}\right|_\alpha >0, \\
 &\displaystyle \left.{\frac{\partial X}{\partial x
}}\right|_\alpha \left.{\frac{\partial A}{\partial \alpha
}}\right|_x - \left({\frac{\partial X}{\partial \alpha
}}\right)_x^2>0.
\end{split}
\end{equation}

Let now as a~result of a~certain external influence the
equilibrium\index{equilibrium, market} of a~market with the
universum is violated. This means that $X$ does not vanish any
more and the parameter $x$ of the equilibrium of the market with
the universum will change. Accordingly, $X$ will change by $\Delta
X_H$. This will lead to the fact that the inner parameter $\alpha$
that determines the equilibrium of the market as such will also
change. In turn, the change of the parameter $x$ will lead to a~new
change of the value of $X$ as compared with the initial
deviation so the resulting deviation at $A=0$ (i.e., under the
restitution of the market equilibrium)\index{equilibrium, market}
will be equal to $\Delta X_K$.

The question is where will the processes that restore the market
equilibrium move the quantity $\Delta X$? It turns out that these
processes lead to the diminishing of the initial deviation, i.e.,
$|\Delta X_H| > |\Delta X_K|$. Indeed,
\begin{equation}
\label{eq128}
\begin{split}
 \Delta X_H &=\left.{\frac{\partial X}
{\partial x}}\right|_\alpha \Delta x, \\
\Bigl.\Delta X_K\Bigr|_{A=0} &=
\left.{\frac{\partial X} {\partial x}}\right|_{A=0}\Delta x.
\end{split}
\end{equation}

The inequality desired is obtained by the method similar to the
one described in the preceding section:
\begin{equation}
\label{eq129}
\left.{\frac{\partial X}{\partial x}}\right|_{A=0}=
\frac{\partial (X, A)}{\partial (x, A)}= \frac{\displaystyle
\frac{\partial (X, A)}{\partial (x, \alpha )}}{\displaystyle
\frac{\partial (x, A)}{\partial (x, \alpha )}}=
\left.{\frac{\partial X}{\partial x}}\right|_\alpha -
\frac{\displaystyle \left.\frac{\partial X}{\partial \alpha
}\right|_x^2}{\displaystyle \left.\frac{\partial A}{\partial
\alpha }\right|_x}.
\end{equation}

Since thanks to equilibrium $\displaystyle \left.{\frac{\partial
A}{\partial \alpha }}\right|_x>0 $, we deduce the result desired.
Taking into account that
\begin{equation}
\label{eq130}
\left.{\frac{\partial X}{\partial x}}\right|_\alpha
\left.{\frac{\partial A}{\partial \alpha
}}\right|_x-\left(\left.{\frac{\partial X}{\partial \alpha
}}\right|_x\right)^{\!2} >0,
\end{equation}
we have
\begin{equation}
\label{eq131}
0<\left.{\frac{\partial X}{\partial
x}}\right|_{A=0}<\left.{\frac{\partial X}{\partial x
}}\right|_\alpha
\end{equation}
or
\begin{equation}
\label{eq132}
\left|\Delta X_K\right|<\left|\Delta X_H\right|,
\end{equation}
i.e., the processes that restore equilibrium\index{equilibrium,
market} inside the market do indeed partly compensate the
influence leading to the destruction of the equilibrium with the
universum. We see that systems of thermodynamic type possess
certain homeostatic properties that try to resist the disturbance
of the equilibrium state. This is the Le Chatelieu principle.

\begin{figure}[ht]\centering
\includegraphics{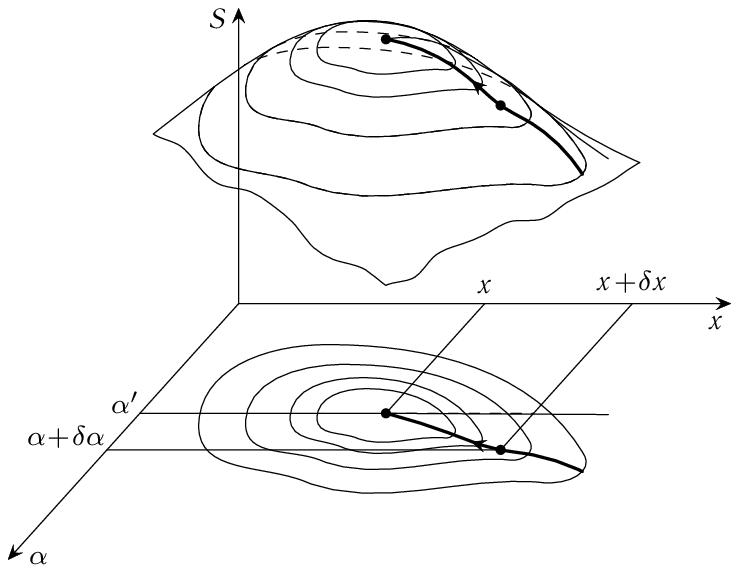}
\caption{}
\label{f1}
\end{figure}

Let us explain the obtained result more graphically. Fig.~\ref{f1}
depicts the situation considered.

After changing $x$ by $\delta x$ the system loses equilibrium.
Nevertheless, the equilibrium begins to be reestablished along the
parameter $\alpha$ which in order to maximize the
entropy\index{entropy} moves into the position $\alpha '$ in which
$\displaystyle \left.{\frac{\partial S}{\partial \alpha
}}\right|_{\alpha =\alpha '} =0.$ Obviously, during this we have
\begin{equation}
\label{eq134}
\left|\delta S(x+\delta x, \alpha ')\right| >
\left|\delta S(x+\delta x, \alpha '+\delta \alpha ')\right|.
\end{equation}

It is clear that the steepness of the surface $S(x, \alpha )$ will
be lesser after a~certain increase of the entropy under the
process of relaxation of the system along the parameter $\alpha$.

We see again that the character of the singularity of the function
S namely the presence of a~maximum, and therefore the convexity of
$S(x, \alpha )$ determines essentially the validity of the
inequalities constituting the essence of the Le Chatelieu
principle.

The Le Chatelieu principle allows one also to derive thermodynamic
inequalities. Thus if for a~parameter $x$ we take the market's
entropy\index{entropy} and for a~parameter $\alpha$ the flow of
goods $V$, then the Le Chatelieu principle\index{principle, Le
Chatelieu} will imply the already known inequalities
\begin{equation}
\label{eq135}
C_P>C_V>0.
\end{equation}

If for $x$ we take the flow of goods to the market and for
$\alpha$ the entropy of the market then the Le Chatelieu principle
implies
\begin{equation}
\label{eq136}
\left.{\frac{\partial P}{\partial V }}\right|_S <
\left.{\frac{\partial P}{\partial V }}\right|_T<0
\end{equation}
In other words, the change of the price under
the change of the flow of goods by a~unit under the constant
entropy is negative and its absolute value is smaller than the
change of the price under the change of the flow of goods by one
under the constant temperature.

Observe that our deduction of thermodynamic inequalities for
economic models {\sl does not differ at all} on the deductions of
thermodynamic inequalities in statistical
physics.\index{statistical thermodynamics}

These inequalities do not depend on the nature of the systems.
They are just another form of expression of a~fundamental
inequality of thermodynamics directly related with the law of
entropy growth, namely that in equilibrium we have $\delta S\leq
0$ plus the assumption on differentiability of thermodynamic
functions.

In other words, both the Le Chatelieu principle\index{principle,
Le Chatelieu} and the thermodynamic inequalities are corollaries
on our hypothesis that, in equilibrium, the system attains the
most probable state and this most probable state embodies the
maximum of a~twice differentiable function.

Essentially, thermodynamics is not a~physical theory. {\bf
Thermodynamics is a~theory of how our knowledge on the possible
states of elementary systems armed with conjectures on a~priori
probabilities of these states determines the most probable
(equilibrium) states of the more complicated compound systems}.

\chapter{Market fluctuations}

\section{Mean values of fluctuations}

Markets fluctuate.\index{market fluctuations} This is well known.
If we observe how market parameters vary with time for example,
consider prices or the amount of goods sold we observe certain
fluctuations of these parameters about their
equilibrium\index{equilibrium, market} states. The reasons for
these fluctuations may be various but for us it is necessary to
differentiate two important types of fluctuations.

The first type of fluctuations is the result of the fact that
markets can rarely if ever be considered isolated. As a~rule,
markets are parts of larger markets and even if the system as a~whole,
i.e., not only the directly observed part but other
connected with it are in equilibrium certain random deviations are
possible which will be the larger the smaller system is being
considered.

The second type of fluctuations are fluctuations due to
speculations and we will study them at the end of this chapter.

We have already seen that the values of the mean quadratic
deviation are inversely proportional to the square root of the
number of the economic agents of the market. In what follows we
will show what kind of theory can be used for calculating mean
quadratic fluctuations. This theory is again identical to the
fluctuation theory known from the statistical physics. But in our
case~--- being applied to economics
--- the fluctuation theory is especially viable because market
fluctuations\index{market fluctuations} is a~quite accessible
procedure and we will see in what follows the mean value of market
fluctuations become related with thermodynamic parameters of the
markets. This means that the measurements of the mean values of
fluctuations can be used to determine thermodynamic parameters of
the markets in particular to define the temperature.

It is hardly needed to explain how important this may be for the
construction of a~theory. We obtain at last in our hands the
measurement tool that can replace special experimental conditions.

Nevertheless, everything said about the artificial reality of the
experiment remains valid. In addition to purely probabilistic
factors related with the peculiarities of the market structure the
fluctuations of the quantities to be measured such as prices and
volumes of the goods will be affected by other non-market factors
such as social, political, demographic and so on.

Our thermodynamic model of the market ignores these extra facts
whereas in reality we cannot get rid of them at best reduce their
influence to a~minimum by selecting specific moments for the
measurements or excluding certain data. As before we face the same
dilemma. In order to actually verify a~theory one needs to create an
artificial reality or, at least, select particular cases which will be
close to an artificial reality so that the influence of the factors
not included in the model will be reduced to a~minimum.

In this sense to measure fluctuations in the given systems at hand
is much more economic way to study economic systems than special
constructing of experimental situations which among other things
can hardly be possible because of an incredibly high price of such
experiments. Thus, let us see how the study of market
fluctuations\index{market fluctuations} can replace such special
experiments. In this section we will assume that market
fluctuations appear by themselves because of the qualities of the
market as such, i.e., we will consider the enforced changes of
macroparameters of the market and land speculations non-existent.
In the next section we will consider the reaction of the system on
the enforced changes. Moreover, what is very essential, we will
only consider fluctuations in a~neighborhood of the
equilibrium\index{equilibrium, market} point. We will see further
that there exist domains of disequilibrium of the markets where a~random
fluctuation may lead to macroscopic and sometimes however
great deviations from the equilibrium~--- a~catastrophe.

It is natural to assume in complete agreement with the man
postulate of the statistical theory that the probability of the
system to be in a~state described by a~macro-parameter $X$ is
proportional to the number of microstates corresponding to this
value of the macro-parameters. In other words, the probability
$W(X)$ of the system to be in the macro-state $X$ is proportional
to the statistical weight\index{statistical weight} $g(X)$ of this
state. Recall that these basic principles of the statistical
theory are never proved. We define the elementary microstates to
be equally probable. This easily implies that the probability of
the system to be in the macro-state $X$ is proportional\footnote{
In (\ref{eq137}), we set the proportionality coefficient equal to
1.} to the exponent of the entropy\index{entropy} of this
macro-state just due to the definition of the entropy as the
logarithm of the statistical weight:
\begin{equation}
\label{eq137}
W(x)= g(x)=e^{\ln{g(z)}}=e^{S(x)}.
\end{equation}

We see therefore that if we are only interested in the relation
between macro and micro parameters of the system, there is no
difference between physical and economic systems.

If we consider a~system in an equilibrium\index{equilibrium,
market} state $X_0$ we can expand the entropy $S(X_{0}+\Delta X)$
in the series with respect to $X$. Since, by definition of the
equilibrium state, the entropy is maximum at we have
\begin{equation}
\label{eq138}
\left.{\frac{\partial S}{\partial X}}
\right|_{X_0}=0, \qquad \left.{\frac{\partial^{2}S}{\partial
X^{2}}} \right|_{X_0}<0,
\end{equation}

Again the properties of the singular point of a~map enable us to
construct a~theory. We see that the probability $W(x)$ of the
system to possess the value $X$ of the macro-parameter that
differs by $\Delta X$ from the equilibrium state $X_0$ is
proportional to
\begin{equation}
\label{eq139}
W(X_{0}+\Delta X)= e^{S_{0}-\frac{1}{2}\frac{\partial^{2}S}
{\partial X^2}\Delta X^2}.
\end{equation}

Since the exponent decreases very rapidly as the argument grows,
the role of $W(X_{0}+\Delta X)$ for large values of $\Delta X$ is
insufficient actually and we can with good accuracy obtain the
normalized constant for the probability distribution integrating
$W$ along $\Delta X$ from $-\infty$ to $\infty$.

Thus we obtain
\begin{equation}
\label{eq140}
\int_{-\infty}^{\infty}dW(X_{0}+\Delta X)=
A\int_{-\infty}^{\infty}e^{-\frac{\Delta
X^2}{2}}\left.{\frac{\partial^{2}S}{\partial X^{2}}}
\right|_{X_0}d\Delta X=1
\end{equation}
or
\begin{equation}
\label{eq141}
W(X_{0}+\Delta
X)=\sqrt{\frac{1}{2\pi}}\left.{\frac{\partial^{2}S}{\partial
X^{2}}} \right|_{X_0}e^{-\left.{\frac{\Delta
X^2}{2}\frac{\partial^{2}S} {\partial X^2}\Delta
X^2}\right|_{X_0}}.
\end{equation}

It is not difficult to deduce from
here the mean square of the fluctuation:
\begin{equation}
\label{eq142}
\overline{\Delta
X^2}=\sqrt{\frac{1}{2\pi}}\left.{\frac{\partial^{2}S} {\partial
X^{2}}} \right|_{X_0}\int_{-\infty}^{\infty}(\Delta
X)^{2}e^{{-\frac{\partial^{2}S} {\partial X^2}\Delta X^2}}d\Delta
X=\frac{1}{\left.{\frac{\partial^{2}S}{\partial X^{2}}}
\right|_{X_0}}.
\end{equation}

Therefore, for the probability of systems deviation from the
equilibrium state,\index{equilibrium, market} we obtain the Gauss
distribution:\index{Gauss distribution}\index{distribution, Gauss}
\begin{equation}
\label{eq143}
W(X)=\frac{1}{\sqrt{2\pi\overline{\Delta
X^2}}}e^{-\frac{\Delta X^2}{2\overline{\Delta X^2}}}.
\end{equation}

Since $\Delta X$ is small and the probability steeply drops as $X$
grows, we can simply find the mean square of any function $f(X)$
by expanding it into the Taylor series and confining to the first
term:
\begin{equation}
\label{eq144}
\overline{\Delta f^2}=\left({\left.{\frac{\partial f}{\partial
X}}\right|_{X=X_0}}
\right)^{\!2}\overline{\Delta X^2}.
\end{equation}

To compute the mean of the product of the fluctuations of
thermodynamic quantities, observe that the mean of the fluctuation
vanishes thanks to the symmetry of the distribution function
relative to the point $\Delta X=0$. The mean of the product of
the fluctuations of independent values $\overline{\Delta \alpha
\Delta b}$ also vanishes since for the independent values we have
$\overline{\Delta \alpha \Delta b}=\overline{\Delta
\alpha}\cdot\overline{\Delta b}=0$.

Now consider the mean of the product of fluctuations, which are
not independent. We will need approximately the same technique of
dealing with thermodynamic quantities that we have already used to
derive thermodynamic inequalities. If a~fluctuation occurs in a~portion
of the market, which is in equilibrium,\index{equilibrium,
market} this means that we can assume the temperature of the
universum and the price constants in the first approximation (for
a fluctuation). This in turn means that the deviations of the flow
of money from the equilibrium will be given by thermodynamic
potential\index{thermodynamic potential} in accordance with
arguments given in \S 6.3\footnote{Here we follow \cite{LL}.}
\begin{equation}
\label{eq145}
\Delta\Phi=\Delta E-T_{0}\Delta S+P_0\Delta V,
\end{equation}
where $T_0$~--- is the equilibrium temperature and $P_0$~--- is
the {\it equilibrium price}.\index{equilibrium, market}

Indeed, the entropy\index{entropy} of the market is a~function on
the money flow. If we perform a~modification of this flow in a~part
of the system then the entropy of the system as a~whole will
change:
\begin{equation}
\label{eq146}
\Delta S=\frac{\partial S}{\partial E}\cdot\left.{\Delta
E}\right|_{P_{0}, T_0}=T_{0}\Delta \Phi .
\end{equation}

The changes of $\Phi$ can be found by expanding $E$ into the
series with respect to $\delta S$ and $\delta V$. Observe here
that the distribution function depends on the total change of the
entropy of the system under the fluctuation whereas $\delta S$ and
$\delta V$~--- are the changes of entropy and the goods flow only
for the separated part of the system.

In the same way as above we have
\begin{equation}
\label{eq147}
\renewcommand{\arraystretch}{1.4}
\begin{array}{l} \mathstrut\displaystyle{\Delta\Phi=\Delta E-T\delta S+P\delta
V}\\
\mathstrut\qquad\displaystyle{=\frac{1}{2}\left({\frac{\partial^2
E}{\partial S^2}\delta S^2+ 2\frac{\partial^{2}E}{\partial
S\partial V}\delta S\delta V +
\frac{\partial^2 E}{\partial V^2}\delta V^2}\right)}\\
\mathstrut\qquad\displaystyle{=\frac{1}{2}\left({\delta
S\delta\left(\left.{\frac{\partial E}{\partial
V}}\right|_{S}\right)}\right)=\frac{1}{2}(\delta S\delta T-\delta
V\delta P)}\end{array}
\end{equation}
(where the first order terms cancel).

Here we see the mathematical meaning of the variation of the
thermodynamic potential\index{thermodynamic potential}
$\Delta\Phi$. It shows how much the money flow deviates from the
tangent plane to the surface of state $E=E(V, T)$.

We can now express this change of $\Delta\Phi$ in various coordinate
systems. Having selected for example, the variables $\delta
V, \delta T$ we can express $\delta S\text{ and }\delta P$ in terms
of $\delta V, \delta T$. After simplifications we obtain
\begin{equation}
\label{eq148}
\delta P\delta V-\delta T\delta S=-\frac{C_V}{T}\delta T^{2}+
\left.{\frac{\partial P}{\partial V}}\right|_{T}\delta
V^{2}.
\end{equation}

The probability of fluctuation under the deviation of the systems
from equilibrium\index{equilibrium, market} is accordingly
proportional to the product of two factors depending on $\delta V$
and $\delta T$
\begin{equation}
\label{eq149}
W(\delta T, \delta V)\approx
e^{-\frac{1}{2T}\left(-\frac{C_V}{T}\delta T^{2}+
\left.{\frac{\partial P}{\partial V}}\right|_{T}\delta
V^{2}\right)},
\end{equation}
i.e., $\overline{\delta V\delta T}=0$.

It is not difficult to compute the mean quadratic values of the
fluctuations by comparing $W(\delta T, \delta V)$ with the Gauss
distribution\index{Gauss distribution}\index{distribution, Gauss}
formula\footnote{See, e.g., \cite{Tr}.}:
\begin{equation}
\label{eq150}
\renewcommand{\arraystretch}{1.4}
\begin{array}{l} \mathstrut\displaystyle{\overline{(\delta
T)^2}=\frac{T^2}{C_V}, }\\
\mathstrut\displaystyle{\overline{(\delta
V^2)}=-T\left.{\frac{\partial V}{\partial
P}}\right|_{T}.}\end{array}
\end{equation}

This gives for the mean value the following value:
\begin{equation}
\label{eq151}
\overline{\delta P\delta V}=\overline{\left({\left.{\frac{\partial
P}{\partial V}}\right|_{T}\delta V-\left.{\frac{\partial
P}{\partial T}}\right|_{V}\delta T}\right)\delta
T}=\left.{\frac{\partial P}{\partial
V}}\right|_{T}\overline{\delta V^2}=-T.
\end{equation}

This relation enables us to measure the temperature of the market
by computing the mean of the product of the fluctuation of the
market by the fluctuation of the flow of goods. Since the
averaging over the ensemble can be replaced by averaging over
time, we obtain the following formula for the empirical computation
of the market's temperature:
\begin{equation}
\label{eq152}
T=-\frac{1}{2T}\lim\int_{0}^{T}(V(t)-\bar{V})(P(t)-\bar{P})dt.
\end{equation}
Both the dependence of the
price on time and the dependence of the flow of goods on time are
accessible, in principle, data, say, for the stock market.
Therefore if we assume that no external factors not determined by
the structure of the market as such influence the prices and the
flows of goods then we have a~means to measure thermodynamic
parameters of markets.

\section{Fluctuations in time}

Let us see how one can consider the dependence of fluctuations on
time in the system led out of the equilibrium state. Observe here
that we may only consider the not too large deviations from
equilibrium states but, on the other hand, not too small
ones.\footnote{Voluminous literature is devoted to the study of
time series corresponding to the market prices, see, e.g.,
\cite{Mi}. We will not discuss here various methods of analysis
and prediction of prices since they are based on universal
mathematical properties of time series and various probabilistic
hypotheses whereas in our approach the market is considered as
a~system with a~specific property~--- equation of state~--- which
takes into account the structure of dependence of the
entropy\index{entropy} of macroparameters of the market.}

If the initial deviation of the equilibrium
state\index{equilibrium, market} is very tiny, the dynamics of the
fluctuations will not differ from the chaotic spontaneous
fluctuations. If, on the other hand, the initial deviation is very
large, one has to take into account the non-linear effects on the
dependence on the speed of the deviation of the quantity under the
study on the value of the initial deviation.

Therefore we will confine ourselves to a~linear case that is we will
assume that in the dependence of the speed with which the quantity
returns to the equilibrium state on the deviation we can ignore all
the in the Taylor series expansion except the first one.

Speaking about practical applications of such an approach for
prediction of behavior of time series, say of the price on the
share and stock markets this means that reasonable predictions can
be only made for short periods of time when the prices are still
capable to return to the equilibrium state but the time spans are
still larger than the value of dispersion (see Fig.~\ref{f2} on which
overbraced are relaxation periods; the time segments $[t_{i_n},
t_{i_k}]$ (subscripts $n$ and $k$ stand for the first letters of
Russian words ``beginning'' and ``end'') are the periods during
which a~prediction can be significant). One can determine the
value of dispersion of $\Delta X$ by means of the arguments from
the preceding section.

\setcounter{figure}{1}

\begin{figure}[ht]\centering
\includegraphics{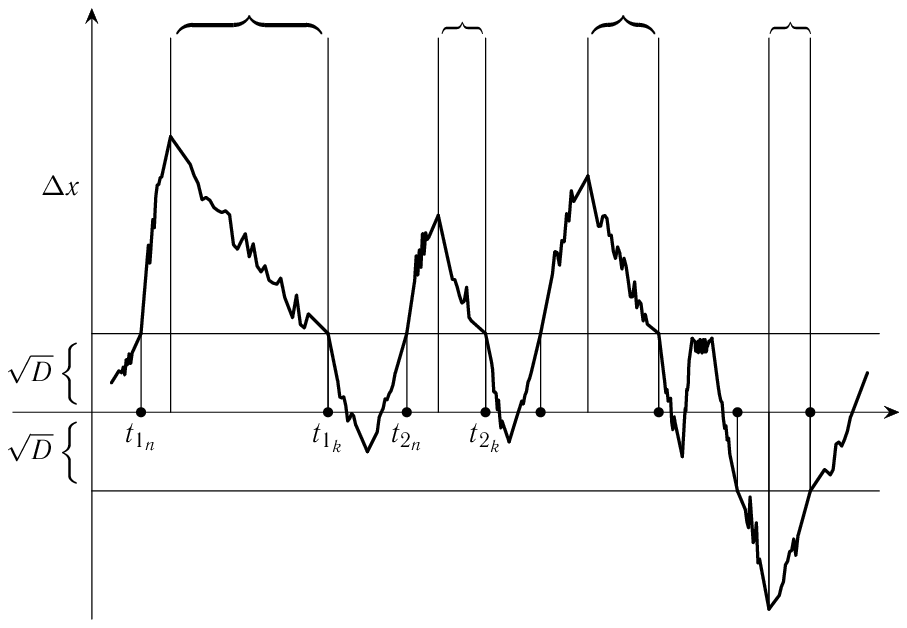}
\caption{}
\label{f2}
\end{figure}

Observe that such a~prediction is possible far from all markets.
As we will see in what follows in order for the purely
thermodynamic approach to work it is necessary that the ``shadow
of future'' does not affect too much the behavior of the market
agents\index{agent} and the existence of a~certain symmetry
between the sellers and the buyers.

In other words, the thermodynamic approach will hardly be
effective for stock exchange, where the fulfillment of both of the
above requirements is hard to imagine but it can certainly be
applicable for the commodity markets.

In order to construct the theory of time fluctuations we have to
introduce an important value called autocorrelation function. It
is defined as the average over the ensemble of the product of the
values of the quantities studied separated by a~fixed time
interval $t_0$:
\begin{equation}
\label{eq154}
\varphi(t_0)=\langle\Delta X(t)\Delta X(t+t_0)\rangle .
\end{equation}

This quantity enables to determine the spectral density of the
fluctuations, that is the probability that the ``frequency'' of
the fluctuation belongs to a~certain interval. Here speaking about
``frequency'' of fluctuations we use a~metaphorical
language\index{metaphor} because in actual fact we have in mind
the existence of processes with a~certain characteristic
relaxation time $t_0$.

The frequency is inversely proportional to the relaxation time:
\begin{equation}
\label{eq155}
\omega=\frac{1}{t_0}.
\end{equation}

In a~more formal presentation, the arguments on a~relation of the
frequency of fluctuations and the relaxation time are as follows.
Consider the Fourier transform of $\Delta X(t)$:
\begin{equation}
\label{eq156}
\Delta X_{\omega}=\frac{1}{2\pi}\int\limits_{-\infty}^{\infty}
\Delta X(t)e^{i\omega t}dt.
\end{equation}

Then $\Delta X(t)$ can be considered as the inverse Fourier
transform of $\Delta X_{\omega}$:
\begin{equation}
\label{eq157}
\Delta X=\int\limits_{-\infty}^{\infty}\Delta X_{\omega}e^{-i\omega
t}d\omega.
\end{equation}

This expression for $\Delta X(t)$ can be substituted into the
definition of the autocorrelation function for $\Delta X(t)$:
\begin{equation}
\label{eq158}
\varphi(t_0)=\left\langle\int\!\!\!\int\limits_{\!\!\!-\infty}^{\!\!\!\infty}
\Delta X_{\omega}\Delta X_{\omega'}
e^{-i(\omega+\omega')t}e^{-i\omega t_0}d\omega
d\omega'\right\rangle.
\end{equation}

Now observe that, in accordance with the main principles of
statistical thermodynamics,\index{statistical thermodynamics} the
averaging over the ensemble can be replaced by averaging over
time.
\begin{equation}
\label{eq159}
\varphi(t_0)=\int\!\!\!\int\limits_{\!\!\!-\infty}^{\!\!\!\infty}\Delta
X_{\omega} \Delta X_{\omega'}
\left({\frac{1}{T}\int\limits_{-T}^{T}e^{-i(\omega+\omega')t}dt}\right)
e^{-i\omega
t_0}d\omega d\omega'.
\end{equation}

In the limit as $T\rightarrow\infty$ the integral in parentheses
becomes an expression for the delta function
$\delta(\omega+\omega')$. Therefore for the autocorrelation
function $\varphi(t_0)$ we get the expression
\begin{equation}
\label{eq160}
\varphi(t_0)=\int\limits_{-\infty}^{\infty}\Delta X_{\omega}^{2}e^{-i\omega
t_0}d\omega.
\end{equation}
or, by performing the Fourier transformation, we obtain an expression
for the spectral density
\begin{equation}
\label{eq161}
X_{\omega}^{2}=\frac{1}{2\pi}\int\limits_{-\infty}^{\infty}\varphi(t_0)
e^{i\omega t_0}dt_0.
\end{equation}
This expression is known as the Wiener-Hinchin\index{Wiener
N.}\index{theorem, Wiener-Hinchin} theorem\footnote{The above
arguments do not follow from this theorem, they only illustrate
the idea. For the deduction, see \cite{W1}. For the role of this
theorem in the study of physical time series, see \cite{W2}.}

Now we can better understand how the frequencies of the
fluctuations and the relaxation time are related. Let the speed
with which the variable $\Delta X$ returns to the
equilibrium\index{equilibrium, market} position (i.e., to $0$)
only depends on the values of this variable itself
\begin{equation}
\label{eq162}
\frac{d\Delta X(t)}{dt}=f(\Delta X).
\end{equation}

Expanding $f(X)$ into the Taylor series and taking into account that
$f(0)=0$ (i.e., in equilibrium the rate of change is equal to $0$) and
selecting all the terms of the expansion except the linear one we
obtain
\begin{equation}
\label{eq163}
\frac{d\Delta X(t)}{dt}=-\lambda\Delta X,
\end{equation}
where $\lambda>0$, i.e., $\Delta X(t)=\Delta X(0)e^{-\lambda t}$.

Substituting this expression for $\Delta X(t)$ into the formula for the
autocorrelation function we get
\begin{equation}
\label{eq164}
\varphi(t_0)=\langle \Delta X^2\rangle e^{-\lambda t}.
\end{equation}

We have to recall here that we

a) neglect the higher terms in the expansion of $\dot{\Delta X(t)}$
with respect to $\Delta X$

b) consider the values of $\Delta X$ greater than the typical
involuntary fluctuations, i.e., ${|\Delta X|<\sqrt{D}}$, where $D$
is the dispersion of the thermodynamic fluctuations of $\Delta X$.

Now we can determine $\lambda$ if we know the autocorrelation
function:
\begin{equation}
\label{eq165}
\int\limits_{-\infty}^{\infty}\varphi(|t_0|)dt_0=\langle\Delta X^2
\rangle2\int\limits_{0}^{\infty} e^{-\lambda t_0}dt_0=\langle
{X^2}\rangle\frac{2}{\lambda}.
\end{equation}

This means that if we have a~time series $\dot{\Delta X(t)}$ we can
having computed the autocorrelation function
\begin{equation}
\label{eq166}
\varphi(t_0)=\lim_{T\rightarrow\infty}\frac{1}{T}\int\limits_{0}^{T}\Delta X(t)
\Delta X(t+t_0)dt.
\end{equation}
and integrating it with respect to time obtain the weight with
which the variable $\Delta X$ returns to
equilibrium,\index{equilibrium, market} i.e., obtain the constant
\begin{equation}
\label{eq167}
\lambda=\frac{2\langle X_2\rangle}{\int_{0}^{\infty}
\varphi(t_0)dt_0}=\frac{2\pi\langle
X^2\rangle}{X_{\omega}X^{2}(0)}.
\end{equation}
where $X_{\omega}^{2}(0)$ is the spectral density at zero
frequency in accordance with the Wiener-Hinchin\index{Wiener N.}
theorem.

Observe that for the exponential autocorrelation function the
spectral density is as follows
\begin{equation}
\label{eq168}
X_{\omega}^{2}=\frac{\lambda}{\pi(\omega^2+\lambda^2)}\langle
X^{2}\rangle .
\end{equation}

For frequencies smaller than $\lambda$, the spectral density is
approximately equal to $\frac{1}{\pi\lambda}\langle X^{2}\rangle$,
i.e., for the frequencies smaller than the inverse relaxation time
of the system, we see that the probabilities for $\Delta X$ to
have such frequency component are approximately equal.

What is the ``physical meaning'' of the above analysis of
thermodynamic fluctuations? Such an analysis is of value when the
system considered has two distinct relaxation times.

One~--- very small~--- to determine an
incomplete\index{equilibrium, market} equilibrium, i.e., such an
equilibrium when it is meaningful to introduce thermodynamic
variables for parts of the system. The other one
--- the large one~--- corresponds to the total equilibrium.

Speaking about markets this means that it is meaningful to
subdivide the market into parts such that each part can be
described by a~thermodynamic model. Speaking practically, this
means that for $\Delta X(t)$ we can take, for example, the
difference of prices at different markets since the relaxation
time of this parameter is greater than the relaxation time on a~particular market.
It would be interesting, of course, to consider
the processes of globalization of the economics from this point of
view.

Observe that to apply the above theory of thermodynamic
fluctuations to the study of price relaxations on markets should
be implemented with utmost caution for the reasons that we
consider in the next section.

\section[``The shadow of future'']{``The shadow of future'' and the collective behavior
at market}

Now we have to consider a~very important question directly related
with the study of market fluctuations\index{market fluctuations}.
How do fluctuations affect the perceptions of the market
agents\index{agent} on the future?

Generally speaking, the market agents have different information
on the situation. In accordance with discussion in Chapter 3, it
is difficult to expect any coordinated behavior of the market
agents anywhere, in particular, in a~neighborhood of the
equilibrium. Indeed, the very lack of such a~coordinated behavior
characterizes the ``extended order'' F.~Hayek wrote
about.\index{Hayek F.} The one who discovers new possibilities,
new types of behavior gets an advantage and the multitude of such
possibilities is unlimited. Moreover it is unknown.

In the case when a~certain stereotype of behavior starts to
dominate one should not expect the growth of ``order''.
Contrariwise, one should expect its destruction. This is precisely
what happens when a~certain idea becomes common for a~considerable
majority of market agents: for example to invest into a~particular
type of activity or company. In this case, the shares quickly
become overvalued and this sooner or later (usually relatively
soon) becomes clear thus influencing a~new wave of spontaneously
coordinated behavior, this time to withdraw money from the
corresponding activity.

Such situations lead not just to market fluctuations,\index{market
fluctuations} but to considerable oscillations and sometimes to a~total
transition of the market. Examples of this type are quite
numerous. It suffices to recall economic catastrophes in Mexico
and South East Asia during the 1990s.

A decisive role in such spontaneously coordinated behavior of the
market agents is played by the ``shadow of future'' that is,
perceptions on a~possible development of the situation. The result
turns out unexpected for the participants because their collective
behavior leads precisely to the very result that each of them
tries to avoid.

A similar effect is well known in so-called non-cooperative games
and is best studied with an example of the game called
``prisoner's dilemma''. The innumerable literature is devoted to
this topic and here is not the place to discuss this problem,
still observe that it is precisely in non-cooperative games
(though in a~somewhat different sense) the radically important
role of the ``shadow of the future'' in the molding of spontaneous
patterns of collective behavior have been singled out\footnote{See
\cite{Ax} and discussion in \cite{SW}.}.

In this case the following problem becomes of interest: how are the
small fluctuations of the system related to patterns of spontaneous
behavior that totally change the market situation? In other words:
what is the role of spontaneously formed collective behavior in the
problem of stability of market economy?

If small fluctuations of market parameters help to form
spontaneous collective behavior that destroys the market
equilibrium then the market becomes evolutionary\index{evolution}
unstable despite of the fact that in ``neoclassical'' sense such
a~market should possess an equilibrium.

Lately similar questions are in the center of attention of
researchers that try to leave ``neoclassical'' orthodoxy and
extend the frameworks of economic studies in particular in
connection with the study of the influence of technical
innovations to economics\footnote{See \cite{AAP}.}. Here we will
confine ourselves to the simple model of ``speculative behavior''
which shows under what conditions the market
fluctuations\index{market fluctuations} can be considered as
thermodynamic ones.

The study of the process of molding of the stock market price is of
particular interest both for creating forecasting models and from
purely theoretical point of view since this price is a~good example
illustrating how a~directed activity of a~multitude of people based on
individual forecasts and decision making leads to a~formation of a~certain
collective variable.

The problem of forecasting stock market prices requires a~very
detailed study of the concrete situation and discovery of a~number
of factors not only of economic but also of political character.
An attempt of construction in mathematical model taking into
account all these factors is doomed to failure. Nevertheless
observe that many of external factors and also a~number of
economic factors (for example the level of actual demand of an
item of goods) may remain constant during a~sufficiently long time
though the prices are subject to constant fluctuations. The reason
for these fluctuations is a~speculative activity. Analysis of
dependence of stock market prices on time shows us that for
sufficiently short lapses of time the nature of fluctuations often
possesses certain common peculiarities. This hints to study
speculative oscillations provided the ``long-ranged'' factors are
constant. This makes it possible to model the modification of
prices making use of the difference in the ``time scale'' for
price fluctuations caused by speculative activities and
oscillations resulting by ``long-ranged'' factors.

Let us abstract from the real conditions of stop market functioning
making several simplifications.

Define the ``mean price''\index{mean price} $\bar{X}(t)$ as the
ratio of the mean amount of a~money $\bar{P}(t)$ spent for the
purchase of the goods per unit of time to the mean value of goods
$\bar{Q}(t)$ sold at the same time:
\begin{equation}
\label{eq169}
\bar{X}(T)=\frac{\bar{P}(t)}{\bar{Q}(t)}.
\end{equation}

Recall that in accordance with Chapter on the free market this
ratio coincides with the marginal price\index{marginal price} that
determines the market equilibrium, where $\bar{P}(t)$ and
$\bar{Q}(t)$ are slowly changing quantities whose value is
determined by the productive powers and other slowly changing
factors.

Speculative activities lead to the change of both $P(t)$ and
$Q(t)$ and this in turn leads to the change of price. The
``instant'' price $X(t)$ depends on $\Delta P(t)$ and $\Delta
Q(t)$:
\begin{equation}
\label{eq170}
X(t)=\frac{\bar{P}(t)+\Delta P(t)}{\bar{Q}(t)+\Delta Q(t)}.
\end{equation}

For short lapses of time we may assume that $\bar{Q}$ and
$\bar{P}$ are time-independent  (this is possible due to the
difference of time scales of the changes between the speculative
and long-ranged factors.

The idea of the ``shadow of future'' discussed above suggests that
the models of molding the market price should radically differ in
structure from traditional mechanical models of equilibrium. The
mechanical models of equilibrium describe a~future state on the
base of our knowledge of the past. Contrariwise, the market price
is essentially formed as a~result of interaction of goal-minded
systems (in other words as a~result of correlation of models of
the future by people taking decisions to purchase a~certain amount
of goods). The market agents\index{agent} act on the base of
predictions they have and therefore the market price depends on
the predictions that the market agents that take decisions stake
to. These predictions may depend not only on the price value in
the past and present but also on their evaluation of the direction
of development of long-ranged factors, on political situation and
various other factors.

It is precisely the fact that the market price is molded as a~result
of forecasts which leads to a~certain unpredictability of
the market prices. Indeed, in order to predict the price one has
to predict the forecasts of each separate market agent.

If we confine ourselves to a~simple assumption that certain extra
amount of goods and money appearing on the market depends on a~possible
profit we can express the market by means of the
following equation
\begin{equation}
\label{eq171}
X(t)=\frac{\bar{P}+\mathop{\sum}\limits_{l}f_l(X(t), \widetilde{X}_l(t+T))}{\bar{Q}+
\mathop{\sum}\limits_{k}\varphi_k(X(t), \widetilde{X}(t+T))},
\end{equation}
where $l$ is the index that characterizes the buyers $k$ is the
index characterizing the sellers, $\widetilde{X}(t+T)$ is the
predicted price for the period $T$ and $\varphi_k$ and $f_l$ are
the functions that characterize the relation of an additional flow
of goods and money on the instant and predicted price.

In this form the equation is too general and not fit for
investigations but it can serve as a~starting point for further
simplifications leading to more tangible equations. Our main
problem will be investigation of the conditions for which we
observe an equilibrium type of price fluctuations~--- oscillations
about a~certain mean value which slowly varies perhaps together
with the volume of goods $Q$ and the volume of money $P$.

Let us simplify as follows:

1) Assume that all the buyers use the same forecast and all the
sellers use the same forecast (though these forecasts are not
necessarily identical);

2) The increase of the offer is proportional to a~possible (predicted)
profit per unit of goods;

3) The increase of demand is also proportional to a~possible profit.

Under these assumptions equation (\ref{eq171}) takes the form
\begin{equation}
\label{eq172}
X(t)=\frac{\bar{P}+\alpha(\widetilde{X}_P(t+T)-X(t))}
{\bar{Q}+\beta(\widetilde{X}_Q(t+\theta)-X(t))},
\end{equation}
where $\widetilde{X}_P(t+T)$ is the buyer's prediction who use the
basis prediction time $T$ and $\widetilde{X}_Q(t+\theta)$ is the
seller's prediction who use the basis prediction time $\theta$.

We should expect that $T$ and $\theta$ may be rather different,
i.e., the market is, generally speaking, asymmetric\footnote{Cf.
Akerloff's\index{Akerloff G.} hypothesis on market's asymmetry,
\cite{Af1}.}.

Let us simplify further. It is natural to assume that the
prediction is determined by the expansion of the price $X(t)$ in
the Taylor series with respect to time and terms higher second
order are neglected. It is difficult to conceive the influence of
the derivatives of the price greater than the second one on human
perception: the eye usually catches the first and second
derivatives from the form of the curve.\footnote{Some, more
perceptive, observers penetrate into even finer details: speaking
about ``the rate of change of inflation'' President
Nixon\index{Nixon R.} implicitly took into account the third
derivative, whereas President Gorbachev\index{Gorbachev M.} once
mentioned in Pravda that ``the tendency of declining of growth
rate of our economy has developed lately'' which demonstrated his
awareness of effects of the fifth, if not sixths, derivative. {\it
D.L.}} Thus the equation (\ref{eq172}) takes the form
\begin{equation}
\label{eq173}
X(t)=\frac{\displaystyle{\bar{P}+\alpha\left(T\dot{X}+
\frac{T^2}{2}\ddot{X}\right)}}
{\displaystyle{\bar{Q}+\beta\left(\theta\dot{X}+
\frac{\theta^2}{2}\ddot{X}\right)}}.
\end{equation}

We have the following alternatives:

a) We may confine ourselves to the first derivatives thus
obtaining the equation
\begin{equation}
\label{eq174}
X(t)=\frac{\bar{P}+\alpha T\dot{X}}
{\bar{Q}+\beta\theta\dot{X}};
\end{equation}

b) We may study the more complicated
equation (\ref{eq173})

Case (a) is rather simple. Resolving (\ref{eq174}) for $\dot{X}$ we
obtain
\begin{equation}
\label{eq175}
\dot{X}=\frac{\bar{P}-\bar{Q}X} {BX-A},
\end{equation}
where $B=\beta\theta$ and $A=\alpha T$.

Certainly one can integrate this equation but we will study it in
a simpler way. The change of variables
$Z=\displaystyle{\frac{B}{A}X}$ leads us to
\begin{equation}
\label{eq176}
F\dot{Z}=\frac{\Phi-Z}{Z-1}\text{, \quad
where\quad}F=\frac{A}{Q}, \quad\Phi=\frac{\bar{P}B}{\bar{Q}A}.
\end{equation}

If $F>0$ (Fig. 3a), that is $A>0$ and $\Phi>1$, we have a~stable
equilibrium at the point
$Z=\Phi\;\displaystyle{\Big(X=\frac{\bar{P}}{\bar{Q}}\Big)}$.


If $F>0$ and $\Phi<1$ (Fig. 3b), then the equilibrium at
${Z=\Phi}\;\displaystyle{\left(X=\frac{\bar{P}}{\bar{Q}}\right)}$
is {\it unstable}. Under a~small increase of the price it steeply
grows up to the value corresponding to
${Z=1}\;\displaystyle{\left(X=\frac{A}{B}\right)}$ and under small
diminishing falls to zero.

\begin{figure}[ht]\centering
\subfigure[]{\includegraphics{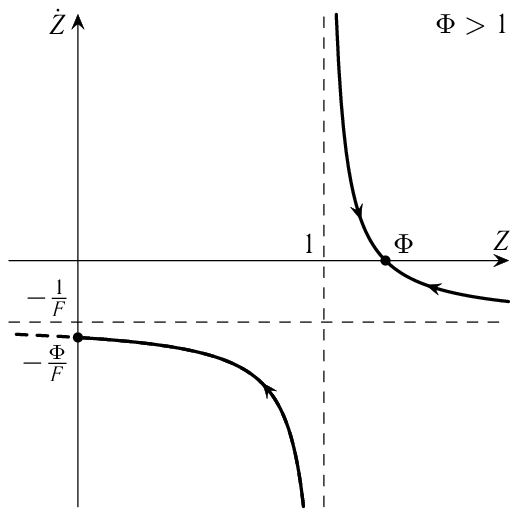}} \hfil
\subfigure[]{\includegraphics{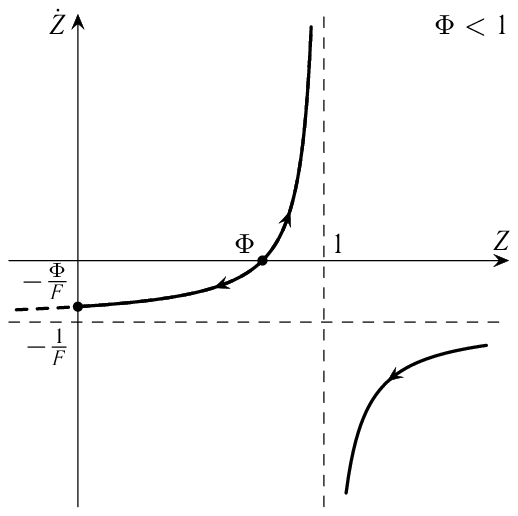}}
\end{figure}

The case $\Phi=1$ is extremely interesting. In this case there is
no equilibrium at all since
$\displaystyle{\dot{Z}=-\frac{1}{F}\;\left(\dot{X}=\frac{Q}{B}\right)}$.

For $B<0$ (Fig. 3c), the price continuously grows whereas for
$B>0$ (Fig. 3d) it falls to zero.

\begin{figure}[ht]\centering
\subfigure[]{\includegraphics{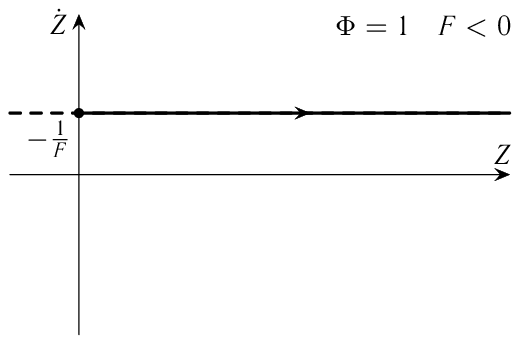}} \hfil
\subfigure[]{\includegraphics{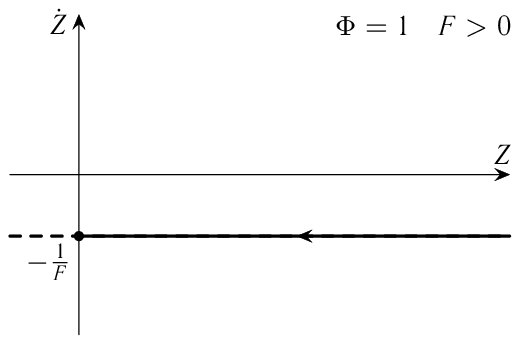}}
\end{figure}

(Obviously only the half-plane
$X>0$ is practically meaningful.) In this situation everything
depends on the nature of the sellers' forecast, i.e., do they
believe that the raising the price will stimulate the production
or one should hide the goods and wait till its price grows up.

Clearly, if the supplies are restricted or if there is a~possibility
to shrink the production, the case $B<0$ is realized.
In other words, in the case of a~monopoly of the seller and
objective restrictions on supplies, the price grows unboundedly.

Apparently, {\bf this model adequately describes the phenomenon of
sudden plummeting of prices during social unrests and wars and
also the inflation in the case of restricted production} (in
particular, the growth of prices of the black market in
1970s--1980s in the countries with ``real socialism'').

Let us now consider another type of prognosis, which takes into
account the second derivative of price. Resolving equation
(\ref{eq173}) for the second derivative we get
\begin{equation}
\label{eq179}
\ddot{X}=\frac{\displaystyle{\bar{P}+\alpha
T\dot{X}-X(Q+\beta\theta\dot{X})}}
{\displaystyle{\beta\frac{\theta^2}{2}X-\alpha\frac{T^2}{2}}}.
\end{equation}

We may study this equation by standard means, see, e.g., \cite{Tr}.
Introduce a~new variable $\dot{X}=Y$ and in the system of equations obtained
eliminate time by dividing $\dot{X}$ by $\dot{Y}$. We get
\begin{equation}
\label{eq180}
\frac{dY}{dX}=\frac{\displaystyle{\bar{P}+\alpha
TY-X(\bar{Q}+\beta\theta Y)}}
{\displaystyle{Y\left({\beta\frac{\theta^2}{2}X-\alpha\frac{T^2}{2}}\right)}}.
\end{equation}

We can simplify this equation by setting
\begin{equation}
\label{eq181}
\displaystyle{k_1=\alpha T, \quad k_2=\alpha\frac{T^2}{2}, \quad
k_3=\beta\theta, \quad k_4=\beta\frac{\theta^2}{2}}.
\end{equation}
We get
\begin{equation}
\label{eq182}
\frac{dY}{dX}=\frac{\bar{P}+\bar{Q}X}
{Y(k_{4}X-k_2)}+\frac{k_{1}-k_{3}}{k_{4}x-k_2}.
\end{equation}

Here we see that if $k_1=k_3$ then the system is an equilibrium at
the point $Y=0$ that is $\displaystyle{X=\frac{\bar{P}}{\bar{Q}}}$
is the expected equilibrium point.

However, if $k_1\neq k_3$ then
$\displaystyle{X=\frac{\bar{P}}{\bar{Q}}}$ is not an equilibrium
point.

This is an astonishing result that shows that under asymmetric
conditions taking into account the second derivative we eliminate
equilibrium. Asymmetric markets\index{market, asymmetric} behave
totally unexpectedly.

So far our considerations resulted in rather unexpected
corollaries: The ``shadow of future'' for the linear forecast
restricts the domain of a~stable equilibrium but even for the
asymmetric forecasts of sellers and buyers it does not totally
eliminate the equilibrium.

Adding the second derivative into the forecast (which amounts,
actually, to professionalisation of the forecast) completely
eliminates equilibrium for the asymmetric markets. In other words,
the market becomes globally unstable.

This deduction is extremely important for our further analysis. It
means that there appears a~possibility to manipulate the market
using restricted resources. In other words, one can turn the
market in a~``heat machine''\index{heat machine} of sorts by
creating a~symmetry in the ways the market agents\index{agent}
perceive the future.

\chapter{``Heat machines'' in economics}

\section{Speculation and ``heat machines''}

An interesting question: can one extract money from the market
making use of its thermodynamic properties?

It is well known that in physical systems one can extract energy
using the difference of temperatures of two systems. The
corresponding device is called a~heat machine. In the simplest
form the performance of a~heat machine is as follows. The working
body is heated by means of a~source of high temperature $T_1$ and
forced to perform a~work and then it is cooled in a~media with
temperature $T_2$ after which the cycle is repeated. The
performance of such a~machine becomes possible thanks to the
dependence of the volume of the body on temperature: expanding the
body is capable to perform work.

In most elegant formulation this works as follows: on the level
surface of the system, say $P=P(S, T)$ one can draw curves of
isotherms $P=P(S, T_0)$ for different $T_0$ and curves of adiabats
$P=P(S_0, T)$ for different $S_0$. On the coordinate net obtained
we separate any curvilinear rectangle bounded by two isotherms
(for $T_1$ and $T_2$) and two adiabats (for $S_1$ and $S_2$).
Moving the system along the boundary of this curvilinear rectangle
we will perform the\index{Carnot cycle} Carnot cycle.\footnote{For
a detailed discussion of the Carnot cycle and its role in
thermodynamics, see, e.g., \cite{So}.} Depending on the direction
of the movement a~work will be either expanded or gained and its
value is equal to
\begin{equation}
\label{eq183}
A=\int TdS=T_1(S_2-S_1)-T_2(S_2-S_1).
\end{equation}

This relation enables to determine the efficiency
of the heat machine, i.e., the part of the heat that can be turned into
work.

Since the expanded heat can be represented as $Q_3=T_1(S_2-S_1)$
this portion is equal to $\displaystyle{\frac{T_1-T_2}{T_1}}$
where $T_1$ is the temperature of the heater and $T_2$ is the
temperature of the cooler. We ask: is it possible to use the
principle of the heat machine, i.e., the idea of turning heat into
the mechanical energy in economics?

In a~sense, the answer is trivial: yes, of course, this is
possible and is performed since long ago. The simplest example of
such an operation is a~purchase of goods where it is plentiful at
low price and selling where it is insufficient at high price. If
the temperatures of the markets are equal the profit obtained is
equal to $V(P_2-P_1)$. This expression corresponds to the
transformation of the difference of pressures obtained through
heating into the mechanical energy.

If we perform sufficiently many such operations then the amount of
goods at market 1 will go down and its price will rise, whereas the
amount of goods at market 2 will grow and its price will go down. We
see that we cannot make a~cycle. The market business results in
leveling the price though a~temporary difference between the prices
can be used to gain a~profit.

Now suppose that we are able to regulate the temperature of the
market. As we saw in Chapter 6, the price depends not only on the
volume of the goods delivered but also on the temperature. Now we
have a~possibility to realize a~simple scheme: cool the market~---
buy goods, heat the market~--- sell goods. The profit will again
be determined by the formula $V(P_2-P_1)$ where $V$ is the volume
of goods bought at temperature $T_1$ and sold at
temperature~$T_2$. But since $P_1=\displaystyle{\frac{T_1N}{V}}$
and $P_2=\displaystyle{\frac{T_2N}{V}}$ the profit resulting in
such operations will be equal to
\begin{equation}
\label{eq184}
D=\frac{1}{2}\left(\frac{T_1N}{V}-\frac{T_2N}{V+\Delta V}\right)
\Delta V=\frac{N\Delta V}{V}\left(T_1-T_2\left(1-\frac{\Delta
V}{V}\right)\right).
\end{equation}

Selling an additional portion of goods will lower the price and the
time during which the profit is gained constitutes only half of the
cycles period (assuming that switching the temperatures is
instantaneous).

In principle, such a~scheme can be realized if we have two markets
with different temperatures and market monopolies. From the
expression for the profit we see that in order to gain profit we
should satisfy the requirement
\begin{equation}
\label{eq185}
\frac{T_1-T_2}{T_2}>\frac{\Delta V}{V},
\end{equation}
i.e., there is a~natural restriction on the amount of the extra
goods $\Delta V$ to the amount of goods on the market $V$. The
scale of speculations is also restricted by the difference of
markets' temperatures.

We see that the nature of cyclic speculations on markets is the
same as that of Carnot cycle.\index{Carnot cycle} The question is
how to ensure sufficiently large difference of temperatures. In
the case where spatially separated markets interact we can make
use of the natural difference in temperatures or the difference in
the levels of delivery of goods on the market which may arise for
multitude of reasons: thus for example, the plants producing
spices do not grow in Europe which in the Middle Ages amounted to
a natural difference in prices at respective markets. To regularly
gain profit one needs a~market monopoly, otherwise the prices
would have been essentially equalized. It was also stupid to
import too many spices to Europe.

The case of two weakly connected markets is trivial.

What is not trivial, we think, is consideration of speculative
operations as a~heat machine. What is also non-trivial is the
possibility to manipulate the market's temperature. Consider the
share market where the price of the share is determined by the
demand and the demand is mainly determined by the firm's
perspectives. The information on the firm's perspectives is a~very
delicate matter and not very reliable one. Therefore the majority
of market agents\index{agent} are governed by the general trend,
i.e., they try to forecast the price on the base of its change.

This is precisely the situation considered in sec. 8.3. In this
case in certain circumstances (for example, for asymmetry of
forecasts of the seller and the buyer) the behavior of the price
may be unstable due to sudden changes in the flow of money or
goods delivery. But it is precisely the amount of money flow that
determines the temperature under the constant number of market's
agents. Under certain conditions this parameter becomes subject to
influences caused by fluctuations. This in turn means that being
able to govern fluctuations that is, using relatively small
resources to sell or buy shares we can steeply change the market's
temperature. In this case it is easy to realize the cyclic heat
machine able to extract money from the share market. The cycle
looks as follows. One sells a~relatively small amount of shares of
a company, small but sufficient to cause a~collective reaction,
i.e., instability. After a~considerable fall of the price one buys
shares causing new instability leading to the price of shares and
the situation returns to square one but the ``hares'' collect the
difference between the total price of shares sold (at high price)
and bought (at low price).

The cycle can be performed in the opposite direction playing ``for
raising''. These differences in the direction of how one
circumvents the cycle precisely correspond to the difference
between the heating machine and the fridge in the Carnot
cycle.\index{Carnot cycle}

In the first case the player gains extra money, in the second one
extra shares. The effectiveness of the operation depends on how
much one can change the temperature of the market: to lower it
(``Bull game'')\index{Bull game}\index{Bear game}\index{game, Bull
Bear} or raise (``Bear game'').

This effectiveness is very dependent of course how close to
instability the market is. Generally, to perform such operations
in order to ``disbalance'' the market and change the temperature
one needs considerable resources. A good example of such
operations are speculations on international markets of currency.
Here it is very interesting to compare the points of view of
politicians of countries whose currency is plagued by such
operations and the actors of the international currency markets.

In 1997, at a~conference of the International Monetary
Fund\index{International Monetary Fund} in Hong-Kong the
Prime\index{Mahatir M.} Minister of Malaysia, M.~Mahatir, called
for necessity to ban or at least restrict speculations on the
international monetary market and accused the actors of this
market in particular, the known American businessman
G.~Soros\index{Soros G.} in intentional damaging Malaysian
currency (as a~result of currency speculations the Malaysian
ringgit dropped 20\% just before that). Mahatir's claim caused
immediately further drop of ringgit by 5\%.

Replying to Mahatir's accusation G.~Soros claimed that
speculations do not affect healthy currencies and only weak or
overvalued currencies can suffer such attacks\footnote{On the
dialog between the Prime Minister of Malaysia Mahatir and
G.~Soros, see Far Eastern Economic Review, 1997, September 25,
October 2, {\bf 160} (39), p. 40.}. It is interesting to analyze
such debates in the light of the above model. The analysis in
Section 8.3 implies that stability domains of the market can
considerably vary the sizes or totally disappear depending on the
nature of forecast used by market agents.\index{agent} For factors
forming the forecast one can take not only the study of the price
changes during a~preceding period but also rumors, statements of
politicians, influential financiers and so on.

Under these conditions there is generally no objective stability of
the market. Its stability depends on the nature of prognoses of the
majority of its agents. But these subjective prognoses are also a~part
of reality and sometimes they are very flexible but can be
extremely conservative. Therefore the result of speculations depends
essentially on the following factors: on the degree the prognoses are
conservative and on their nature (what exactly is subject to be
forecast) and finally on the amount of goods used to perform a~speculation
(this is in close connection with the prognoses:
financiers believe that there is no sense to try to disbalance a~stable market).

Obviously, the larger resources an agent or a~group of market agents
starting to play possesses, the higher probability of their success.
However, to precisely predict the result is difficult since in order
to predict it one should know very well the nature of prognosis of the
other participants in the market play and potential resources of the
one who is attacked.

But for the case of speculations on currency markets these factors are
usually very well known. G.~Soros manifestly knows what he speaks
about saying that without being sure in success there is no sense to
start the operation. There is however, an asymmetry in potential
losses of the sides. Having started an unsuccessful speculation the
financier risks but a~little: having not obtained the effect desired~---
the steep change of the market's temperature~--- one can return to
square one without essential losses. But for the country whose
currency is being attacked the stake is the amount of national wealth
and possible its evaluation (through the cross-currency exchange rate)
becomes subject to the nature of prognosis of the dealers of the
currency market, i.e., on subjective factors that governments can
influence it seems much less than large-scale financiers.

To predict a~possible behavior of the government is not an easy task
either. We know from history long periods when many governments
refused to make their currencies freely convertible. Therefore too
great successes of financiers in the usage of heat machines in
international finances may result in sharp reaction of interested
governments.

The above model of the market shows a~possibility to use ``heat
machines''. But one should understand that financial heat machines
are external devices in relation to the market and not its parts.
This actually is one of the principal deductions we obtained as a~result
of application of the notion of heat machine to economics.

The market can well exist without financial heat machines: they are
not necessary components of the market.

They are intellectual superstructures over the market that enable one
to extract profit not from production of goods or selling them but
from cyclic operations, i.e., away to extract money from the market
agents who are already sufficiently poor, that is, do not have
possibilities sufficient to influence the market situation.

Unfortunately this is precisely the position occupied by the
population of various countries, the population having on their hands
paper money issued by the government.

Certainly the role of financial ``heat machines'' in globalization
of economic processes is huge. Thanks to them and also to the
world market we establish a~certain analogue of a~global economic
equilibrium.\index{equilibrium, market} To create a~certain ``code
of behavior'' in this domain of economic activity may become
necessary at least to avoid the growth of economic populism which
threatens the freedom of market relations.

\section{The government and the economics}

Let us see what are the possibilities of the government concerning
means to extract resources from the economic system in our
thermodynamic approach to economics.

For the country it is vitally important to have a~reliable pattern of
taxation. A natural way to extract resources is to tax transactions
and immovable property. In the case of immovable property and also
transactions controlled by the government (for example, when the goods
cross the border of the country) there are no problems. The problem
appears when the collection of resources is based on taxation on
non-controllable transactions. Such a~scheme turns out to be
extremely ineffective.

The most notable example is the income tax.\index{tax, income} In
order to collect this tax we should have a~possibility to control
at least in principle transactions of tens of millions of people.
So here the principle of ``Maxwell's Demon''\index{Maxwell's
Demon} starts to work: it is impossible to control the
transactions (i.e., get the corresponding information) without
spending some resources.

The firm must possess a~documentation on transactions, whereas to
force private persons to lead such documentation is practically
impossible and to verify all the transactions is rather difficult
since the number of potential sources of income is practically
infinite.

The verification of the income also costs money, its own price of
transaction. In order for this verification to be effective its
probability should be rather high. This means that to perform
these verifications full-scale (for tens of millions of tax
payers) one should spend huge amount of money with a~low
probability to find tax evasions. Transactions not declared are
usually never registered. In the countries with a~considerable
portion of ``black'' or ``gray'' economy such a~method of tax
collection is extremely inefficient and amounts to huge losses for
the government because both ``black'' and ``gray'' transactions by
definition cannot become taxable: to declare above them will mean
at least the loss of a~source of income for the tax payer in
future.

Since the price of verifications depends on their number it is
obvious that the effectiveness of taxation is inversely proportional
to the number of subject of taxation and this is the main argument
in favor of refusal of income tax\index{tax, income} and replacing
it by the value added tax\index{tax, value added} or turnover
tax\index{tax, turnover} and the tax\footnote{Income tax was
introduced in merchant societies (e.g., Venice and the Netherlands),
so small in size that rich people knew each other and could perceive
the size of each other's wealth. As the society grows, a~Demon
enters into play.} on immovables. Obviously, this being implemented,
the effectiveness of taxation will be much higher.

Why then despite the obvious drawbacks of this institution the
majority of countries still collect the income tax?

Arguments on social justice can hardly be a~sufficient
justification. It is not difficult to correct the taxes by means
of a~value added tax\index{tax, value added} collecting it from
the luxury items at highest rate.

Provided the argument on social justice can, to an extent, be
applicable for the explanation of the practice of income tax with
the countries with democratic government (say, in the light of
economic illiteracy of the majority of populace) it is difficult
to find reasons for the effectiveness of this argument in
authoritarian governments.

Among other things, the collection of income tax\index{tax,
income} is doubtful from the liberal law system's point of view
since it actually abandons the presumption of innocence principle
during accusations in tax evasion and contradicts the principle
fixed in many democratic constitutions according to which the man
does not have to witness against himself or herself.

This is why in the USA (say) the tax payer has a~right to refuse to
fill in the tax declaration (and in this case the tax is being
computed by taxation organs on the base of their own estimation of the
subject's expenditures).

The inefficiency of the income tax striping off the government the
incomes from the ``black'' and ``gray'' sectors of economics and a~possibility
to replace it by the value-added tax became subject of
serious discussions lately in the USA in relation with the growth
of these sectors of economics (especially drugs selling).

The above analysis justifies another interpretation of
continuation of the practice of income tax\index{tax, income}
collecting: it can be considered as a~means of a~political and
economic control over subjects. It is precisely the totalitarian
nature of this institution which is it seems the main argument in
favor of its existence.

It does not give much money to the government but it establishes
who is rich and who is poor~--- the information of important
political flavor. It is not by accident that the income tax was
discovered in Venice by the ``democratic oligarchy'' notorious by
its sophisticated means of control over citizens' behavior.

From the point of view of our thermodynamic model of economics
income tax collecting is equivalent to absolutely unthinkable and
as the study of the paradox of ``Maxwell's Demon''\index{Maxwell's
Demon} show simply non-realizable way to separate system's energy.
Indeed, just imagine a~construction of a~robot which takes from
each of the particles of gas a~percentage of its kinetic energy
and in turn depends on energy.

This metaphor\index{metaphor} (which as we have seen above with
the examples of the study of economic processes by means of
statistical thermodynamics\index{statistical thermodynamics} is
much more than just a~metaphor) makes inefficiency of income tax
absolutely manifest. Observe, however, that various governments
retaining this economically inefficient and costly institute for
political goals use it in order to extract money from the populace
much more effective economic methods we discussed in the preceding
section: manipulation with the monetary market, valuable papers
and shares.

Since the central banks of the countries are the largest
reservoirs of the resources, it is not difficult for them to
realize various heat machines which enable them (of course in the
limits bounded by the possibilities of total destruction of the
economics) extract considerable resources from the economic system
which represents interrelated markets at different temperatures.
Since some of these markets, in particular share markets, are very
easy to control, that is to change their temperature  almost does
not require any financial input, it follows that to control them,
only mass media or just controlled ``leakages of information''
from the decisive organs are usually sufficient.

The possibilities of governments to extract distributed resources
from economic systems are practically unlimited up to a~total
destruction of economics.

This is one of the main arguments which stimulates the civil
society usually through the means of parliamentary control to
rigidly curb the activity of their governments. Empirically this
is testified by the practice of creation of central banks
supposedly independent on governmental decisions and also (in the
cases when due to the peculiarities of political culture to ensure
independence of the central bank is difficult) the creation of
currency boards that tie the national currency to one or several
stable foreign currencies in order to exclude possibilities of
political manipulations in economics performed by central banks
under the pressure of the respective governments.

\chapter[The limits of rationality]{The limits of
rationality: the thermodynamic approach
and~evolutionary theory}

\section{Rationality and uncertainty in economics}

In the preceding chapters we have shown how to construct a~mathematical
theory of economic equilibrium\index{equilibrium,
market} without the notion of utility.\index{utility} In this
approach we have to answer: what is rationality in economics?

If the prices are formed not as a~result of maximization of
utility but due to the fact that certain states of the system turn
out to be far more probable than the other states (have greater
entropy)\index{entropy} what is the role of human decisions in
economics? To what extent these decisions can rely on a~rational
analysis at all? What can one know about the economic situation?

In order to answer these questions we have to introduce a~very
important distinction concerning the types of knowledge on the
situation and the types of economic decisions. The first ones are
subdivided into ``micro-knowledge'' and ``macro-knowledge'' and
the second ones into ``micro-decisions'' and ``macro-decisions''.
Without making this distinction it is impossible to discuss the
problem in general since the character and possibilities of
application of ``micro-knowledge'' and ``macro-knowledge'' are
totally different and the difference in uncertainties the subject
of economic activity encounters with in the domain of
``micro-decisions'' and ``macro-decisions'' are cardinal.

The owner of a~shop or manager of a~small firm in their search for
acceptable deals is confronted with uncertainty of prices.

The prices are different in different places and since there are
many possible sellers the uncertainty of the situation is related
first of all with the fact that not the whole information on the
market is available. Somewhere perhaps nearby there is a~big lots
seller capable to sell at a~price lower than the ones known to the
shop's owner but to find this seller in the chaos\index{chaos} of
market is sometimes very difficult.

The ``micro-knowledge'' is first of all the knowledge about such
perhaps rare cases, the knowledge where and when one should turn
in order to obtain the goods at a~low price and to sell it at high
price. Such knowledge is based on the connections and is a~result
of participation in an informal informational networks. However
perfect the formal system of market information would be the
informal contacts and operative knowledge will always give an
advantage at least because the information cannot be instantly
included into the formal commonly available net and even it is,
one has to be dexterous in extracting it. There is always a~certain
time lag between the moment of availability of a~possible
deal and the general spread of this information whereas the
resources for making a~deal may be exhausted faster than the
official information becomes available. The macro-knowledge is an
understanding of the global situation on the market, the knowledge
of general tendencies in the developing of prices, availability of
resources, possible consequences of the decisions that influence
the market situation as a~whole. The character of uncertainty for
the ``macro-knowledge'' and ``macro-decisions'' is totally
different than on the ``micro'' level. In many ways the economic
``macro-knowledge'' is determined by the ideas on the character of
the economic equilibrium,\index{equilibrium, market} in other
words, the equilibrium metaphors\index{metaphor} used in the
analysis of the economic situation.

The character of rationality differs accordingly with the
differences in the character of knowledge on the micro and macro
levels. Lately, in the theoretical economics, the study of the
rationality problem became one of the central topics, see \cite{HH}.

The insufficiency of the model of the rational choice to explain
economic phenomena becomes more and more obvious especially in
connection with the growing interest to the study of economic
institutions, in particular, property rights. In his time,
R.~Coase\index{Coase R.} \cite{Co} pointed out that the economic
equilibrium depends on the prices of transactions whereas the
prices of transactions are directly related with the property
rights, see \cite{Wi, NoI}. Therefore the work of
Coase\index{Coase R.} initiated in economics an interest to the
study of social institutions. It is not difficult to find out that
institutionalized processes of decision-making are regulated not
by a~rational choice between alternatives with the help of
utility\index{utility} functions but certain routine rules to a~large extent traditional.

The discovery of this fact required a~serious modification on the
point of view of the nature of human activity in economics. If not
all human activity is determined by utility functions then how to
construct equilibrium models? To answer this question in the
framework of the neoclassical approach is very difficult, if
possible at all.

Thus in economic theory in addition to the ``instrumental
rationality'' related with utility one more type of
rationality\index{rationality, instrumental}\index{rationality,
procedural} appears (the ``procedural'' in terminology of
Hargreaves--Heap).\index{Hargreaves--Heap S.} H.~
Simon\index{Simon H.} intensively studied the procedural
rationality both in his works on theoretical economics \cite{Si}
and in connection with his activity related with artificial
intelligence. H.~Simon connected the appearance of procedural
rationality with the restriction of computational possibilities of
humans (theory of ``bounded rationality''). In other words, being
unable to compare all the possible alternatives H.~Simon relies on
the procedures formed from experience and they become conditional
agreements that form the structure of social institutes.

Chess play is a~good metaphor\index{metaphor} for an explanation
of this situation: one cannot evaluate and compare all the moves
allowed and one is forced to use the standard schemes, debuts,
difficult combinations, general positional principles, and so on.
It seems that the problem is more serious here than the near
restricted ability for calculations. We think we have to admit
that life is not a~play with fixed rules. The list of alternatives
is open: the alternatives of actions can be created by our mind.
But in this case there should appear at least one more type of
rationality, which we would like to call ontological rationality.
There should exist certain rules that regulate the inclusion of
alternatives into the list and in principle regulating generating
alternatives, see \cite{Sco, SB1}.

This can be only performed if we rationalize the fact that the
world in the perception of a~subject possesses a~certain ontology.
In other words, there exist certain ways to perceive what is real
and what is essential and should be included into the
consideration. There should also exist mechanisms for generating
alternatives. Here obviously metaphors and examples are of huge
importance, see \cite{Sco}.

Apparently Hargreaves--Heap\index{Hargreaves--Heap S.} had
something close to this conception introducing the notion of
``expressive rationality''. He wrote that the expressive
rationality is defined by the universal human interest in
understanding the world in which we live. It is thanks to our
purposefulness we have to give sense to the world: the world must
be rationally described if we want to act in it. We have in mind
the necessity of a~cosmology which answers the question on the
meaning of this, on the foundations for that, on the
interrelations of an individual and the society and so on, see
\cite{HH}.\pagebreak

We believe that the conception of ontological
rationality\index{rationality, ontological} better grasps the
problem concentrating attention on the most principal question in
the theory of social sciences: where and how the human takes the
alternatives from which he or she chooses?

The assumption that the alternatives are given externally introduced
in the theory of rational choice does not correspond to reality and
rudely bounds the sphere of human freedom.

We believe that from the positions of cognitive analysis of the
decision-making process one can justify the existence of three
types of rationality. The first phase of this process~--- the
formulation of alternatives~--- is an activity which possesses its
own logic and does not reduce to cataloging of alternatives
already known or enforced by the environment. It is connected with
ontological rationality.\index{ontological rationality}

The second phase of the process~--- a~procedural development of
alternatives (something similar to the plans at general headquarters
compiled in case of possible conflicts) corresponds to procedural
rationality.

The third phase~--- the evaluation of constructed and procedurally
formed alternatives with the help of a~given\footnote{For the
social sources that ``give'' or predestine the hierarchy of
values, see, e.g., E.~Berne, {\em Games People Play : The basic
handbook of transactional analysis}. Ballantine Books, 1996, 216
pp.} hierarchy of values~--- corresponds to the theory of rational
choice (instrumental rationality).

Of course this is a~very simplified scheme. In reality the phases
coexist, a~cyclic return to a~previous phase is possible and so on.
But such a~scheme gives at least a~framework for a~much deeper
analysis of human activity in general and economic activity in
particular than the theory of rational choice which has eliminated the
two most important phases of the process of decision-making~--- the
ontological and procedural ones.

The presence of such a~scheme makes it possible to explain how the
thermodynamic approach works in economics. The most essential is
the fact that various subjects not only possess a~different choice
of alternatives (and some of these alternatives are better and
some are worse from the point of view of optimization criterion)
but tend to discover new alternatives.

Under conditions of institutional restrictions imposed by
historical tradition, political pressure of interested groups, and
so on, only the process of competition at large ensures the
equilibrium\index{equilibrium, market} in the sense we introduced,
that is the equilibrium as the lack of flows between the parts of
the system.

The presence of institutional restrictions in economics stunned
the eco\-no\-mists of the neoclassical school for a~long time. It
is interesting to note that R.~ Coase\index{Coase R.} in his well
known paper \cite{Co} poses a~seemingly strange question: {\bf why
do firms exist?}

This question
is sudden only for the one who perceives the world through the
neoclassical ontology. Indeed, why a~part of transactions became integrated
inside of
corporations instead of being performed through market relations?

The answer to such questions required a~reconsideration of the
methodology of economic studies. Together with the problem of
fundamental uncertainty the interest to which was initiated by
works by Knight\index{Knight F. H.} \cite{Kn} and
Keynes\index{Keynes J. M.} \cite{Ke} and also the accent on the
importance of the study of {\it human actions}\index{human
actions} under the uncertainty conditions, the action made by all
the representatives of the Austrian school\index{Austrian school}
after von Mises\index{von Mises L.} \cite{Mis} the institutional
analysis forced theoreticians to pass from the study of
equilibrium\index{equilibrium, market} to the study of
evolution\index{evolution} of economic institutes. The main idea
of this passage is to try to discover on the level of evolutionary
process something we cannot discover in the activity of an
individual market agent\index{agent} (realization of
utility's\index{utility} maximality) because of the uncertainty
and institutional restrictions.

\section{Rationality and evolution}\index{evolution}

As a~result of mathematization of the neoclassical economics the
``invisible hand''\index{invisible hand} practically disappeared
from the economic theory being reduced to the existence theorem
for a~vector of prices. The powerful metaphor\index{metaphor} of
self-regulating of the market by means of individual interests
which convinced A. Smith\index{Smith A.} that in economics, unlike
political sciences, a~principle of spontaneous equilibrium acts
provided an ontological justification of the economic theory for
more than a~century. A gradual dissolving of this metaphor in
abstract mathematical constructions could not but worry
theoreticians. F.~Hayek\index{Hayek F.} returned, as we saw above,
to the idea of ``invisible hand'' \cite{HRS} suggesting to
consider the evolutionary process as such an ``invisible hand''.

In 1950, A.~Alchian\index{Alchian A.} published a~paper \cite{Alc}
in which on the model level he showed how the evolution can
replace human rationality. Al\-chi\-an's initial position was that
market uncertainty devalues instrumental rationality. He suggested
to consider the process of natural selection as a~means that
determines which type of business activity survives and which does
not. As a~filter of evolution, Alchian suggested to consider the
value of profit as a~result of economic activity provided the
successful patterns of business activity are copied by other
participants of the economic process which guarantees
proliferation of the corresponding pattern.

Alchian's work continues to provoke
animated discussions until now \cite{Lo}.

Certain positions of his work were criticized: in particular, the
possibility to reproduce the pattern \cite{Wnt} the assumption on
sufficiency of the positive profit as a~filter of evolution
\cite{Da} but general very high evaluation of his work is based on
the belief that Alchian\index{Alchian A.} indeed managed to return
to economics the strong version of the ``invisible
hand''.\index{invisible hand}

In this way again now on the level of modern science we obtain a~mechanism
that establishes a~rational order by a~means not
depending on the degree of rationality of particular participants
of the process. This order is not a~result of somebody's plan and
is achieved by decentralized activity. Alchian's works
demonstrated great possibilities of the information
theory\index{information theory} in economic analysis. One of his
followers recently wrote:\vspace{1mm plus 1pt}
\begin{thetext}
{``{\sl My own understanding of economics was under heavy influence
of Alchian's\index{Alchian A.} ideas. I would have formulated the
base of his teaching in the phrase: ``Everything in economics is
information theoretical...''}.'' \cite{V}}
\end{thetext}

Alchian's\index{Alchian A.} ideas heavily influenced the
development of institutional economic theory in works of
O.~Williamson,\index{Williamson O.} D.~ North\index{North D.} and
others \cite{Wi}.

Indeed, despite an obvious attractiveness of the idea to take into
account in the economic theory of the prices of transactions,
property rights, and so on, it was totally unclear how to perform
this in the framework of the neoclassical model of the rational
choice. The difference of the institutionalized behavior from the
behavior in the frameworks of the models of the rational choice is
in the practical difficulty to ascribe any utility\index{utility}
value to institutional procedures based on conventions. Even if
one performs this inside the intra-institutional behavior it is
still totally unclear how to relate such a~utility with the
effectiveness of the institution as a~whole.

The idea to consider evolutionary process as a~global
understanding of sort which guarantees rationality by selecting
and eliminating the rules of behavior unable to compete gave a~theoretical
foundation for the institutional analysis comparable
with its force of conviction with the neoclassical equilibrium
theory but essentially surpasses it in applicability to study real
economic institutions. The institutional scope in theoretical
economics created a~considerable revival of the interest to the
ideas of the ``Austrian school''\index{Austrian school}
practically completely forgotten in 1950--60s\footnote{See, e.g.,
\cite{NoI}. Observe that a~revival of interest to the Austrian
school of economics is related with the Nobel prize in economics
awarded to F.~Hayek.\index{Hayek F.}}

Nevertheless, the evolutionary approach did not quite solve the
problems that the economic analysis of market processes faced when
took into account such factors as the restriction of information,
restrictions in behavior related with the cultural tradition, taking
into account of the prices and transactions, and so on. The heart of
the problems consists in the utmost labor consuming feature of the
modeling of evolutionary processes and feasible to perform only for
very simple systems. Moreover, for solutions of such problems there
is no developed analytical tools. Mainly evolutionary models are
being demonstrated by computer simulations. This does not preclude
one to make important theoretical deductions but the poverty of the
analytical apparatus clearly manifests the weakness of the approach.

The evolutionary analysis deals with micro-knowledge and
micro-decisions but cannot say practically anything on
macro-knowledge and macro-decisions\footnote{For a~discussion of
potential of the evolutionary theory,\index{evolutionary theory}
see \cite{Ho}.}. Meanwhile it is impossible to deny the role of
macro-decisions. It goes without saying that a~certain spontaneous
order\index{spontaneous order} is being developed in the process
of evolutionary selection. But what to do in the cases when one
has to ``correct'' the activity of the ``invisible hand'' which
depends itself on certain macro-parameters, such as for example
financial and taxation law, the practice of licensing of certain
types of economic activity, and so on?

The evolutionary theory\index{evolutionary theory} can hardly help
in this case\footnote{Cf. the discussion of potentialities of
evolutionary theory in \cite{Ho}.}. The uncertainty problem and
insufficiency of information on macro-level exists nevertheless
and is represented perhaps at a~greater scale than on micro-level.
Our book is devoted to the development one more, thermodynamic,
way to introduce the ``invisible hand'' to economics.

In Chapter 4 we have discussed the relation between the mechanical
and thermodynamic versions of the ``invisible hand''. What is the
relation between the ``invisible hand'' of the evolution and the
``invisible hand'' of thermodynamics?

\section{Evolution and thermodynamics}\index{evolution}

In order to apply the evolutionary approach to the study of
economic processes, one has first of all to have a~clear structure
of the evolutionary theory. This theory contains two extremely
important but weakly related aspects. The first aspect concerns
the birth of innovations in the system and the second one the
mechanism that selects innovations.

In the theory of biological evolution, the first aspect caused a~huge
amount of disputes. The distinction between the principal
versions of the evolutionary theory\index{evolutionary theory}~---
Darwinism\index{Darwinism} and Lamarkism\index{Lamarkism}
--- is related with the different understanding of this aspect of
the theory. In Darwin's theory\index{Darwin's theory} innovations
are totally random whereas in Lamark's theory\index{Lamark's
theory} they are the result of education. At present we have no
direct experimental testimonies in favor of inheritance of
features obtained. But this does not mean at all that Darwin's
theory has no difficulties.

The main difficulty of Darwin's theory is how to explain
appearance of a~complicated construction as a~result of random
mutations: such a~construction consists of elements which can be
only created by separate independent mutations but it gives a~preference
in the selection only when all the necessary elements
are already present and assembled into a~functioning mechanism
such as a~hand or an eye.

We do not have to discuss these problems here since thankfully the
economic theory does not have to speak about a~birth of a~new
structure (this is a~result of human activity) but the selection
problem stands in full height. Essentially the selection problem
is a~purely thermodynamic problem. One has to determine what is
the selection of the ``organisms'' or the ``nutritional niches''
in order to understand for example how the restriction of the
number of ``nutritional niches'' affects the distribution of the
``organisms'' or the appearance of new ``organisms'' with
different properties modifies the general distribution over the
``nutritional niches''.

It is not difficult to see that in the process of selection
evolutionary stable states are being created and we can identify
them with the equilibrium states of thermodynamic systems. Since
in the evolutionary approach to the economic theory there is no
need to introduce notions of generations, inheritance and genes
\cite{Ho} the evolutionary theory\index{evolutionary theory}
applied to economics becomes much simpler than in biology.

Here we only speak about the mechanism that filters institutional
innovations. As A.~Alchian\index{Alchian A.} observed, this is
a~pure problem of information theory\index{information theory}
\cite{Alc}. But due to the practical identity of the information
theory and thermodynamics the problem of innovation filtrations
becomes a~problem of search for an equilibrium\index{equilibrium,
market} distribution function for economic ``organisms'' over
certain ``micro-parameters'' determined by the structure of the
evolutionary filter. Consider one more example how this idea
works. Suppose we have $N$ firms each of which being characterized
by a~certain annual profit $\varepsilon$. Define the structure of
the filter as follows: the profit should be positive. This is
sufficient in order to construct the distribution of firms
according to their incomes under the equilibrium~--- that is, at
temperature $T$ and the migrational potential\index{migration
potential} $\mu(T)$.

The meaning of the approach consists in a~way to define the
distribution function of the firms according to their incomes
having given the spectrum of possible values of the income and
assuming that each of these values can be ``populated'' by any
number of firms and also taking into account that the system is in
equilibrium.

The equilibrium\index{equilibrium, market} is understood here in
the sense that if we separate the system into parts according to
the parameters not related with the study of income distribution
(for example, considering geographical location and assuming that
this parameter and the income are independent) then the
distribution function is preserved under such partition of the
system. In this case the preservation of the number of
``organisms'' or the income is inessential. We only need an
empirical assuredness in the existence of the invariant
distribution function for the corresponding equilibrium parameters
in our case - the temperature and migrational potential. We can
consider this system as a~number of systems that appear and
disappear during an infinite number of equal time intervals. If we
observe the stability of the distribution function of the
``organisms'' over the parameters of the filter this is an
equilibrium.

To analyze such a~system the technique of the large statistical
sum\index{statistical sum} can be applied. In this case separating
one of the states of the income and computing for it the large
statistical sum we obtain
\begin{equation}
\label{eq186}
Z=\mathop{\sum}\limits_{n=0}^{\infty}\lambda^{n}e^{-n\frac{\varepsilon}{T}}=
\frac{1}{1-\lambda e^{-\frac{\varepsilon}{T}}}.
\end{equation}
We consider the spectrum of
possible states consisting of positive equidistant quantities
$\varepsilon_n=n\varepsilon_0$. We have selected the discrete
version of this spectrum to simplify the solution of the model
problem.

For the probability function of a~particular state of the income
being filled we get
\begin{equation}
\label{eq187}
\langle
n(\varepsilon)\rangle=\frac{\sum\limits_{n=0}^{\infty}nx^n}
{\sum\limits_{n=0}^{\infty}x^n}.
\end{equation}
where $x=\lambda e^{-\frac{\varepsilon}{T}}$.

Computing this expression we obtain for (\ref{eq187}) identical to
the expression for the Bose-Einstein
distribution\index{Bose-Einstein distribution}\index{distribution,
Bose-Einstein} function in the statistical physics
\begin{equation}
\label{eq188}
\langle
n(\varepsilon)\rangle=\frac{1}{e^{\frac{\varepsilon-\mu}{T}}-1}.
\end{equation}

Recall that $T$ is an equilibrium parameter defined from the
equation
\begin{equation}
\label{eq189}
\frac{\partial S(E)}{\partial\varepsilon}=\frac{1}{T}.
\end{equation}

We can compute the number of ways to fill in the income level of
the systems in large placing $N$ firms in order with the fixed
level of income and therefore can compute the temperature and the
migrational potential\index{migration potential} $\mu$ determine
it as earlier from the condition
\begin{equation}
\label{eq190}
\sum\limits_{\varepsilon}\langle
n(\varepsilon, \mu)\rangle=N.
\end{equation}

We have obtained a~very interesting distribution function. It is
well known that in the systems with Bose-Einstein statistics an
effect of ``Bose-condensation''\index{Bose-condensation} is
observed. Namely, at low temperatures the particles condensate on
the base level of the energy of the system. It is not difficult to
see that in our case the same will happen.

We obtain the following expression for the occupation of the
zero-th level of income:
\begin{equation}
\label{eq191}
n(0, T)=\frac{1}{e^{-\frac{\mu}{T}}-1}=N_0.
\end{equation}

For $T=0, \quad N(0, T)=N$ we can obtain the value of migrational
potential $\mu$ on $T$
\begin{equation}
\label{eq192}
\mu(T)=-T\ln\left(1+\frac{1}{N}\right).
\end{equation}
or
\begin{equation}
\label{eq193}
\lambda=1-\frac{1}{N}\quad\left(\lambda=e^{\frac{\mu}{T}}\right).
\end{equation}

Having known the occupation functions of the levels we can
compute how many particles (firms) will occupy the non-zero levels
of income:
\begin{equation}
\label{eq194}
N_1=\sum\limits_{n=1}^{\infty}{\frac{1} {
e^{\frac{(\varepsilon_{n}-\mu(T))}{T}}-1}}.
\end{equation}

 Now we can obtain the value of the temperature of ``Bose
condensation'' starting from the fact that at this temperature the
number of particles on the base level of the system is equal to
the number of particles on the excited levels:
\begin{equation}
\label{eq195}
N_0(T_0, \mu)=N_1(T_0, \mu)=\frac{N}{2}.
\end{equation}

We have two equations to determine two parameters: $T_0$ and
$\mu_0$ where $\mu_0$ is defined in terms of the occupation of the
base level
\begin{equation}
\label{eq196}
\mu_0=-T\ln\left(1+\frac{2}{N}\right).
\end{equation}
and $T_0$ is found from the equation
\begin{equation}
\label{eq197}
\frac{2}{N}= \sum\limits_{n=1}^{\infty}{\frac{1}
{e^{\frac{n_T\varepsilon_0}{T_0}+\ln\left(1+\frac{2}{N}\right)}-1}}.
\end{equation}

To estimate this expression we can replace the sum by the
integral assuming $\varepsilon_0$ very small and in principle
compute $T_0=\varphi(N, \varepsilon_0)$.

Further having calculated the mean income $E$ of the system we may
obtain the dependence of the temperature on the value of the
income for given $N_0$. This procedure means that the
equilibrium\index{equilibrium, market} condition can be expressed
in terms of parameters $T, \mu$ equally well as in terms of
parameters $E, N$. We will not do this here since the model is too
rough.

Here only one deduction of the model is essential for us, namely
at a~finite non-zero temperature the lowest level of income will
be occupied by a~``microscopic'' portion of the firms, i.e., in
the asymptotical limit for $N$ large we have
\begin{equation}
\label{eq198}
\frac{N_0}{N}\approx O(1).
\end{equation}
Now suppose that conditions slightly changed for example one has
to pay an additional tax for each transaction. This means that the
income of each firm will diminish by a~fixed value. We obtain the
well-known effect of a~``crash'', an essential portion of existing
firms cease to exist. Observe again that the main peculiarity of
the considered model is the Bose-Einstein occupation function of
the values of the spectrum of possible income.

We see that under certain conditions the system becomes unstable
with respect to small modifications of external parameters though
it is in equilibrium. Under a~small increase of the price of
transactions the ``macroscopic'' number of firms dies out.

This is just one example of how to use statistical
thermodynamics\index{statistical thermodynamics} for analysis of
the performance of a~filter of the evolutionary process.

Here we just wanted to show that the survival in the evolutionary
theory\index{evolutionary theory} can be described in the frames
of the thermodynamic model and thus establish a~relation between
the evolutionary and thermodynamic approaches to the description
of economic processes.

\chapter{Conclusion}

In this book we tried to show that a~new
metaphorical\index{metaphor} frame is possible in the theoretical
economics the frame which preserves in the most strong form the idea
of spontaneous order\index{spontaneous order} or ``invisible
hand''\index{invisible hand} and at the same time provides with
richer analytical possibilities than neoclassical theory of
equilibrium or evolutionary modeling.

The thermodynamic approach shows that macro-parameters in the
economic system become related by the equation of state, i.e., lie
on a~surface whose form is determined by a~Pfaff
equation.\index{Pfaff equation}\index{equation, Pfaff} In order to
able to say how the macro-parameters of the system will change
under variation of one of them one has to know this equation.

Generally speaking, a~success in attempts to move economics in
a~certain direction by changing the macro-parameters will be only
achieved if we know the equation of the state. The system with
a~sufficiently exotic equation of state (and apparently the economic
system with non-stabilized markets) may as a~result of seemingly
rational attempts to modify its state arrive not where a~``rational''
politician pushes it but to quite other state.

We believe that this corollary is of huge importance for
macro-economics. Uncertainties one has to deal with in the process
of macro-decisions are much more complicated than uncertainties of
the same system on a~micro-level as follows from the thermodynamic
approach to the economics. The rationality of macro-decisions
becomes ``tied'' in much stronger sense than the one H.~Simon had
in mind. And the problem is not only in the restricted ability of
humans to calculate, but in our ignorance of the constraints
imposed on the system.

Once again, the deduction one can make from the above analysis:
without knowing the equations of the state of the system we cannot
speak about rational macro-decisions.

In the absence of the knowledge of the equation of state one can
only speak about empirical decisions in macro-economics. At
present the equations of the state of economic systems not only
are unknown but such vital parameters that determine the state of
economic systems as temperature and migrational
potential\index{migration potential} are not being taken into
account.

We think that economic macro-theory is now on approximately the
same level the study of heat processes was during the period of
reign of the phlogiston theory, that is before the laws of
thermodynamics were discovered.

The economics is being globalized\index{globalization} and this
globalization provides new unexpected possibilities for
construction of gigantic heat machines in economics which enable
those who controls international financial flows to work with
national economic systems in approximately the same way as what
worked with the steam engine. Recall that the steam engine was
invented long before the thermodynamic relations became subject of
scientific analysis.

In the process of macro-economic decisions two dangers work. One
--- the main one~--- is the wide application by politicians by the
standard procedures that do not take into account the ``equation
of the state'' of national economies and what is remarkable, the
politicians pay no responsibility for the damages caused by such
actions, whereas there is no evolutionary effective mechanism for
selection of macro-decisions on the national level. Institutional
mechanisms of democracy are too weak and work satisfactory as a~rule
only in places where politicians have worked out a~certain
procedural practice of working with economics anyway.

On the other hand, the procedures of using ``heat machines'' in
the sphere of international finances are subject to ruthless rules
of evolutionary selection based on the profitability of the
economic activity and therefore are effective but totally void of
any control from those whose well-being actually depends on the
procedural manipulations in the globalized world economics.

I hope that the approach presented in this book will help at least to
a small extent diminish the uncertainty and increase the degree of
rationality of macro-economic decisions based not on the ideological
dogmas but as a~result of establishing an interrelation of micro- and
macro-analysis in theoretical economics.


\vspace*{2cm}

\thispagestyle{empty}

\renewcommand{\thefigure}{A\arabic{figure}}

{\appendix

\renewcommand{\thechapter}{\arabic{chapter}}

\chapter{Physics as a~tool in sociology}



\section{Social changes and tacit knowledge}
It is well known that under certain circumstances sudden
transformations that change the most important parameters of
social structures can take place in human societies. Sudden
revolutionary changes may touch the relation of the members of the
society to the religion, or economic, or political system. Until
now there are no models except, perhaps, the catastrophe theory
able to describe such processes however rigorously.

Despite of the fact that an uncountable amount of literature is
devoted to revolutions of all types (social and intellectual) the
nature of the jump-like modification of the mass conscience cannot
be grasped by purely functional models. One can, of course, guess
that further exploitation might ensue in a~revolt, and that the
widening of corruption in the church hierarchy might lead to
success of a~new religious trend. Such happenings, however, do not
always occur and to determine the precise moment of time or a~concrete
situations when such jump-like changes of mass patterns
of behavior is beyond the limits of the possibilities of any
theoretical analysis.

The most serious obstruction for such mass changes is the
conservative nature of the social patterns of behavior. The
institutional structure of the society is targeted precisely to
maintaining such conservativeness, see \cite{N1}. At the same
time, the factor that obstructs any institutional changes is the
fact that the knowledge determining the functioning of social
institutes is a~two-level one.

In addition to formal rules abided by the members of the society
--- the rules that formulate and restrict the human behavior\index{human behavior}~---
there is a~huge volume of non\defis formalized knowledge, and it is the
latter that makes the functioning of the visible social forms
actually possible. Since these informal rules (tacit knowledge)
are rooted very deeply, they are seldom an object of reflection,
and therefore are subject to conscience changes with more
difficulty than rules understood, see \cite{S2}, \cite{P3}.

\section{The paradox of changes}

The above described situation leads to the paradox of changes:
rational (meaning: understood) changes of formal social institutes
do not touch the existing deep tacit knowledge that constitute
their basis. But, on the other hand, without changing the formal
rules how can the tower of tacit knowledge be created, the tower
that should be a~pillar of these changes?

This paradox of the hen and the egg is the stumbling point of any
however developed evolutionary theory\index{evolutionary theory}
(the theory of biological evolution is, perhaps, the most graphic
example).

The moment two ontological changes become objects of the study of
an evolution theory (for example, in biology: the phenotype and
genotype) the problem of relation between them and the
interrelation between changes that take place on these two levels
become subjects of heated discussions. Here, I think, lie the
roots of incessant discussion between neo-Darwinists and
neo-Lamarcists in the theory of biological evolution, see
\cite{K4}.

Negation by neo-Darwinism of the principal possibilities of
directed influence of the environment on the genotype leads to
logical difficulties in the theory, whereas at the same time there
are no examples of such an influence.

In the theory of social evolution, there are, fortunately, no such
logical difficulties that make the development of the evolutionary
theory\index{evolutionary theory} in biology so painful: the
directed influence of the environment on the tacit knowledge is
quite obvious. However, the presence of interaction in both
directions between the two ontological levels of social reality
does not, nevertheless, eliminate the paradox of changes.

Heated debates on the possibility to use the theory of rational
choice in order to explain revolutionary changes, such as the one
at the seminar organized by Swedish Collegium of Future Studies in
Social Sciences in 1995, is one of many examples. If the selection
of rules of behavior is subject to a~rational choice, then how to
describe the mass support of revolutionary changes that lead to
equally large-scale worsening of social conditions?

Ethnic conflicts similar to recent conflict in Bosnia are subject
to a~rational explanation even less. One can, of course, try to
interpret such phenomena on the base of national spirit in the
spirit of theoreticians of nationalism. But from the
methodological point of view such theories encounter serious
counterarguments, see, e.g., \cite{K5}.

It seems vitally necessary to have a~theory that unifies phenomena
of social evolution and avoids both the logical difficulties of
the theory of rational choice and the additionally non-observable
entities.

\section{The Interaction Space}
We assume that the reader got acquainted himself or herself with
ideas of application of thermodynamics to economic systems in the
main text. The only difference is that transaction costs in social
networks can be considered in the same way as a~flow of money to
the market.

In physical systems of many particles, the interaction takes place
in a~space. The interaction depends on the distance and all
reasonable mathematical models take this into account. All
interactions occur somewhere. The space in physics is the place
that rooms the changes. Some difficulties in creation of
mathematical models of social systems are related first of all
with the fact that, obviously, social interactions occur not in a~physical space.

Of course, the actors are placed in a~physical space, but does
their interaction depend on the distance between them?

Manifestly, for very large distances, the possibility to interact
depends on the level of communication technology. But it is
precisely this dependence that robs the physical space of social
sciences of the meaning it has in physics. If the possibility to
interact depends on technology, then the physical space ceases to
be universal. Ten kilometers in mountains is more than 100 km in
the steppe or along the sea with the sufficient means of
transportation provided.

The structure of social space in this example depends on the means
of transportation. Therefore the natural desire of the researcher
to consider the physical space as the one that hosts social
interactions is inapplicable. If the physical space does not fit
to host social interactions, what can we offer as the space of
social interactions?

The answer to this question is one of the main objectives of this
Appendix. And with all its simplicity the answer seems to be
rather unexpected. If the object of a~social study is the social
network, each relation of which is characterized by a~transaction
cost, why shouldn't we consider a~graph of a~social network that
is the set of vertices--agents\index{agent} joined by
edges--interactions?

The problem is immediately solved. Thus, a~formally defined graph
is a~typological structure insensitive to the actual distance
between the vertices. One can express such graphs differently but
as a~mathematical object any two of its presentations are
indistinguishable, see Fig. \ref{fA1}.

\begin{figure}[ht]\centering
\includegraphics{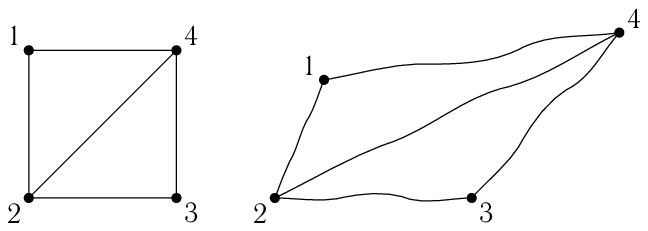}
\caption{}
\label{fA1}
\end{figure}

Therefore, for a~model of a~social system, we can consider the
totality of agents with certain patterns of behavior and connected
into a~social network represented by a~graph.

To the edges of this graph we ascribe transaction costs depending
on the patterns of behavior at the vertices of the graph. Nothing
prevents us from expanding the ontology of the model and
considering that the costs depend not only on the patterns of
behavior but also on the patterns of tacit knowledge. Such an
approach to the definition of the space of interaction admits a~very
important generalization.

Suppose we know the construction of the stable relations in the
social network. Does it mean that the social agents\index{agent}
do not interact outside these stable channels?

Obviously not. Therefore we must have a~possibility to introduce
transaction costs between the agents not directly connected in the
social network. How is this done?

The answer is again very simple. Having the graph of stable
relations we may introduce the distance between the vertices
taking for this distance the least number of edges that form a~path
that connects these two vertices. It is not difficult to see
that thus defined integer distance satisfies the axioms of the
metric space. Therefore, having a~graph that represents a~social
network, we can introduce an interaction that depends on the
(integer) distance in a~totally classical way.

If we consider the common for all social agents patterns of inner
states as analogues of the spin\index{spin model} of physical
particles, then for the space determined by the graph of social
network we obtain a~formal description equally identical to the
description of a~physical system by means of the Hamiltonian of
the interaction. If we want to study the properties of such a~system at equilibrium,\index{equilibrium, market} it suffices to
compute the statistical sum\index{statistical sum} as a~function
on the temperature:
\begin{equation}
\label{eq199}
Q=\mathop{\sum}\limits_{\sigma}\left\langle e^{-H/T}\right
\rangle,
\end{equation}
where $H=\mathop{\sum}\limits_{ij}V(\sigma_{i}\sigma_j)$ and were
$\sigma$ run over all possible states of ``mental patterns''.

Considering the expression for the heat capacity of such a~system,
i.e., the ratio of the change in transaction cost to the change in
temperature in terms of the derivatives of the statistical sum, we
may find a~point of phase transition in a~social system in exactly
the same way as one does this for physical systems. Now we have a~means to obtain mathematically absolutely non-trivial corollaries
describing behavior of the social systems under the influence of
significant external social and economic forces, the structure of
the social networks, and the level of economic prosperity.

Finally, we have obtained a~possibility to regularly construct and
investigate by mathematical methods highly non-trivial models of
social systems, predict behavior of such models in various
situations and compare these predictions with experimental
historical data, and search for known and previously unknown
effects.

\section{Phase Transitions in Social Systems}

Social and political studies under such formulation of the problem
begin to correspond to natural sciences at the level of
probability. We have, of course, taken into account general
considerations with respect to the nature of social models. Since
we are interested in the thermodynamic limit, i.e., the behavior
of the system when its size (the number of\index{agent}
agents-nodes) grows without bound, then it is clear that such a~thermodynamic
limit exists only for some types of social systems.

To obtain a~thermodynamic limit of the model, we should be able to
expand the graph without changing its macrostructure. There exist
several simple, but important from the point of view of
description of social networks, graphs of such type. One such is,
for example, the graph of the $n$-dimensional lattice (Fig.~\ref{fA2}\,a),
the graph of a~regularly branching tree (Fig.~\ref{fA2}\,b), the complete
graph of all interactions (Fig.~\ref{fA2}\,c). Such graphs can be enlarged
without changing their type.

\begin{figure}[ht]\centering
\includegraphics{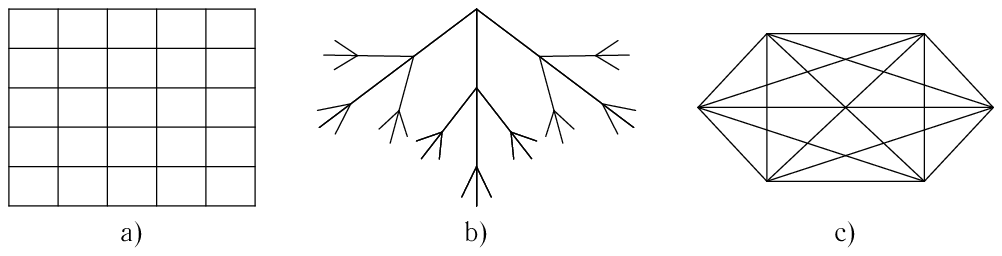}
\caption{}
\label{fA2}
\end{figure}

Let us make one more important remark. The paradox of this
approach that eliminates the physical space from the models of
social networks and replaces it by the space of a~graph of social
interaction is very interesting from the point of view of
tractability of statistical models.

The point is that a~number of exactly solvable statistical models
(in particular, the spin\index{spin model} statistics of the {\it
Boethe tree}) do not have any corollary in physics.

To realize such models, one needs not a~three-dimensional but an
infinite-dimensional physical space.

Conversely, in the study of statistical behavior of social
networks, these exactly solvable models are very meaningful. It
suffices to observe that the {\it Boethe tree} is just the graph
describing the structure of a~regular social hierarchy (Fig.~A2b).

The question of the existence of a~phase transition in a~social
hierarchy~--- i.e., its collapse or essential inner transformation
--- is one of the principal problems in the study of social
changes. And, within the framework of the above described model,
this problem has an exact analytical solution.

In what follows we consider several similar results and their
interpretation. Now consider several simple network models.

First, consider a~social network in which each\index{agent} agent
has a~fixed and equal for all agents number of neighbors in the
network and in which there are only two patterns of behavior
(agent's states). We may further assume that only nearest
neighboring agents interact. Assume also that the transaction
costs for all pairs of interacting neighbors are equal and only
depend on the combination of two possible patterns of the agents'
behavior:
\begin{equation}
\label{eq200}
V(\sigma_{i}, \sigma_j)=J\sigma_{i}\sigma_j,
\end{equation}
where $\sigma_{i}$ and $\sigma_j$ can be equal to either 1 or
$-1$. The system obtained is known in physics as the Ising model.

Observe that the Ising model can describe two essentially
different situations:

\medskip

(1) an analogue of the spin\index{spin model} system or

(2) the case of replacement.

\medskip

In the first case, each of the agents\index{agent} can change the
pattern of behavior.

In the second case, there are two types of agents incapable to
change their own patterns of behavior but capable to replace each
other in the nodes of graph. The second case, where the
interaction between the agents of two types ($A$ and $B$) is given
by three values of transactional expenditures $V_{AA}$, $V_{BB}$
and $V_{AB}$, is reduced to the Ising model but the statistical
sum\index{statistical sum} in the Hamiltonian of the interaction
accrues by the term corresponding to an external field.

Certain important conclusions concerning the social network
represented by the replacement solution can be made immediately as
a result of its formal reduction to the spin model.\index{spin
model} It turns out that the quantity
\begin{equation}
\label{eq201}
2J=\frac12(V_{AA}+V_{BB})-V_{AB}
\end{equation}
is a~very essential parameter. If this parameter is negative, then
the agents of types $A$ and $B$ will tend to randomly replace each
other in the network whereas, if it is positive, then under the
decrease of the total costs (the sum of all transaction costs) the
network will fragment into clusters consisting of agents of type
$A$ and agents of type $B$.

In a~real social system (in which types $A$ and $B$ are ethnic or
social groups), ethnic or social cleansing correspond to this
process.

Now consider a~social network in which each agent\index{agent} has
$2n$ neighbors. The most easy way to realize such a~network is to
form of an $n$-dimensional lattice. Simple thermodynamic arguments
allow one to determine whether in this case the one-phase state
(when both patterns of behavior are the same) is stable or not.

First, consider the case of a~one-dimensional network. If, in such
a system, we allow domains with different patterns of behavior,
then we allow the changes of the level of transaction costs. The
appearance of one borderline  causes the total change of the
transaction costs by the doubled constant of interaction between
two neighbors.

At the same time the borderline can occur at $N$ distinct places,
where $N$ is the length of the network. Therefore the change of
the free energy\index{free energy} can be estimated as
\begin{equation}
\label{eq202}
\Delta F=\Delta E-T\Delta S=2J-T\ln N,
\end{equation}
so $\Delta F$ becomes negative for sufficiently long net. The
negative value of the change of the free energy in the process of
creation of the disordered state means that this ordered state is
thermodynamic unstable (this argument is due to Peierls, see
\cite{P6}).

Somewhat more sophisticated arguments prove that, in the network
represented by a~two-dimensional lattice, the domains with
orientation opposite to that of dominant one are unstable at low
temperatures.

The same applies also for lattices of higher dimensions.

In other words, the dissident clusters in the scenario described
by the Ising model of social networks in dimensions $>1$ are only
stable at temperatures above a~critical one, and therefore in this
case a~phase transition is possible.

Though very rough, this model catches one very essential aspect of
the situation. The diminishing of the temperature in the network
causes fragmentation of the model. Therefore the tolerance of the
system to dissidents is a~corollary of a~sufficiently high level
of well-being.

It is very well known empirically that democratic regimes tend to
exist stably under the level of per capita income exceeding
three--four thousand dollars. It is interesting to try to relate
this empirical fact with phase transitions in network models.

With the help of Peierls arguments we can also evaluate the
stability of clusters in regular transitions (i.e., in nets
corresponding to Boethe lattices). The results obtained here are
rather astonishing: The branches of the hierarchical tree are
unstable with respect to reorientation. But this is not the case
for the inner domains of the hierarchical tree in which, for
sufficiently low temperatures, the ``dissident'' clusters cannot
occur. Once again the social projection of these corollaries is
not without interest. It is very well known that under lower
temperature conditions (that is when the total costs diminish)
rigid hierarchies tend to appear. This rigidity of hierarchies
melts with the heightening of the temperature that is with the
growth of income. The history of collapses of autocratic regimes
illustrates that the dissident groups are usually formed on the
middle levels of autocratic hierarchies (the branches fall off)
and the social changes by means of coups occur much more often
than people-driven revolutions.

It goes without saying that the Ising model of transactional
interactions is too rough, if not rude, to describe social
networks. It is not difficult, however, to improve it by
considering the mean field approximation (\cite{Z7}). Under such
an approach the interaction of each agent\index{agent} with its
nearest neighbors is also influenced by a~mean field of
interactions with the more distant agents; let this interaction be
self-consistent.

The obtained approximate description of the system is known as the
Curie-Weiss model,\index{Curie-Weiss model}\index{model,
Curie-Weiss} which also has a~phase transition for a~sufficiently
low temperature. In other words, in terms of the social theory the
result consists in the fact that for the low total income there is
a tendency towards a~uniform behavior.

\section{Certain Conclusions}

In this note we have no possibility to analyze in detail a~larger
number of statistical models of social networks. To do so, though
not truly difficult, would require a~much larger amount of work.

We have no possibility to discuss interesting mathematical details
of the models of phase transitions in social nets, either. The
level of mathematical apparatus necessary for a~reasonable
discussion of such models far exceeds the level of complexity of
the mathematical apparatus usually applied in social studies.

To understand the inner quantitative peculiarities of such models,
it is natural to use modern representations of the spectral
analysis of operators in functional spaces, Feynman integrals, and
certain basic notions of the modern quantum field theory and
statistical mechanics, such as renormalization group and scaling.
\footnote{The nature of the mathematical machinery useful in the
study of statistical theory of social networks can be appreciated,
for example, with the help of the books \cite{S8, W9}.}

Such a~situation naturally poses several serious questions
pertaining to the realm of the theory of science.

First of all, the very fact that thermodynamic approaches and,
speaking more broadly, the methods of statistical mechanics have
not been, so far, widely used in the social theory is astonishing.
As in disordered physical systems, in social systems, the
distributions of a~certain conserved parameter among the elements
of the system is of huge importance, which guarantees the success
of thermodynamic approach to the study of such systems.

Furthermore, in social systems, the parameters of order can be
naturally introduced which allows one to apply physical models of
phase transitions for the description of social systems. The
utility\index{utility} of application of statistical models of
social systems, in spite of their roughness, can hardly cause any
doubt.

The perspective of development of such methods is, nevertheless,
considerably impaired by the social structure of science and
higher education.

It seems to me that the general level of mathematical education at
the universities preparing researchers in the domain of social
sciences does not correspond to the level necessary for the active
work with models of such sophistication.

On the other hand, the interest to work in the domain of social
studies among experts with sufficient mathematical training~---
theoretical physicists and mathematicians~--- is restrained
precisely due to the widespread understanding among theoretical
physicists and mathematicians of dullness of social sciences, that
is impossibility for the representatives of the natural and
precise sciences to apply their intellectual potential in its
totality.

One of the main tasks of this note was to demonstrate how faulty
these prejudices are. We believe that at the moment the social
sciences represent one of the most promising domains for
application of the enormous stock of methods and models worked out
in mathematics, theoretical and mathematical physics, especially
lately.

\begin{thebibl}{MMM}

\bibitem[1]{N1}
North D. {\em Institutions, Institutional Changes and Economic
Performance}. Cambridge University Press, London, 1991.

\bibitem[2]{S2}
Simon H. {\em Models of Bounded Rationality}, M.I.T. Press,
Cambridge Mass., 1982.

\bibitem[3]{P3}
Polanyi M. {\em The Tacit Dimension}, Anchor, 1967.

\bibitem[4]{K4}
Koestler A., Janus. {\em The Summing Up}, Picador, London, 1979.

\bibitem[5]{K5}
Kassiner E. {\em The Myth of the State}, Yale University Press,
New Haven, 1968

\bibitem[6]{P6}
Peierls R., Quelques propri\'et\'es typiques des corps solides.
(French) Ann. Inst. Henri Poincar\'e, v.5, 1935, 177--222; see
also Ziman, J. {\em Models of disorder. The theoretical physics of
homogeneously disordered systems}. Cambridge University Press,
Cambridge-New York, 1979. xiii+525 pp

\bibitem[7]{Z7}
Ziman J., {\em Models of disorder}. The theoretical physics of
homogeneously disordered systems. Cambridge University Press,
Cambridge-New York, 1979. xiii+525 pp.

\bibitem[8]{S8}
{\em Stability and phase transitions} (Russian) Translated from
the English by S.~P.~Malyshenko and E.~G.~Skrockaya. Mir, Moscow,
1973. 373 pp. This collection contains translations into Russian
of four lectures. The first three were presented by F.~J.~Dyson,
E.~W.~Montroll and M.~ Kac at the Brandeis University Summer
Institute in Theoretical Physics in 1966 [{\em Statistical
physics: phase transitions and superfluidity}\/ (Brandeis Univ.
Summer Inst. in Theoret. Phys., 1966), Vols. 1, 2, Gordon and
Breach, New York, 1968], the fourth by M.~E.~Fisher at the
International School of Physics ``Enrico Fermi'' in Varenna in
1970 [Proceedings of the International School of Physics ``Enrico
Fermi'' (Varenna, 1970). Course 51: Critical phenomena, Academic
Press, New York, 1971].

\bibitem[9]{W9}
Wilson K., Kogut J., The renormalization group and
$\epsilon$-expansion, Physics reports. 12C no.2 (1974) 75--199

\end{thebibl}



\chapter{Economic dynamics as geometry}

If the macro-parameters of an economic system are functionally
dependent, and the surface of state is differentiable, then the
differentials of the macro-parameter are related by a~system of
Pfaff equations.\index{Pfaff equation}\index{equation, Pfaff}

In fact, we may consider a~general system described by parameters
$x, \ldots , z$. If we change the state of this system, small
changes may be considered as differentials $dx, \ldots , dz$.
Generally speaking, if the system is constrained by internal
links, the differentials of parameters are not independent. They
are related by linear equations of the form
\begin{equation}
\label{eq203}
A(x, \ldots  , z)dx +B(x, \ldots  , z)dy +  \ldots   = 0.
\end{equation}

We can find a~number of such Pffaf equations that single out our
system.

Such a~general case is too complicated for analysis. Let us
restrict ourselves with a~simpler case with one Pfaff equation and
a number of functional dependencies between parameters of our
system:
\begin{equation}
\label{eq204}
f(x, \ldots  , z)=0,\quad  g(x, \ldots  , z)=0, \ldots   .
\end{equation}

We can change the system of coordinates and minimize the number of
independent parameters of the Pfaff equation. The smallest such
number is called the {\it class} of the Pfaff equation. The class
of one equation is always odd (\cite{R1}).

We have just obtained a~system completely identical to the
generalized Hamiltonian system studied by Dirac \cite{Di2} (see
also Faddeev \cite{F3}, Pavlov \cite{P4}). Formally, all
functional constraints are equivalent and each one may be
considered as a~Hamiltonian function.

In the analysis of generalized Hamiltonian system, time plays an
outstanding role. In our case, we do not need to distinguish the
role of one parameter. We may choose one constraint $ f(x, \ldots
, z)=0$ equation on the surface defined by the constraints
(\ref{eq203}), see \cite{R1}, Chap.~IX. It is possible to relate
a~vector field to each field of characteristic directions. We may
act by this vector field on any function $g(x, \ldots  , z)$. The
result is called the {\it Jacobi},\index{bracket,
contact}\index{bracket, Jacobi} or {\it Legendre}, or {\it
contact, bracket} $\{f, g\}_{K.b.}$ (see \cite{R1}, Chap. IX).

The difference between the Jacobi and Poisson brackets consist of
the fact that the Poisson brackets are defined on an
even-dimensional space with a~bivector whose Schouten bracket
(\cite{Gr0}) with itself vanishes (if the bivector is
non-degenerate, it is equivalent to a~non-degenerate closed
differential 2\defis form that determines a~symplectic space). The
Jacobi brackets are defined not for differential form but for
a~Pfaff equation in the space with odd dimension.

To find a~$k$-dimensional integral surface on the surface singled
out by the equations of functional constraints is only possible if
all Jacobi brackets between the constraints vanish on this surface
\cite{R1}, Chap.~IX. If some of these brackets do not vanish, we
must include the result of bracketing into the set of constraints
and calculate Jacobi brackets between all the elements of the new
set. After several iterations we may come to one and only one of
the following results:

\medskip

(a) The number of constraints exceeds $k+1$, so no integral
surface of the Pfaff equation exists.

(b) The process terminates and we get a~surface on which we can
find an integral surface of our Pfaff equation. In absence of
singularities this surface simply consists of $k$-dimensional
integral surfaces of the Pfaff equation.\index{Pfaff
equation}\index{equation, Pfaff}

\medskip

Despite the simplicity of the argument the consequences could be
very important for the economic systems. {\bf We need a~check of
coherence of constraints imposed on the system}. This means that,
in the case of economic system, control actions cannot be
arbitrary. Any additional constraint imposed on the system must be
checked on compatibility with other constraints, it should not
destroy the very existence of a~solution of the system of
equations.

Consider a~hypothetical example. Imagine that, in a~certain
country, the currency board is established. This means that an
additional link is created between the country's economy and the
world economy. Could we be sure that such an additional link would
not destroy the equilibrium of the country's economy?

One more remark. It is known that, for some equations of
generalized Hamiltonian systems, there exist certain hidden,
``non-physical'' variables \cite{Di5}. This means that, under
certain conditions, there exist functions of parameters of the
system that could be changed without visible effect. A
contemporary physical theory (gauge theory) attributes a~great
importance to the role of such cases. It is rather interesting to
investigate what kind of effects could be connected with that type
of equations. We may conclude here with one very general and
challenging statement: The integral surface of a~system of Pfaff
equations has some geometrical features. It is possible to define,
for example, different connections on this manifold, using the
natural frame bundle, or to define characteristic classes. All
such geometrical objects may be interpreted as dynamic objects, if
dynamics is introduced properly.

We may change the direction of inference and interpret
equilibrium\index{equilibrium, market} economical dynamics as
geometry.

Of course, this is only general definition of a~very complex
program of future studies, but I am convinced that the realization
of such a~program could change significantly our understanding of
economic processes.

\subsection*{Editor's remarks} {\bf The idea of locality.} Basically, when
we describe our world by mathematical equations, we may use two
distinct types of theories: local and non-local ones. Non-local
theories are described by integral or integro-differential
equations.

Locality of a~theory means that any change of the system's
parameters of a~point is defined by its infinitesimally small area
i.e., the long-distant interaction is absent. In such a~case, the
mathematical model of the system is represented by differential
equations (perhaps, of arbitrary order, non-linear, etc.).

Any system of differential equations may be reformulated as
a~system of Pfaff equations \cite{BCG}. Some of these systems
determine an integrable distribution, others determine
non-integrable distributions; Hertz called such distributions {\it
nonholonomic}. A branch of mathematics I am working in
(representation theory) provides with tools to analyze~--- at
least, in principle~--- on a~qualitative level solvability of
a~given system of differential equations and stability of its
solutions, if any exist.

This book shows how to represent economic theories as nonholonomic
systems.

{\bf How to solve differential equations}. I could never
understand why engineers (or any other customer who needs to solve
differential equations in earnest, in real life) never use
criteria for formal integrability of differential equations
Goldschmidt suggested in late 1960's (for an exposition, see
\cite{BCG}).  As a~byproduct, these criteria yield an approximate
solution.  It is just inconceivable that these criteria are only
good for nothing but theoretical discussions and no type of
differential equations is ``convenient'' enough for these criteria
to be implemented in real computations.

The following facts constituted, perhaps, the stumbling block:

\medskip

(a) psychological threshold: these criteria are formulated in
terms of Spencer {\it cohomology}~--- a~new and scary term, not
clear if worth learning;

(b) even now no efficient code for computing cohomology exists
(although the problem obviously involves lots of repeated
computations and sparse matrices, none of these features was ever
exploited); at the time Goldschmidt suggested his criteria the
situation was much worse;

(c) the criteria embraced only ``one half'' of all equations.

\medskip

To explain (c), recall that a~theorem of S.~Lie (\cite{KLV})
states that all differential equations are of the two types: for
one, all its symmetries are induced by point transformations, for
the other one, all its symmetries are induced by contact
transformations. Goldschmidt's criteria fit only differential
equations whose symmetries are induced by point transformations,
they did not embrace differential equations whose symmetries are
induced by contact transformations. Now, recall that the contact
structure is the simplest of {\it nonholonomic} structures.

The first nonholonomic examples were from mechanics: a~body
rolling with friction over other body.  Among various images that
spring to mind, a~simplest is that of a~bike, or just a~ball,
rolling on an asphalt road.  At the point of tangency of the wheel
or the ball with asphalt the velocity is zero.  This is a~{\it
linear} constraint.

More generally, a~manifold (no dynamics) is said to be {\it
nonholonomic} if endowed with a~non-integrable distribution (a
subbundle of the tangent bundle).  A famous theorem of Frobenius
gives criteria of local integrability: the sections of the subbundle
should form a~Lie algebra. One often encounters {\it non-linear}
constraints: switching the cruise control of your car ON you single
out in the phase space of your car a~distribution of spheres over
the configuration space.

On my advice this book is appended with a~paper by Vershik with
first rigorous and lucid mathematical formulations of nonholonomic
geometry and indications to various similar structures in several
unexpected, at the time Vershik's paper was written, areas (like
optimal control or macro-economics, where nonlinear constraints
are also natural). Vershik summarizes about 100 years of studies
of nonholonomic geometry (Hertz, Carath\'eodory, Vr\u anceanu,
Wagner, Schouten, Faddeev, Griffiths, Godbillon; to his list we
should add that Internet returns hundreds of thousands entries for
``nonholonomic'', and its synonyms (anholonomic,
``sub-Riemannian'', ``Finsler'', ``cat's problem'' and
``autoparallel''; there seems to be more, actually, nonholonomic
dynamical systems than holonomic ones; finally ``supergravity'' is
also a~nonholonomic structure, albeit on {\it super}manifolds).

\begin{thebibl}{MMM}

\bibitem[Di1]{Di2}
P. A. M. Dirac, The Hamiltonian form
 of field dynamics. Canadian
J.  Physics {\bf2}, No. 1  (1951), 1--23.

\bibitem[D2]{Di5}
P. A. M. Dirac, {\it Lectures on Quantum Mechanics}. Yeshiva
University, New York (1964).

\bibitem[F]{F3}
L. D. Faddeev,  Feynman integral for singular Lagrangians. Teor.
Matem. Fizika  {\bf 1}, No. 1 (1969), 3--18.

\bibitem[Gr]{Gr0}
Grozman P., Classification of bilinear invariant operators on tensor
fields.  Functional Anal. Appl. 14 (1980), no. 2, 127--128; for
details, see \texttt{arXiv: math.RT/0509562}

\bibitem[KLV]{KLV}
Krasilshchik, I. S.; Lychagin, V. V.; Vinogradov, A. M. {\em
Geometry of jet spaces and nonlinear partial differential
equations}. Translated from the Russian by A. B. Sosinsky.
Advanced Studies in Contemporary Mathematics, 1.  Gordon and
Breach Science Publishers, New York, 1986.  xx+441 pp.

\bibitem[P]{P4}
V. P. Pavlov, Dirac's bracket. Teoret. Mat. Fiz.  {\bf 92}, No. 3
(1992),  451--456.

\bibitem[R]{R1}
P. K. Rashevsky, {\it Geometrical Theory of Partial Differential
Equations.} OGIZ, Moscow--Leningrad, 1947. 354 pp. (Russian)

\end{thebibl}


\chapter{Mathematics of nonholonomicity (A.\,M.~Vershik)}

\section
{Dynamics with constraints}\index{Vershik A. M.}

The purpose of this article\footnote{This is an edited (by me, {\em
D.L.}), mainly, as far as English is concerned, version of the
namesake article published in Yu.~Borisovich, Yu.~Gliklikh (eds.),
{\em Global Analysis~--- Studies and Applications. I.}, Lecture
Notes in Mathematics 1108, 1984, 278--301. I tried to return this
difficult to read version of English to match the original Russian
transcript of Vershik's inspiring lecture at Voronezh Vinter school:
 Vershik, A. M. Classical and nonclassical dynamics with
constraints. (Russian) In: Yu.  G. Borisovich and Yu.  E. Gliklikh
(eds.)  {\em Geometry and topology in global nonlinear problems,
23--48, Novoe Global. Anal.}, Voronezh.  Gos.  Univ., Voronezh,
1984. The article expounds the joint paper of A.~Vershik and
L.~Faddeev (Lagrangian metrics in invariant form. In: Problems of
Theoretical Physics. vol. 2, Leningrad, 1975 = Selecta Math. Sov.
1:4 (1981), 339--350) and gives a~mathematical description of
several aspects of nonholonomic manifolds, i.e., manifolds with
a~nonintegrable distribution. A similarity of the constrained
dynamics for the linear constraints with fields of cones in the
optimal control \cite{22} is observed. I also tried to update
references and eliminate typos. I am very thankful to
Springer-Verlag for the permission to publish another translation of
this important review. {\em D.L.}} is to give a~detailed and, to
a~large extent, self-contained account of results, to raise a~number
of questions on dynamics with constraints on the tangent bundle of
a~smooth manifold. The classical problems of this kind are the
problems of nonholonomic mechanics, nonclassical ones~--- the
problems of optimal control and economic dynamics. The investigation
of these topics from the point of view of global analysis was
started fairly recently (see \cite{1, 2, 3}). Such a~treatment needs
a~detailed study of geometry of the tangent bundle, connections and
other notions necessary for general Lagrangian mechanics and,
particularly, for the theory of nonholonomic problems. The present
article continues the investigations of geometry of the tangent
bundle and dynamics on it. For standard facts from the geometry of
manifolds and the Riemannian geometry, see \cite{13, 18, 19}.

\ssec{0.1}{Lagrangian dynamics} The Lagrangian formalism is based on
a~procedure that allows one to invariantly construct a~special (see
(\ref{eq205}) vector field given an arbitrary $C^2$-smooth function
on the tangent bundle (this function is called the {\it
Lagrangian}). Formally, this construction uses only two canonical
objects present in the tangent bundle of any manifold: the {\it
principal tensor} and the {\it fundamental vertical field}. These
structures are invariantly defined in \cite{1, 2, 3}.

The Lagrangian mechanics studies the structure of trajectories of
special vector fields~--- their integrability, stability, integrals,
and so on~--- as dependent on the Lagrangian.

Such entities and principles of classical mechanics as forces,
virtual displacements,\index{displacement, virtual, see virtual
displacement} variational principles can be completely described in
terms of geometry of the tangent bundle. In addition to that, there
is a~number of geometric notions insufficiently used so far but,
probably, important for mechanics, e.g., the notion of connection.

Recall here the following fundamental result due to Levi-Civita
(later elaborated by Synge et al.) which connects mechanics with
geometry:
\begin{thetext}
{\sl any mechanical system with quadratic Lagrangian moves by
inertia along the geodesics of the corresponding Riemannian
manifold.}
\end{thetext}

Levi-Civita also defined the Riemannian connection. In \cite{1, 2}
(see \S 2) this theorem was generalized to the non-Riemannian
connections that appear in the nonholonomic case.

The translation of Lagrangian mechanics into the language of
geometry was initiated in the works \cite{1, 2, 3}, but it is not
entirely completed.  For an invariant formulation of {\it
d'Alembert's principle}~--- the most general local principle of
mechanics  also valid for dynamics with constraints, see \cite{2}.

Observe that the Hamiltonian mechanics in an invariant form
(symplectic dynamics) gained much more attention than the Lagrangian
one. This is quite understandable, because the symplectic structure
is a~universal object of analysis and geometry (see \cite{4}). Yet,
from the point of view of mechanics, examples of symplectic dynamics
are scantier, such dynamics have no equivalent for certain
mechanical notions (force, constraint, and several more).

Apart from that, the presence and necessity of two parallel
formalisms (the Lagrangian and Hamiltonian ones) in the quantum
theory indicates that they might be unable to completely replace
each other on the classical level, either. In this article we
primarily use the Lagrangian formalism, passing to symplectic
geometry only for reduction and examples (see~\ref{s3}).

\ssec{0.2}{The dynamics with constraints} At least three branches of
science~--- nonholonomic mechanics, optimal control and dynamics of
economics~--- lead to necessity of considering the following
generalization of Lagrangian dynamics:
\begin{thetext}
{\sl given a~submanifold (e.g., a~subbundle) or a~distribution
(a~field of subspaces) in the tangent bundle $TQ$ of a~manifold $Q$,
construct a~dynamics so as the trajectories could not leave the
given submanifold, or, equivalently, the vector field would belong
to the given distribution.}
\end{thetext}
In some problems (e.g., of optimal control) the value of vector
field at each point might belong to a~submanifold with boundary or
even with corners.

These problems were almost nowhere considered from the point of view
of global analysis and coordinate free differential geometry, save
the study of nonholonomic dynamics in \cite{2}. The initial aim of
the article \cite{2} was to comprehend nonholonomic mechanics in an
invariant form and to revise Lagrangian mechanics in conformity with
this comprehension.

The fields of cones or polytopes in the tangent bundle (see \cite{4,
5, 22}) are important in optimal control and economic dynamics.

There exist two different constructions of dynamics with
constraints, each of construction leading to reasonable mathematical
problems. A choice between the two possibilities lies outside
mathematics.

Let a~field of submanifolds in the tangent spaces be given either as
the set of zeros or as the set of non-positivity of a~collection of
functions and, additionally, let there be given a~Lagrangian or,
more generally, a~ functional. Then one can:

\medskip

1) consider the conditional variational problem of minimizing
a~certain functional (action, time, and the like) provided the
trajectories belong to the given submanifold, and, with the help of
an appropriate version of the Lagrange method, derive the
Euler-Lagrange equation, which in other words can be expressed, as
we will see, as a~vector field,

2) consider the projection of the vector field that corresponds by
1) to the Euler-Lagrange equation of the unconditional problem (on
the whole tangent bundle) at every point of the given submanifold
onto the tangent space to the submanifold at this point, and again
obtain a~vector field.

\medskip

Both these vector fields tangent to the submanifold of constraints,
are {\it special} (\ref{eq205}), and therefore determine a~dynamics
with constraints. But, generally speaking, these fields do not
coincide. In the first case, the constraints are ``in-built'' in the
Lagrangian, in the second case, only the reactions of constraints
take effect.

It is the second construction this should be used for description of
the movement of mechanical systems. In a~number of special cases
(holonomic constraints, Chaplygin's case~--- a~special case of
linear constraints, and so on), both constructions lead to the same
vector field. (Among a~considerable number of works on nonholonomic
mechanics, we mention here \cite{7, 8, 9, 10}, where one can find
further references. I know no work on this topic that uses an
invariant approach except \cite{2}, \cite{23}.)

Faddeev and me shown in \cite{2} that the second construction
corresponds to the general d'Alembert's principle, and, in this
situation, {\bf no variational principle corresponds
to~it}.\footnote{Boldface is mine: this important observation might
explain the in-built (and, perhaps, impenetrable) obstacles in
quantizing supergravity since the Minkowski superspace is
nonholonomic. {\it D.L.}}

Of course, the principle is a~postulate verified by practice, and by
no means a~theorem.

An implicit abusing of the two approaches favored confusions in the
foundations of nonholonomic mechanics until recently. Starting with
the first works of classics, these confusions were present, in some
form, in almost all mathematical textbooks on Calculus of
Variations.

For example, in mathematics, the conditional Lagrange problems with
non-integrable conditions on derivatives are called {\it
nonholonomic}. This usage might give the reader an idea that the
nonholonomic mechanical problems are conditional variational
problems with non-integrable constraints. As we observed above, this
is not true. (Of all the textbooks I know, the difference between
the ``problems with nonholonomic constraints'' and ``problems of
nonholonomic mechanics'' is distinguished in \cite{24} only.)

The first construction is employed in optimal control and other
applications of dynamics with constraints. However, it is also
possible to use the second construction.

Many technical difficulties appearing in the treatment of dynamics
with constraints are due to the absence or insufficient usage of the
adequate geometric apparatus. We mean here, first of all, the
coordinate-free theory of distributions and connections. With these
notions instead of vague ``quasi-coordinates'' and sophisticated
procedures of exclusion, the account becomes simple and the whole
dynamics turns into the source of new lucid mathematical problems
whose solution will promote new effective applications.

\ssec{0.3}{Contents and main results} In this article, we give
a~detailed account of the theory of the problems with constraints on
the tangent bundle. In \ref{s1}, we briefly, with certain
innovations and simplifications, recall the results from \cite{1,
2}, i.e., we give an invariant derivation of the Euler-Lagrange
equations for an arbitrary Lagrangian and an invariant study of
equations with constraints.

In \ref{s2}, we analyze the systems with a~quadratic Lagrangian (the
Newton equations), and the systems with linear constraints. For this
purpose, we define a~new geometric object~--- the {\it reduced
connection} on a~subbundle of the tangent bundle of a~Riemannian
manifold. To my mind, this object corresponds to the old coordinate
notion of ``nonholonomic manifold'' which appeared in a~number of
geometric works written in the 1930s--40s (V.\,V.~Wagner, Schouten,
Vr\u anceanu). The geodesic flow for this connection is the main
object to study.

The most interesting is the case of an {\it absolutely nonholonomic
constraint}: the corresponding phase space is then endowed with
a~new (non-Riemannian) metric, and the {\it Hopf-Rinow theorem}
(\ref{hrth}) in the usual formulation is false for this metric.

Further in \ref{s2} we consider the general problem with linear
constraints and quadratic Lagrangian, and give a~complete proof of
the following theorem: \vskip 2mm
\begin{center}
\begin{minipage}[l]{.8\textwidth}
{\sl The gliding of a~system with quadratic Lagrangian and linear
constraints occurs along the geodesics of the reduced connection in
the subbundle that the constraints single out.}
\end{minipage}
\end{center}

\vskip 1mm

In a~less precise form this theorem is given in \cite{1, 2}.

\ref{s3} is devoted to the main example~--- the group theory
(symmetry study) of the problems with constraints. We formulate the
problems of rolling without sliding as group-theoretical ones and
consider gliding. The main statement of this section is a~reduction
theorem analogous to that of symplectic dynamics. We reduce the
problem to dynamics on the dual of a~Lie algebra. It is interesting
to find out the cases of its integrability.

In \ref{s4}, we briefly consider non-classical problems for the
fields of indicatrixes. The peculiarity of these problems is that
the constraints are given by inequalities; in each tangent space,
they single out submanifolds with boundary or with corners. In
problems of optimal control (stated in terms of differential
inclusions) and of dynamics in economics, there appear analogous
fields of convex subsets in the tangent spaces~--- analogs of
distributions (linear subspaces) in the classical case, see \cite{5,
6, 21}.

Here it is necessary to use the extremal posing of the problem,
because the constraints work in an active form (through control) and
not through reactions.

These problems are also very close to the problem of infinite
dimensional convex programming. In the simplest case, this
phenomenon was noted in \cite{5, 11}.

We also formulate a~number of open problems.

\section[The tangent bundle and Lagrangian
mechanics]{Geometry of the tangent bundle and Lagrangian
mechanics}\label{s1}

\ssec{1.1}{Background} Let $Q$ be a~smooth (always $C^\infty$)
connected manifold without boundary, $TQ$ its tangent bundle,
$\pi\colon TQ\tto Q$ the canonical projection, let $T^2Q = T(TQ)$
and $d\pi\colon T^2Q\tto TQ$.

The tangent spaces at the points $q\in Q$ and $(q, v)\in TQ$ are
denoted by $T_{q}$ and $T_{q, v}$, respectively. A vector field $X$
on $TQ$ is called {\it special}\index{vector field, special} if
\begin{equation}
\label{eq205} d\pi(X_{q,v}) = v.
\end{equation}
The term ``special vector field'' is a~synonym of the terms
``virtual displacement''\index{virtual displacement} in Mechanics
and ``second order differential equation''\index{differential
equation, second order} in Calculus.

The vertical tangent vectors, i.e., the vectors tangent to the fiber
$T_{q} \subset TQ$ over $q$ of the vector bundle $TQ$, form the
subspace $\tilde T_{q, v} \subset T_{q, v}$. Evidently, $\tilde
T_{q, v}$ can be identified with $\tilde T_{q}$ and, by the same
token, with $T_{q}$. Thus, we have the following canonical
monomorphism
\begin{equation}
\label{eq206} \gamma_{q, v}\colon T_{q}\tto T_{q, v}
\end{equation}
and a~$(1, 1)$\defis tensor
\begin{equation}
\label{eq207} \tau_{q, v} =\gamma_{q, v}d\pi_{q, v}
\end{equation}
on $TQ$.

The tensor field $\tau=\{\tau_{q, v}\mid (q, v)\in TQ\}$ is called
the {\it principal tensor field}\index{tensor field, principal} on
$TQ$. This field, as a~map of vector fields, annihilates all the
vertical vector fields (because they are annihilated by $d\pi$) and
only them. The range of $\tau$ is the subbundle consisting of
vertical vector fields.

The coordinate form of $\tau$ is as follows:
\begin{equation}
\label{eq208} \tau\left (a\pder{q}+b\pder{v}\right)=a\pder{v}.
\end{equation}

The dual tensor field $\tau^*$ acts on forms:
\begin{equation}
\label{eq209} \tau^*(a dq + bdv) = bdq.
\end{equation}
The range and the kernel of $\tau^*$ coincides with the bundle of
horizontal $1$\defis forms on $TQ$ (i.e., the forms that annihilate
the vertical fields).

The vertical field on $TQ$ with coordinates
\begin{equation}
\label{eq210} \Phi_{q,v}=\gamma_{q, v}v=\sum v_{i}\pder{v_{i}}
\end{equation}
will be called the {\it fundamental}\index{vertical field,
fundamental} one.

As is easy to see,
\begin{equation}
\label{eq211} \text{a field $X$ is special if and only if $\tau
X=\Phi$.}
\end{equation}
Indeed, in local coordinates, $X_{q,v} = \sum v_{i}\pder{q_{i}}+
...$ and $(\tau X)_{q,v} \sum v_{i}\pder{q_{i}}=\Phi$.

With these two notions~--- $\tau$ and $\Phi$~--- we formulate all of
Lagrangian mechanics in invariant way.\footnote{L.~D.~Faddeev and me
defined the tensor $\tau$ in 1968. A report on invariant
construction was made for the Leningrad Mathematical Society in May
1970 (see Uspekhi Mat. Nauk, 1973, v. 28, no. 4, p. 230). Later I
learned about the book \cite{3}, published in French in 1969 and
translated into Russian in 1973, where $\tau$, called a~``vertical
endomorphism'', was considered and where the invariant form of the
Euler equations (without constraints) was given. The work \cite{2}
has been accomplished in 1971 and was sent to press in early 1973.}

Let $L$ be a~Lagrangian, i.e., a~linear on fibers smooth function on
$TQ$ (i.e., a~1\defis form on~$Q$). Then $\tau^* (dL)=
\pderf{L}{v_{i}}dq_{i}$ is a~horizontal 1\defis form called the {\it
momenta field of the Lagrangian}.

In mechanics, the horizontal 1\defis forms describe forces, and the
integral of such a~form along an integral curve of a~special field
is the work performed by the force along this curve.

The {\it Lagrangian $2$\defis form}\index{Lagrangian $2$\defis form}
is $\Omega_{L} = d(\tau^* (dL))$ (cf. \cite{3}, Ch. 11, \S 1); in
coordinates, we have:
\begin{equation}
\label{eq212} \Omega_{L}=\frac{\partial ^2L}{\partial v_{i} \partial
v_{j}}dq_{i}dv_{j}+ \frac{\partial ^2L}{\partial v_{i} \partial
q_{j}}dq_{i}dq_{j}.
\end{equation}

The image of this 2\defis form under the Legendre
transformation\index{Legendre transformation} (provided the Hessian
$\frac{\partial ^2L}{\partial v_{i} \partial v_{j}}$ is
non-degenerate) is the canonical 2\defis form $\sum dp_{i}dq_{i}$ on
the cotangent bundle. The {\it
energy}\index{energy}\index{Hamiltonian} (or the {\it Hamiltonian})
is
\begin{equation}
\label{eq213} H_{L} = dL(\Phi) -L,
\end{equation}
and the {\it Lagrangian force on the virtual displacement} (special
field) $X$ is
\begin{equation}
\label{eq214} \Omega_{L}(X, \cdot)-dH_{L}.
\end{equation}
This is a~horizontal 1\defis form, its value on the vector field $Y$
can be interpreted as the work of the Lagrangian force along $X$ on
$Y$.

In our terms, {\it d'Alembert's principle}\index{d'Alembert's
principle}\index{principle, d'Alembert} (the principle of virtual
displacements) is formulated as follows:\vskip 2mm
\begin{center}
\begin{minipage}[l]{.8\textwidth}
{\sl On the vector field that determines the real trajectories of
motion, the Lagrangian force is equal to the exterior force
$\omega$. In particular, if the exterior force vanishes, then so
does the Lagrangian force}:
\begin{equation}
\label{eq215} \Omega_{L}(X, \cdot)-dH_{L}=\omega.
\end{equation}
\end{minipage}
\end{center}

In coordinates, the {\it Euler-Lagrange
equations}\index{Euler-Lagrange equations}\index{equations,
Euler-Lagrange} are:
\begin{equation}
\label{eq216} \frac{d}{dt}\pderf{L}{\dot q_{i}}=
\pderf{L}{q_{i}},\quad\text{where }\; i= 1, \dots , \dim Q.
\end{equation}
If the 2\defis form $\Omega_{L}$ is non-degenerate, then one can
easily find the vector field $X_{L}$ itself (i.e., solve the
equations with respect to second derivatives) in a~Hamiltonian form:
\begin{equation}
\label{eq217} X_{L} = \Pi_{L}(dH_{L}),
\end{equation}
where $\Pi_{L}$ is the bivector given by the expression
\begin{equation}
\label{eq218} \Omega_{L}(\Pi_{L}(\omega), Y)=\omega(Y).
\end{equation}
The fact that the field $X_{L}$ is special follows from local
formulas (\cite{2}).

The form $\Omega_{L}$ is non-degenerate if and only if $\det
(\frac{\partial ^2L}{\partial v_{i} \partial v_{j}})\neq 0$.

Eq. (\ref{eq217}) is the equation of motion written in the
Hamiltonian form. The matrix $\Gamma_{L}=(\frac{\partial
^2L}{\partial v_{i}
\partial v_{j}})$ determines a~quadratic form on the fiber and is
connected with $\Pi_{L}$ by the relation\footnote{Cf. \cite{2},
where ``horizontal forms'' on p.~132, 7-th line from below, should
be replaced by ``quotient space modulo horizontal forms'' and where
$\tau^*$ means what $\tau$ means here.}
\begin{equation}
\label{eq219} \Gamma_{L}^{-1}= \Pi_{L}\tau^*=-\tau\Pi_{L}.
\end{equation}

The case where $L$ is a~positive definite quadratic form on the
fibers of $TQ$, i.e., $L$ is a~Riemannian metric on $Q$, is of
special interest. In this case formula (\ref{eq215}) can be reduced
to the form of the Newton equations by means of the Riemannian
connection on $Q$ (see \ref{s3}).

Observe that the Lagrangian can be considered as a~closed (rather
than exact) 1\defis form on $TQ$. All the arguments are valid for
this case, but the Hamiltonian is defined only locally. This theory
is being actively studied in recent years \cite{12}.

\ssec{1.2}{Constraints} Consider a~dynamical system\index{dynamical
system} on a~smooth manifold $Q$. A {\it
constraint}\index{constraint} is a~distribution $S\subset TQ$, and a
{\it dynamical system on $Q$ with constraint $S$} is a~dynamical
system for which the velocities at every point $q\in Q$ belong to
$S$. Accordingly, the most general notion of constraint in mechanics
(and in the theory of the second order equations in general) is
defined as a~subbundle in $T^2Q$.

It is often more convenient to consider the codistributions, i.e.,
the subbundles of $T^*Q$. Then the above distribution $S$ can be
regarded as the annihilator of a~codistribution.

A {\it constraint} on the phase space $TQ$ is a~codistribution
$\theta$ on $TQ$. A {\it dynamical system concordant with the
constraint} $\theta$ is a~special vector field $X$ on $TQ$ such that
$\theta(X)=0$ at every point. In coordinates:
\begin{equation}
\label{eq220} \theta = \Span(a_{ik}dq_k + b_{ik}dv_k\mid i=1, \dots,
m).
\end{equation}

We do not require here the codistribution to be integrable, but in
the majority of \lq\lq usual" examples they are: $\theta$ is given
as the linear span of differentials of a~set of functions on $TQ$,
these functions (or rather their zeros or the level sets) determine
a~constraint in the usual sense of this word. Such a~constraint can
be called a~{\it functional} one. In this case all vector fields and
forms are considered only on the level sets of these functions
rather than on the whole $TQ$ without particular reservations.

A classical example of this kind: linear constraints~--- functions
on $TQ$ linear in velocities:
\begin{equation}
\label{eq221}
\renewcommand{\arraystretch}{1.4}
\begin{array}{l}
\theta= \Span\left (\varphi_{i}(q, v) =
\mathop{\sum}\limits_{k}a_{ik}v_{k}\mid i= 1, ..., m\right) =\\
\Span\left(\mathop{\sum}\limits_{k}a_{ik}dv_{k} +
\mathop{\sum}\limits_{k, j}\pderf{a_{ik}}{q_{j}}dq_{j}\mid i= 1,
..., m\right).
\end{array}
\end{equation}
The corresponding subset of $TQ$ is the subbundle formed  by the
codistribution
\begin{equation}
\label{eq222} \Span\left(\mathop{\sum}\limits_{k}a_{ik}dv_{k}\mid i=
1, ..., m\right).
\end{equation}
The principal tensor field permits one to define the notion of {\it
constraint reactions}. By that we mean the horizontal codistribution
$\tau^*\theta$, where $\theta$ is the constraint.

Any 1\defis form belonging to $\tau^*\theta$ is called a~{\it
constraint reaction force}. A constraint is said to be {\it
admissible} if $\dim \tau^*\theta= \dim \theta$, i.e., if the
codistribution $\theta$ has no horizontal covectors at any point
(recall that the kernel of $\tau^*$ is the space of horizontal
1\defis forms).

A constraint is said to be an {\it ideal} one if it annihilates the
fundamental vector field $\Phi$.

\sssbegin{1.2.1}{Statement} {\em 1)} If a~constraint is admissible,
then there exist a special vector fields that satisfies this
constraint.

{\em 2)} If a~constraint is an ideal one, then the $1$\defis forms
that represent constraint reactions vanish on cycles lifted to $TQ$
from $Q$ (i.e., ``do no work''). \end{Statement}

\begin{proof} 1) Since the special fields $X$ are exactly those for which
$\tau X= \Phi$, we have to establish solvability of the linear
system
\begin{equation}
\label{eq223} \tau X= \Phi,\quad \theta (X)=0.
\end{equation}
Let
\begin{equation}
\label{eq224}
\renewcommand{\arraystretch}{1.4}
\begin{array}{l}
X=\mathop{\sum}\limits_{i} v_{i}\pder{q_{i}}+\sum f_{i}\pder{v_{i}};\\
\theta=\Span\left(\mathop{\sum}\limits_{i}
(\theta_{ki}^1dq_{i}+\theta_{ki}^2dv_{i})\mid 1\leq k\leq m\right);\\
\tau^*\theta=\Span\left(\mathop{\sum}\limits_{i} \theta_{ki}^2dq_{i}\mid 1\leq k\leq m\right).\\
\end{array}
\end{equation}
Since $\dim \tau^*\theta = \dim \theta$, it follows that $\rk
(\theta^1 , \theta ^2) = \rk \theta ^2$; but
\begin{equation}
\label{eq225} \theta(X) =\mathop{\sum}\limits_{i}
(\theta_{ki}^1v_{i}+\theta_{ki}^2f_{i}) = 0;
\end{equation}
hence the system
\begin{equation}
\label{eq226} \mathop{\sum}\limits_{i}\theta_{ki}^2f_{i}=
\mathop{\sum}\limits_{i} \theta_{ki}^1v_{i}\quad (k = 1, ..., m)
\end{equation}
is solvable (for $f$). It is easy to see that, in this case, the
number of linearly independent special vector fields is $\geq \dim
Q-\dim \theta$.

2) Let $\xi$ be a~cycle in $Q$ (i.e., $\xi\colon S^1\tto Q$, where
$S^1$ is the circle, is a~continuous map), let $\tilde \xi$ be its
lift to $TQ$. Then
\begin{equation}
\label{eq227} \int_{\tilde \xi}\tau^*\theta=\int_{S^1}\langle
\tau^*\theta, \dot\xi\rangle=\int_{S^1}\langle\theta,
\tau\dot\xi\rangle=\int_{S^1}\langle \theta, \Phi\rangle=0. \qedhere
\end{equation}
\end{proof}

Let a~distribution $\theta$ be given as a~system of
Pfaff\index{Pfaff equation}\index{equation, Pfaff} equations for $X$
(for some functions $\varphi_i$ on $TQ$):
\begin{equation} \theta = \Span(X\in \Vect(Q)\mid
\label{eqpfeq} d\varphi_i(X)=0\;\text{ where }\; i\in I).
\end{equation}

\sssbegin{1.2.2}{Statement} If the functions $\varphi_i$ are
homogeneous in $v$ of homogeneity degree one, then the constraint
$\theta$ is an ideal one. \end{Statement}

\begin{proof} It follows from a~theorem of Euler that, on the
zero set of the collection of functions $\varphi_i$, where $i\in I$,
we have
\begin{equation}
\label{eq228} \
\renewcommand{\arraystretch}{1.4}
\begin{array}{l}
\mathop{\sum}\limits_{j}
v_{j}\pderf{\varphi_i}{v_{j}}=\lambda \varphi_i, \\
\text{ and hence $\mathop{\sum}\limits_{j}
v_{j}\pderf{\varphi_i}{v_{j}}=0$ on the set of zeros of the
$\varphi_i$}, \end{array}
\end{equation}
i.e., $\langle \varphi_i, \Phi\rangle=0$.
\end{proof}

Now let us proceed to derivation of the equations of constrained
dynamics (cf. \cite{2}). Let $\alpha= \{\alpha_i\}_{i\in I}$ be
a~set of 1\defis forms that define a~distribution on $Q$, let $L$ be
the Lagrangian, let $H_{L}$ and $\Omega_{L}$ be the corresponding
Hamiltonian and the Lagrangian 2\defis forms (see above). As before,
we proceed from the d'Alembert principle:
\begin{equation}
\label{eq229} \Omega_{L}(X, \cdot) - dH_{L}=\omega,
\end{equation}
where $\omega$ is the constraint reaction force that causes the
vector field $X$ to be found be compatible with the constraints
\begin{equation}
\label{eq230} \alpha_i(X) = 0\;\text{ for $i\in I$}.
\end{equation}
\nopoint \end{proof}

\sssbegin{1.2.3}{Theorem} If the Lagrangian $L$ is non-degenerate,
and the Hessian $\left( \frac{\partial^2}{\partial v_i\partial
v_j}\right )$ is positive definite at all points $(q, v)$, then, for
every admissible constraint $\alpha$, there exists a~special vector
field $X$ concordant with constraints (\ref{eq230}) and satisfying
d'Alembert's principle (\ref{eq215}), where $\omega$ is a certain
constrained reaction force.
\end{Theorem}

\begin{proof} Let us solve the system composed of (\ref{eq229}), (\ref{eq230}) for $X$ and
$\omega$. Since $\Omega_{L}$ is non-degenerate, there exist an
antisymmetric 2-field $\Pi_{L}$ of maps from the space of 1\defis
forms to that of vector fields defined from
\begin{equation}
\label{eq231} \Omega_{L}(\Pi_{L}(\rho), Y)=\rho(Y)\;\text{ for any
1\defis form }\rho.
\end{equation}
Then we have
\begin{equation}
\label{eq232} X = \Pi_{L}(dH_{L} +\omega),
\end{equation}
and by (\ref{eq230}) we have (for $i\in I$)
\begin{equation}
\label{eq233} \langle \alpha_i, \Pi_{L}(dH_{L} +\omega)\rangle = 0
\end{equation}
or
\begin{equation}
\label{eq234} \langle \alpha_i, \Pi_{L}(dH_{L})\rangle= -\langle
\alpha_i, \Pi_{L}(\omega)\rangle\; \text{ and }\;X=X_{L}
+\Pi_{L}(\omega),
\end{equation}
where $X_{L} = \Pi_{L}(dH_{L})$.

By the definition of constraint reaction we seek $\omega$ in the
form $\tau^{*}\rho$, where $\rho\in\alpha$, i.e,
\begin{equation}
\label{eq235} \langle \alpha_i , X_{L}\rangle= -\langle \alpha_i ,
\Pi_{L}(\tau^{*}\rho)\rangle.
\end{equation}
Since the constraint $\alpha$ is admissible, $\tau^{*}$ preserves
its dimension. As the Hessian, and hence $\Pi_{L}$, is positive
definite, the restriction of $\Pi_{L}$ onto $\tau^{*}\alpha$ is also
positive definite, and therefore non-degenerate. The determinant of
the system of equations $(\ref{eq235})$ for $\rho$ is nonzero at
every point:
\begin{equation}
\label{eq236} \rho=
\mathop{\sum}\limits_{j=1}^m\lambda_{j}\alpha_{j},\quad
\mathop{\sum}\limits_{j=1}^m\lambda_{j}\langle
\alpha_{j}(\Pi_{L}(\tau), \alpha_i\rangle=-\alpha_i(X_{L}).
\end{equation}
Here the $\lambda_{j}$ are the desired coefficients of expansion~---
the Lagrange multipliers.
\end{proof}

\begin{Remarks} 1) The vector field desired is the sum of a~special
vector field for the constraint-free problem and an additional
vector field $\Pi_{L}(\omega)$, the latter field being vertical,
since $\omega$ is a~horizontal 1\defis form. Thus, one can obtain
the field for the constrained problem from the field of the
constraint-free problem by means of a~projection. This projection
consists in adding a~certain vertical field (the horizontal
component being unchanged) whereupon the field becomes tangent to
the constraint. This projection depends on the Lagrangian (see
\ref{s2}).

2) If the Hessian, even a~non-degenerate one, were not positive
definite, its restriction on a~certain codistribution would,
possibly, be not of the maximal rank, i.e., the determinant of
system (\ref{eq236}) could vanish for a~certain constraint.
Similarly, if the constraint were not admissible, the rank of the
system could drop for a~certain Lagrangian. In this sense, both
conditions of the Theorem are necessary for the existence of
a~solution.

3) One can allow violation of these conditions on submanifolds of
$TQ$ of lower dimensional (and this is inevitable for certain
problems). In this case the vector field desired can have singular
points.

4) A coordinate expression of equations (\ref{eq236}) is
\begin{equation}
\label{eq237} \frac{d}{dt} \pderf{L}{\dot
q_{i}}-\pderf{L}{q_{i}}=\mathop{\sum}\limits_{j=1}^m
\lambda_{j}\alpha_{j}^i, \quad i= 1, \dots , m.
\end{equation}
Here one finds the $\lambda_{j}$ from the conditions of concordance
of the solution to the constraints:
\begin{equation}
\label{eq238} \mathop{\sum}\limits_{j=1}^m \alpha_{j}^i
dq_{i}+\tilde \alpha_{j}^i d\dot q_{i}= 0,\quad\text{where }\; j= 1,
..., m,
\end{equation}
where $\alpha_{j}= (\alpha_{j}^1, ..., \alpha_{j}^n,
\tilde\alpha_{j}^1, ..., \tilde\alpha_{j}^n) $ are the coordinates
of the constraining forms in variables $q_1, ..., q_n$, $v_1, ...,
v_n$ and the $d\dot q_i$ are taken from (\ref{eq237}). In the case
of functional constraints (exact forms $\alpha$), equations
(\ref{eq238}) turn into functional relations
\begin{equation}
\label{eq239} \varphi_{j}(q, \dot q) =0,\quad\text{where }\; j= 1,
..., m,
\end{equation}
i.e., into common constrained equations. In particular, for linear
homogeneous functional constraints, they turn into the conditions
\begin{equation}
\label{eq240} \mathop{\sum}\limits_{i=1}^na_{j}^i(q)\dot q_{i} =
0,\quad\text{where }\; j= 1, ..., m.
\end{equation}
\end{Remarks}

Once again observe that (\ref{eq237}) and (\ref{eq238}) are not,
generally speaking, the Euler-Lagrange equations for any conditional
variational problem: the Lagrange multipliers enter the right-hand
side of the Euler equation, rather than the Lagrangian. Thus,
universally adopted motion equations of nonholonomic Lagrangian
mechanics are derived from the general d'Alembert principle by means
of invariant structures in the tangent bundle.

\section{Riemannian metric and reduced
connection}\label{s2}

\ssec{2.1}{Various classical formulations for Riemannian manifolds}
If $2L = g$ is a~Riemannian metric on $Q$, i.e., a~quadratic
symmetric positive definite form on $T_q$ for every $q$, and, by the
same token, a~quadratic on fibers function on $TQ$, then the vector
field of the dynamical system with the Lagrangian $L$ admits many
descriptions. In textbooks, one usually proves the equivalence of
these descriptions implicitly. Let us recall the most important
formulations, several simplifications having been fixed previously.
Let
\begin{equation}
\label{eq241} L = H_{L} =\frac12g.
\end{equation}
Then $T^*Q$ can be identified with $TQ$ by means of the
metric\footnote{One can also identify $TQ$ with $T^*Q$ for arbitrary
non-degenerate Lagrangians, too (the Legendre\index{Legendre
transformation} transformation), but it is important that in our
case this identification is linear on fibers, hence, it preserves
linearity of constraints.}, and $\Omega_{L}$ is the canonical
2\defis form on $TQ$. The manifold $TQ$ is also a~Riemannian one
with the metric $G$ defined on $T^2Q$ from the decomposition
$T_{q,v} = T_{q,v}^{\text{vert}}\oplus T_{q,v}^{\text{hor}}$ (as
Euclidean spaces), where $T_{q,v}^{\text{vert}}$ and
$T_{q,v}^{\text{hor}}$ are the vertical and horizontal subspaces of
$T_{q,v}$, respectively (they exist due to existence of the
Riemannian connection). Each of these subspaces is canonically
isomorphic to $T_{q}$ (thanks to the isomorphisms $d\pi \colon
T_{q,v}^{\text{hor}}\simeq T_{q}$; and $T_{q,v}^{\text{vert}} =
\tilde T_{q,v}$, see \ref{s1}), hence, each of them is endowed with
the form $g_{q}$. The explicit block form of the tensors $\tau$ and
$\tau^*$ (with respect to the decomposition $T =
T^{\text{vert}}\oplus T^{\text{hor}}$) is very simple:
\begin{equation}
\label{eq242} \tau=\begin{pmatrix} 0&1_{a}\cr 0&0\end{pmatrix},
\quad \tau^*=\begin{pmatrix} 0&0\cr 1_{b}&0\end{pmatrix}, \;\text{
where $a=\dim T^{\text{vert}}=b=\dim T^{\text{hor}}$}.
\end{equation}
The horizontal forms (fields) are connected with vertical forms
(fields) by a~fixed isometry, because every vector from $T_q$ has
the horizontal lifting in $T_{q,v}$, and
$T_{q,v}^{\text{vert}}\simeq T_{q}$.

Finally, the matrix form of $\Pi_{L}$ is $\begin{pmatrix} 0&1_a\cr
-1_a&0\end{pmatrix}$. All these facts follow from elementary
geometry of Riemannian spaces.

The vector field $X_{L}$ corresponding to the motion with the
Lagrangian $L=\frac12g$ can be described by any of the following
principles:

\medskip

{\bf A}) $X_{L}$ is a~Hamiltonian field in $TQ$, i.e.,
$X_{L}=\Pi_{L}(dg)$, i.e., $\Omega_{L}(X_{L}, \cdot) = dg(\cdot)$.

The field $X_{L}$ is horizontal with respect to the Riemannian
connection because it determines the geodesic flow. Hence, its
integral curves permit another description:

{\bf B}) Integral curves of the field $X_{L}$ are the curves
$x(\cdot)$ in $Q$ satisfying the Newton equation $\nabla_{\dot
x}\dot x = 0$ (in coordinates, $\ddot x^k=\Gamma_{ij}^k\dot x^i\dot
x^j$) lifted to $TQ$.

{\bf C}) The field $X_{L}$ is the field corresponding according to
(\ref{eq217}) to the Euler equation for the variational problem of
minimum length (the {\it principle of least action}).

\medskip

The equivalence of the formulations of principles A and B is an
important fact usually proved by comparing formulas or by
establishing an equivalence with principle C. However, A and B are
local principles (d'Alembert's\index{d'Alembert's
principle}\index{Gauss principle} and Gauss's, respectively), and
their nature is distinct from that of principle C.

Ordered as generality diminishes, these principles are: A, C,
B\footnote{The principle B seems to have more generality than
formulated here. I think that the objects like connection and
covariant derivative exist for a~more general class of Lagrangians
than Riemannian metrics.}.

\ssec{2.2}{Linear constraints and the reduced connection} Now, let
us proceed to the specialization of equations from \ref{s1} for the
case of the Riemannian metric and linear constraints. In mechanics,
one usually considers the linear constraints. This accounts for
linearity of constraints in the majority of nonholonomic applied
problems (rolling, etc.). Moreover, the possibility of realizing
non-linear constraints in mechanical problems has been an open
problem for a~long time, cf. \cite{9}.

Let $Q$ be a~Riemannian manifold with metric $g$, let $\beta$ be
a~codistribution on $Q$, and let $\beta^\perp$ be the distribution
annihilating $\beta$. From here on we make no distinction between
distributions and codistributions since the metric allows us to
identify 1\defis forms with vector fields. However, we determine
every constraint by its annihilator, hence, we denote the
distribution of admissible vectors as $\beta^\perp$. The set of
pairs
\begin{equation}
\label{eq243} T^\beta Q = \{(q, v) \mid \beta_q(v) =0\}
\end{equation}
 is a~subbundle of $TQ$. The subbundle
$T^\beta Q$ determines a~constraint and is described by the system
of equations
\begin{equation}
\label{eq244} \varphi_i(q, v) := \langle a_i(q), v\rangle = 0, \quad
i = 1, ..., m,
\end{equation}
where $a_i(q)$ are 1\defis forms forming a~basis of the space of
sections of $\beta$ at $q$. The forms that correspond to our
constraint are determined by the set of differentials of the
functions $\varphi_i$. We do not need these forms thanks to the
following statement (true for arbitrary manifolds).

\sssbegin{2.2.1}{Statement} Reactions of the linear constraints
defined by a~codistribution $\beta$ on~$Q$ form the codistribution
$(d\pi)^*\beta$ on $TQ$, see sec.~\ref{ss1.2}. The linear
constraints are admissible and ideal ones. \end{Statement}

\begin{proof} For $i=1 , ..., m$, we have
\begin{equation}
\label{eq245} \tau^*(d\varphi_i) =\tau^*\{\partial_{q}a_idq +
a_idv\} = \{a_idq\} = (d\pi)^*a_i.
\end{equation}
The ideal nature of constraints follows from homogeneity in $v$;
admissibility is evident.
\end{proof}

Now, define connections in $T^\beta Q$. By the definition, any
connection in $T^\beta Q$ is an adjoint connection in the principal
fiber bundle $B^\beta Q$ of partial frames, where the fiber over
$q\in Q$ is the space of all frames in the subspace
$\beta^\perp_{q}$, and the structure group is $GL(\rk\
\beta^\perp)$. Since $T^\beta Q$ is a~vector bundle, the connection
can be determined by a~covariant derivative for the fields
compatible with $\beta^\perp$.

First, let us prove the following general lemma (without using the
metric).

\sssbegin{2.2.2}{Lemma} Let $Q$ be an arbitrary manifold with
a~linear connection determined by a~covariant derivative $\nabla$.
Given a~subbundle $T^\beta Q$ in $TQ$, and, for every fiber $R_{q}$
(a subspace of $T_{q}$), a~projection $F_{q}\colon T_{q}\tto R_{q}$
smoothly depending on $q$. Then $$ \widetilde \nabla_X(Y):=F
\nabla_X(Y)\quad\text{ for any $X, Y\in \Gamma(T^\beta Q)$} $$
determines a~connection in the subbundle.\end{Lemma}

\begin{proof} Let us verify that the properties of connection are satisfied for
$\widetilde \nabla$ (see \cite{13}). Of the four properties, only
the following one is nontrivial:
\begin{equation}
\label{eq246}
\renewcommand{\arraystretch}{1.4}
\begin{array}{l}
 \widetilde \nabla_X(\lambda Y)=(X\lambda ) Y+\lambda \tilde
 \nabla_X(Y)\;\text{ for any function }\lambda.
\end{array}
\end{equation}
This property holds since
\begin{equation}
\label{eq247}
\renewcommand{\arraystretch}{1.4}
\begin{array}{l}
F \nabla_X(\lambda Y)=X(\lambda Y))+F\lambda (\nabla_X(Y)=\\
\lambda\widetilde \nabla_X(Y)+X\lambda\cdot FY
\end{array}
\end{equation}
and $FY = Y$ for any $Y\in T^\beta X$. Hence, $\widetilde \nabla$ is
a~connection in $T^\beta X$.\end{proof}

Let $Q$ be a~Riemannian manifold and $\beta$, $\beta^\perp$
distributions on $Q$. At every point $q$, define the orthogonal
projection $F_q\colon T_q\tto \beta^\perp_q$ by means of the
Riemannian metric on $Q$, and the connection $\widetilde \nabla:
=\nabla^\beta$ by Lemma~\ref{ss2.2.2}. The connection $\widetilde
\nabla$ (on $T^\beta Q$) is said to be a~{\it
reduced}\index{connection, reduced} one.

\sssbegin{2.2.3}{Theorem} In local coordinates, the reduced
connection $\widetilde \nabla$ in the subbundle $T^\beta Q$ is
expressed as:
\begin{equation}
\label{eq248}
\nabla^\beta_{X_i}X_j=\mathop{\sum}\limits_{k}\Gamma_{ij}^kX_k,
\quad i,j,k = 1, \dots, m,
\end{equation}
where the $\Gamma_{ij}^k$ are the Christoffel symbols of the
Riemannian connection but $X_i, X_j, X_k$ are the linearly
independent (coordinate) fields from $\beta^\perp$
only.\end{Theorem}

\begin{proof} Since $F$ is the orthogonal projection onto
$\beta^\perp$, eq. (\ref{eq248}) follows from the expression for the
covariant derivative.\end{proof}

\sssbegin{2.2.4}{Remarks} 1) If the distribution $\beta^\perp$ is an
involutive one, then $\nabla^\beta$ is the induced Riemannian
connection on the fibers of the bundle determined by $\nabla^\beta$.

2) The reduced connection can be extended to a~(non-Riemannian)
connection in $TQ$, but {\bf there hardly exists any {\it canonical}
extension}.\footnote{Boldface is mine. Here, Vershik was too
pessimistic, as one can see in \cite{L}. {\it D.L.}}

3) The expression
\begin{equation}
\label{eq249} \nabla^\beta_{X}(Y)- \nabla^\beta_{Y}(X) -[X, Y] = F
([X, Y]) -[X,Y]
\end{equation}
does not vanish if $\beta^\perp$ is not involutive. On the other
hand, the symbol $\Gamma_{ij}^k$ is symmetric. This means that one
can not define torsion by formula (\ref{eq249}). {\bf It well may be
true that the torsion for $\nabla^\beta$ can not be defined at
all}.\footnote{Boldface is mine. Fortunately (this is an important
place), Vershik was too pessimistic here also, as one can see in
\cite{L}. {\it D.L.}}

4) The curvature form is rather sophisticated\footnote{It is not if
defined as in \cite{L}. {\it D.L.}} and is studied insufficiently. A
most interesting problem is to find the holonomy groups for
$\nabla^\beta$.\end{Remarks}

\ssec{2.3}{Inertial motion with quadratic Lagrangian and linear
constraints} Now, apply the results of \ref{s1} and subsec.
\ref{ss2.1}, \ref{ss2.2} to the initial problem.

\sssbegin{2.3.1}{Theorem} Let $Q$ be the position space, $L=
\frac12g$, where $g$ is a~Riemannian metric on $Q$. Let $T^\beta Q
=\{(q, v) \mid \beta_{q}(v) = 0\}$ be the subbundle corresponding to
a~distribution $\beta$. Then the equations of motion for the
dynamical system with the Lagrangian $L$ and linear constraints
$\beta$ are equations of the geodesics for the reduced connection
$\nabla^\beta$.\end{Theorem}

\begin{proof} By Theorem \ref{ss1.2.3} the field  desired exists in $T^\beta Q$
because the constraint is admissible; it is given by the formula
$X_{L} = X+\Pi_L(\omega)$, where $X$ is the vector field for the
system without constraints (geodesic field), and $\omega$ is
a~1\defis form from the codistribution of the constraint reactions.
This means that $\omega$ is a~covector (hence, vector) field on $Q$
lying in $\beta$. Thus, $\Pi_L(\omega)$ is a~vertical field. It is
chosen so that the field $X_L$ be concordant with the constraint,
i.e., the vertical projection of $X_L$ lies in the image of
$\beta^\perp$.

On the other hand, the connection $\nabla^\beta$ can be represented
as
\begin{equation}
\label{eq250} \nabla^\beta= F \nabla= \nabla+ S,
\end{equation}
where $S$ is a~$(1,2)$\defis tensor (the difference between any two
covariant differentiations is a~$(1,2)$\defis tensor), i.e., $S =
F^\perp\nabla$, where $F^\perp$ is the projection on $\beta$. Hence,
any horizontal vector of the connection $\nabla^\beta$ differs from
a~horizontal vector of the connection $\nabla$ in a~vertical vector
lying in $\beta$ and is concordant with the constraint. The
decomposition into vertical and horizontal components is unique, and
hence the special horizontal field of the connection $\nabla^\beta$
at the points $(q, v)\in T^\beta Q$ and the field $X$
coincide.\end{proof}

\sssbegin{2.3.2}{Corollary} The equations of motion for our problem
can be represented as
\begin{equation}
\label{eq251} \nabla^\beta_{\dot X_i}\dot X_j=0
\end{equation}
or
\begin{equation}
\label{eq252} \ddot X^k=\mathop{\sum}\limits_{k}\Gamma_{ij}^k\dot
X^i\dot X^j, \quad i, j, k = 1, ..., m=\dim\beta^\perp,
\end{equation}
where the coordinates are chosen so that $\dot X^1, \dots ,\dot X^m$
form a~basis in the fiber of the distribution $\beta^\perp$. If
there are exterior forces, potential, etc., then one makes a~change
in $(\ref{eq251})$ in the usual way:
\begin{equation}
\label{eq253} \nabla^\beta_{\dot X_i}\dot X_j=\omega_{1},
\end{equation}
where $\omega_{1}$ is the vector field of exterior forces (the
gradient in the potential case). \end{Corollary}

If the distribution $\beta^\perp$ is involutive, then (\ref{eq251})
is the equation of geodesics in the fiber of the bundle, and if the
distribution is geodesic (i.e., its fiber is completely geodesic,
see \cite{13}), then the solution of (\ref{eq251}) coincide with the
solution of the constraint free problem.

\sssbegin{2.3.3}{Remark} As we have said already, for the dynamical
problem, one usually uses the field $X_{L}$ (in our terms: the field
of geodesics) only. Connections were not generally used even for
constraint-free problems. As one can see from above, their roles are
rather essential.
\end{Remark}

\ssec{2.4}{Non-holonomicity and the Hopf-Rinow
theorem}\index{Hopf-Rinow theorem}\index{theorem, Hopf-Rinow} Of the
most interest is the case when the involutive hall of the
distribution $\beta^\perp$ generates the whole bundle. By the
Frobenius theorem this happens when the brackets of the vector
fields from $\beta^\perp$ generate the whole Lie algebra of vector
fields on $Q$.

\sssbegin{2.4.1}{Problem} Describe the distributions of dimension
$m$ in general position and involutivity of these distributions.
(One can suppose that, for $m>1$, the distribution generates the
whole $T_q$ for generic points $q\in Q$, and the dimension of the
involutive hull diminishes at singular points.)
\end{Problem}

\sssbegin{2.4.2}{Statement}{\em(The Hopf-Rinow theorem, cf.
\cite{13}, Ch. 3, \S 6, and \cite{JJ})}\label{hrth}\index{theorem,
Hopf-Rinow} The connection $\beta^\perp$ on a~complete Riemannian
manifold is complete.\end{Statement}

\sssbegin{2.4.3}{Corollary} If the distribution $\beta^\perp$ is
nonholonomic, then, for any two points $q_{1}, q_2\in Q$, there
exists a~continuous curve connecting these points and consisting of
pieces of the geodesics of the connection $\nabla^\beta$ . In other
words, the geodesic flow is transitive.\end{Corollary}

At the same time not every two points can be immediately connected
by a~$\nabla^\beta$-geodesic.\footnote{For the most lucid example,
see \cite{Poi}. {\it D.L.}} The following interesting problem
arises.

On a~Riemannian manifold $Q$ with a~nonholonomic distribution
$\beta^\perp$, define the following new metric:
\begin{equation}
\label{eq254} d_\beta(q_{1}, q_2)=\inf\{l(\tau)\mid \tau(0)=q_{1},
\;\tau(1)=q_{2},\; \dot\tau\in\beta^\perp\},
\end{equation}
where $\tau$ is a~smooth curve, $l(\cdot)$ the length with respect
to the initial Riemannian metric. This infimum is not attained on
the geodesics of the connection $\nabla^\beta$, and hence the
classical {\it Hopf-Rinow theorem} is not valid in this case.

\sssbegin{2.4.4}{Problem} Describe intrinsically the $d_\beta$-type
metrics on Riemannian manifolds. (Compare with the ``space of
geodesics'' defined by Busemann \cite{14}.)\end{Problem}

An example of this kind was considered by Gershkovich \cite{21}.

\section{Group mechanics with constraints}\label{s3}

\ssec{3.1}{Problem of rolling and its group model} The most popular
example of a~nonholonomic problem is the following one: two bodies
(e.g., a~ball and a~plane) move (from force of inertia or in
a~field) so that the linear velocities of both bodies at the point
of contact coincide (no sliding). The reader interested in the
traditional technique of derivation of the constraint equations is
referred to special literature. Here we describe (apparently for the
first time) the group model of this problem in a reasonably general
formulation. The group example is a~central one here, as in the case
of ordinary dynamics.

Let $G_{1}$ and $G_2$ be two arbitrary connected Lie groups
determining position of each body in a~moving coordinate system,
$G_1^o$ and $G_2^o$ their stationary subgroups (provided the point
of contact is fixed).  The coincidence of velocities at the point of
contact is expressed as an isomorphism $s\colon \fg_1/\fg_1^o\simeq
\fg_2/\fg_2^o$, where $\fg_i$ is the Lie algebra of $G_i$. Now
consider $Q= G_1\times G_2$ as the position space and, in $(TQ)_e =
\fg_1\oplus\fg_2$, single out the subspace $\fM$ spanned by
\begin{equation}
\label{eq255} (a_1, a_2) \text{ such that }s\circ t_1(a_1) =
t_2(a_2),
\end{equation}
where $t_i \colon\fg_i \tto \fg_i^o$ are the canonical projections.
Clearly, $\fM$ is a~linear subspace of $\fg_1\oplus\fg_2$. By
transferring this subspace to all points of $G= G_1\times G_2$ by
means of left translations we obtain a~distribution (linear elements
with coinciding velocities at the points of contact). This is the
principal model.

Now one can consider a~Lagrangian (left-invariant for inertial
motions) and apply the methods  developed in \ref{s2} to form the
equations of motion. For a~more general scheme, one does not need to
specialize the group as the direct product of two (or more) groups.
The most natural form of our scheme is the following one.

Let $G$ be an arbitrary connected Lie group, $\fM$ a~subspace in the
Lie algebra $\fg$ (usually, a~complement to a~subalgebra). The
distribution $\beta^\perp$ is the distribution of the left
translates of $\fM$ in $TG$. Such a~model will be called a~{\it
group} one.

\sssbegin{3.1.1}{Example} 1) $G = SO(3)$, $\fg = \fo (3)$. This is
the rotation of a~solid body with a~fixed point and a~zero linear
velocity at it.

2) $G = SO (3)\times \Ree^2$, $\fg = \fo (3) \oplus \Ree^2$,
$\fM=\{(a, h) \mid \pi(a)=h\}\subset \fg$, where $\pi\colon \fo(3)
\tto \fo(3)/\fo(2)\simeq\Ree^2$. This is a~description of a~rolling
ball on a~rough plane. \end{Example}

\ssec{3.2}{Reduction}  For the case of a~left-invariant Lagrangian,
we can reduce the proposed group model to a~system on the Lie
algebra. Observe that the problems with constraints are not
symplectic, i.e., the  fields that appear do not preserve,
generally, any 2\defis form, hence, the question about integrals and
reduction for these systems should be considered separately.
Reduction of the order of these systems is less than that of
symplectic systems. The following statement refers to a~general
system with constraints.

\sssbegin{3.2.1}{Statement}{\em (See \cite{2}.)} If the constraint
$\alpha$ is an ideal one, then the energy $H_L$ is a~motion integral
for system with the Lagrangian $L$.
\end{Statement}

\begin{proof}
\begin{equation}
\label{eq256}
\renewcommand{\arraystretch}{1.4}
\begin{array}{l}
 \frac{d}{dt}H_L = X(dH_L) =\Omega_L(\Pi_L(dH_L), X) =\\
\Omega_L(\Pi_L(dH_L), \Pi_L(dH_L))
+\Omega_L(\Pi_L(dH_L),\Pi_L(\omega))=\\
\omega(\Pi_L(dH_L)=\omega(X_L)=\tau^*\rho(X_L)=\\
\rho\tau(X_L)=\rho (\Phi)=0.\qquad \qedhere
\end{array}
\end{equation}
\end{proof}

\sssbegin{3.2.2}{Theorem} Let the position space be a~Lie group $G$,
and the (linear) constraints  and the (quadratic) Lagrangian of the
system are left-invariant. Then the phase flow of the system is the
fiber bundle whose base is the set of flows whose vector fields are
the projections of the Euler fields (without constraints) on the
subspace of constraints in the Lie coalgebra, and the fiber is the
set of conditionally periodic flows on the group. \end{Theorem}

\begin{proof} We identify $TG$ and $T^*G$ by means of the Lagrangian
(or,  for semi\Defis simple groups, by means of the Killing form).
The field $X$ that describes the motion of the system commutes with
the left-invariant fields $A$ on $G$. Indeed, by hypothesis we have
\begin{equation}
\label{eq257} {}[X,A] = [X_L, A] + [X, \Pi_L(\omega)] =0.
\end{equation}
Hence, the partition into orbits of the natural (Hamiltonian) action
of the group $G$ in $T^*G$ is invariant under the flow of the field
$X$. Thus, the base of the fiber bundle is $T^\beta
G/G\simeq\fM\subset \fg^*$, where $\fM$ is the subspace determined
by the constraints and the fiber is the group~$G$. The flow on the
fiber is determined by the motion on orbits, i.e., on the group, and
commutes with the left translations, and hence it reduces to the
flow on left cosets with respect to a~maximal torus, the latter flow
being conditionally periodic on each of these cosets. The
trajectories are determined by their initial vectors at a~certain
point, e.g., at the unity element of the group.

Proceed now to the base. Recall that for the constraint-free systems
one can define the moment map $dH \colon T^*G\tto \fg$, see, e.g.,
\cite{4}. On $\fg^*$, we have the Euler equation:
\begin{equation}
\label{eq258} \dot a~= \{dH(a), a\},
\end{equation}
where $a\in\fg^*$ and $\{\cdot, \cdot\}$ is the Poisson bracket. For
the constraint-free system, the Euler equation determines the motion
on the base.

By Theorem \ref{ss3.2.2}, in our case the vector field defined on
the subspace of constraints $\fM$ is the orthogonal projection of
the field corresponding to the Euler equation on
$\fM\subset\fg=\fg^*$. Thus, the equation on $\fM$ takes the form
\begin{equation}
\label{eq259} \dot a~= P\{dH(a), a\},
\end{equation}
where $P$ is the orthogonal (with respect to the Lagrangian)
projection of $\fg$ onto $\fM$. \end{proof}

\sssbegin{3.2.3}{Remark} 1) Actually, we asserted that the
projection of the initial field onto constraints commutes with the
group factorization. But, unlike the generalized Noether theorem,
here one can not assert that the elements from the center of the
enveloping algebra are integrals of motion.\footnote{The reduction
of Hamiltonian systems under the regular actions of Lie groups
(generalizations of the Noether theorem) was considered and
rediscovered by many researchers. For Lie groups, I defined it as
early as 1968 (the reports were made at seminars in Leningrad and
Moscow State Universities and at a~conference in Tsakhkadzor in
1969). The well-known 2\defis form discovered by A.~A.~Kirillov in
1962 (\cite{4}) naturally appears from the canonical form in $T^*G$
in the process of reduction. A list of literature on this topic (by
no means complete) is given in the book \cite{4} and in the surveys
\cite{20}, the latter being devoted to the symmetries of more
complicated nature.} Hence, the system obtained on $\fM$ does not
preserve, generally, the  orbits of the coadjoint action.

2) If the Lagrangian is not quadratic, the reduction is also
possible, but in this case the Legendre
transformation\index{Legendre transformation} on the fiber of $TG$
is non-linear, the quotient space $T^*G/G$ has no linear structure,
and the reduced system is rather sophisticated.
\end{Remark}

\sssbegin{3.2.4}{Example} Consider the problem of rolling by inertia
of two $n$-dimensional solid bodies. In this case, we have (in
notations of sec. \ref{ss3.1}):
\begin{equation}
\label{eq260}
\renewcommand{\arraystretch}{1.4}
\begin{array}{l}
 G_{1}= G_2 = SO(n),\quad G_1^o = G_2^o = SO(n-1),\quad G=
 G_1\times G_2;\\
\fg =\fg_1\oplus \fg_2 = \fo(n) \oplus\fo(n), \quad \fg ^o:=\fg_1^o
= \fg_2^o =
\fo(n -1),\\
\fM= \diag ~\fg^o \oplus(\fg_1^o\oplus\fg_2^o),\\
\text{where $\diag ~\fg^o = \{(a, a) \mid a\in \fg^o\}\subset
\fg_1\oplus \fg_2$.} \end{array}
\end{equation}
Hence,
\begin{equation}
\label{eq261} \fM=\{(a+ b, a+ c)\mid a, b, c\in \fo(n-1)\}.
\end{equation}
Let $P$ be the projection onto $\fM$. (In our case, the orthogonal
projection $P$ does not depend on the inertia tensors.) Then, for
$i, j= 1, ..., n- 1$, we have
\begin{equation}
\label{eq262} P(X_{ij}^1, X_{ij}^2)=\begin{pmatrix}
X_{ij}^1+X_{ij}^2& X_{ij}^1\cr -X_{ij}^1&0 \end{pmatrix} +
\begin{pmatrix} X_{ij}^1+X_{ij}^2& X_{ij}^2\cr -X_{ij}^2&0
\end{pmatrix}.
\end{equation}
Let $L_1, L_2$ be the inertia tensors of the bodies. Then the
equations of motion in $\fM$ have the form
\begin{equation}
\label{eq263}
\renewcommand{\arraystretch}{1.4}
\begin{array}{l}
 \dot a^{(1)} = P((a^{(1)})^2I_1 -I_1(a^{(1)})^2),\\
\dot a^{(2)} = P((a^{(2)})^2I_2 -I_2(a^{(2)})^2),
\end{array}
\end{equation}
for any $(a^{(1)}, a^{(1)})\in\beta$. The projection $P$ connects
both equations.

I do not know to qualitatively describe the  motion for this system.
\end{Example}

In the same manner one can take another stationary subgroup (e.g.,
$SO(n-k)$ that corresponds to the Stifel manifold; however, I do not
know whether the constraint-free problem is integrable or not).

\section{Non-classical problems with constraints}\label{s4}

The following problems arise in optimal control and economics
dynamics.

\ssbegin{4.1}{Problem} Let $Q$ be a~manifold, $TQ$ its tangent
bundle, and we are given a~field $B$ of submanifolds $B(q)$ (with
boundary or corners) in fibers $T_q$ of $TQ$. We have to minimize
a~certain functional of the boundary value problem, the tangent
vectors to admissible curves lying in the given submanifold:
\begin{equation}
\label{eq264} \inf \{F(q(\cdot)) \mid \dot q \in B(q) \subset T_q;
\; q(a) = \bar q, \ q(b) =\bar{\bar q}\} .
\end{equation}
The problem in this formulation is said to be {\it problem in
contingencies}\index{problem in contingencies} (differential
inclusions).  \end{Problem}

One passes to this formulation from the traditional one in the
following way: $B(q)$ is the set of right-hand sides of differential
equations when controls run over the admissible region.

Usually, $B(q)$ are convex solids, in particular, polyhedral sets
(cones or polytopes). In connection with this it is important to
consider the theory of fields of such sets as a~generalization of
the theory of distributions (involutivity, singularities, etc.). For
another motivation (connected with other problems), see \cite{6,21}.

\ssbegin{4.2}{Problem} Consider the following principle and apply it
to these problems.\end{Problem}
\begin{center}
\begin{minipage}[l]{.8\textwidth}
{\sl One can construct the field corresponding to the Euler equation
for the constraint\Defis free problem and then project this field
onto the constrain manifold as we did it above.}
\end{minipage}
\end{center}

\vskip .2cm

In the interior points, the field does not change under this
projection, but jumps (switches) can appear on the boundary.
Constraints of this kind are not ideal in general.

\ssbegin{4.3}{Problem}  Find out problems of optimal control for
which the field desired is obtained by the above principle.
\end{Problem}

Mostly, the field of sets $B(\cdot) \subset TQ$ is defined by means
of inequalities. For example, let $B(\cdot)$ be the field of
polyhedral sets $B(q) =\{v\mid A(q)v \leq b(q)\}$, where $b(\cdot)$
is a~map from $Q$ to $\Ree^m$ and $A(\cdot) \in \Hom(\Ree^n,
\Ree^m)$. In this case, we have the following general problem:
\begin{equation}
\label{eq265} \inf \{\Phi(q(\cdot)) \mid A(q)\dot q\leq b(q), \; q
\in Q\}.
\end{equation}
Thus, this is a~field of problems of linear (if $\Phi$ is linear) or
convex programming in the tangent bundle $TQ$. Similar problems were
considered in the linear theory of optimal control and also (for $Q
= \Ree^1$ or $Q=[a, b]\subset \Ree^1$) in the theory of continuous
economic models.

\ssbegin{4.4}{Problem} For the above problems, relate Pontriagin's
optimality principle with the duality theorem of convex programming.
Conjecturally, these two statements are equivalent, as one can
easily verify for the case when $A(\cdot), b(\cdot)$ are constant on
$Q = \Ree^n$.  \end{Problem}

\ssbegin{4.5}{Problem}  A problem close to the classical ones from
\ref{s2} and \ref{s3} consists in finding a~minimum in the following
situation. Let $Q$ be a~Riemannian (or Finsler) manifold, $B(\cdot)$
a~field of convex sets (polytopes, spheres, etc.) in $TQ$. It is
convenient to suppose that $B(q) $ is central-symmetric for all $q$.
One should find the shortest (in the sense of metric) curves that
are admissible for the restrictions $q \in B(q)$. \end{Problem}

The obtained new metric (non-Riemannian) seems to be similar to the
metric from~\ref{s2}. Manifolds with these metrics possess, perhaps,
unusual properties. (I did not see this formulation of the problem
in the literature, but, possibly, similar questions were stated and
solved. There is a~considerable number of works that indirectly
refer to the question discussed. We confine ourselves to the
reference to the monograph \cite{11}.)

Study metrics on smooth manifolds appearing in these problems, see
sec.~\ref{ss2.4}, cf. \cite{21}.

\ssbegin{4.6}{Problem}   In non-classical dynamics (as well as in
arbitrary dynamics), the most interesting example of a~problem with
constraints is the group one. Let a~Lie group act transitively on a
manifold $M$, and let the field of constraints be invariant under
this action. If the functional is invariant too, then one can,
conjecturally,  reduce the problem (in the same manner as in
\ref{s3}) to the Lie algebra if $M$ is a~group.
\end{Problem}

The difficulty appearing here is that the reduction permits one to
study only the Cauchy problem rather than the boundary value problem
(same as in \ref{s3}). This circumstance is not substantial for
classical mechanics without constraints, because usually (e.g., for
Riemannian complete manifolds) the set of solutions of the Cauchy
problem includes solutions of all boundary value problems (the
Hopf-Rinow theorem). Here this is not the case and one should search
for a~solution considering not only the result of reduction but the
whole fiber bundle.

\ssbegin{4.7}{Problem}  Describe the reduction for group problems of
optimal control and the structure of the corresponding fiber bundle.
\end{Problem}

\begin{thebibl}{9999}


\bibitem[A]{4}
Arnold V., {\em Mathematical Methods of Classical Mechanics},
Graduate Texts in Mathematics, 60. Springer, New York, 1997. xvi+516
pp.

\bibitem[BC]{18}
Bishop R.L., Crittenden R.J., {\em Geometry of Manifolds}, Academic
Press, New York. 1964.

\bibitem[B]{14} Busemann H. {\em
The Geometry of Geodesics}, Academic Press, New York, 1955.

\bibitem[Ch]{7}
Chaplygin S.A., {\em Studies on Dynamics of Non-Holonomic Systems},
Moscow, 1949 (in Russian); [Selected works. Gas and fluid mechanics.
Mathematics. General mechanics] Annotations and commentaries by S.
A. Hristianovi\v c, L. V. Kantorovi\v c, Ju. I. Ne\u\i mark and N.
A. Fufaev. Biographical sketch by M. V. Keldy\v s. Documentary
biographical chronology, and complete bibliography of the works of
S. A. \v Caplygin by N. M. Semenova. Nauka, Moscow, 1976. 495 pp.

\bibitem[D]{8}
Dobronravov V. V., {\em Foundations of Mechanics of Non-Holonomic
Systems}, Moscow, 1970 (in Russian).


\bibitem[Ger]{21}
Gershkovich V., Two-sided estimates of metrics generated by
absolutely nonholonomic distributions on Riemannian manifolds.
(Russian) Dokl. Akad. Nauk SSSR 278 (1984), no. 5, 1040--1044.
English translation: Soviet Math. Dokl. 30 (1984), no. 2, 506--510

\bibitem[Gl]{23}
Gliklikh Yu., Riemanian parallel translation in nonlinear mechanics.
In: Yu. Borisovich, Yu. Gliklikh (eds.), {\em Global Analysis~---
Studies and Applications. I.}, LN in Mathematics, Springer, 1108,
1984, 128--151

\bibitem[G]{3}
Godbillon C., {\em G\'eometrie differ\'entielle et m\'echanique
analytique}, Hermann, Paris, 1969.

\bibitem[G]{10}
Gohman A.V. {\em Differential-geometric foundations of the classical
dynamics of systems} Izdat. Saratov. Univ., Saratov, 1969. 93 pp.
(in Russian).

\bibitem[JJ]{JJ}
Jost J., {\em Riemannian Geometry and Geometric Analysis}, Springer,
Berlin, 2002, XIII+455 pp.

\bibitem[KN]{13}
Kobayashi S., Nomizu K. {\em Foundations of Differential Geometry},
vol. 1, Interscience, New York, 1963.

\bibitem[L]{L}
Leites D., The Riemann tensor for nonholonomic manifolds. Homology,
Homotopy and Applications, vol 4 (2), 2002, 397--407;
\texttt{arXiv:math.RT/0202213};

Grozman P., Leites D., The nonholonomic Riemann and Weyl tensors for
flag manifolds; Theor.Mathem.Phys, 2007, 153:2, 186–-219
\texttt{arXiv:math.DG/0509399}



\bibitem[NF]{9}
Neimark Yu., Fufaev N.A., {\em Dynamics of Non-Holonomic Systems},
Nauka, Moscow, 1967 (in Russian).

\bibitem[Nov]{12}
Novikov S., Periodic solutions of Kirchhoff equations for the free
motion of a~rigid body in a~fluid and the extended
Lyusternik-Shnirel\cprime man-Morse theory. I. (Russian)
Funktsional. Anal. i Prilozhen. 15 (1981), no. 3, 54--66; id.,
Variational methods and periodic solutions of equations of Kirchhoff
type. II. (Russian) Funktsional. Anal. i Prilozhen. 15 (1981), no.
4, 37--52, 96. English translation: part I, Functional Anal. Appl.
15 (1981), no. 3, 197--207 (1982); part II, ibid. 15 (1981), no. 4,
263--274 (1982)

\bibitem[P]{20}
Perelomov A., Integrable systems of classical mechanic and Lie
algebras. Systems with constraints. ITEP-116, preprint, Moscow, 1983

Olshanetsky, M. A.; Perelomov, A. M.; Reyman, A. G.;
Semenov-Tyan-Shanski\u\i, M. A. Integrable systems. II. (Russian)
Current problems in mathematics. Fundamental directions, Vol. 16
(Russian), 86--226, 307, Itogi Nauki i Tekhniki, Akad. Nauk SSSR,
Vsesoyuz. Inst. Nauchn. i Tekhn. Inform., Moscow, 1987.

Perelomov, A. M. {\em Integrable systems of classical mechanics and
Lie algebras}. Vol. I. Birkh\"auser, Basel, 1990. x+307 pp.

\bibitem[Poi]{Poi}
Poincar\'e, H., Les id\'ees de Hertz sur la M\'ecanique (Revue
g\'en\'erale des Sciences, t. 8, 1897, 734--43. In: {\em \OE uvres}.
Tome VII. (French) [Works. Vol. VII] Masses fluides en rotation.
Principes de m\'ecanique analytique. Probl\`eme des trois corps.
[Rotating fluid masses. Principles of analytic mechanics. Three-body
problem] With a~preface by Jacques Levy. Reprint of the 1952
edition. Les Grands Classiques Gauthier-Villars. [Gauthier-Villars
Great Classics] \'Editions Jacques Gabay, Sceaux, 1996. viii+635 pp.

\bibitem[Sm]{24}
Smirnov V., {\em A course of higher mathematics}, vol. IV, p. 1,
Moscow, 1974 (in Russian).

\bibitem[SS]{19}
Stenberg S., {\em Lectures on differential geometry}. Second
edition. Chelsea Publishing Co., New York, 1983. xviii+442 pp.

\bibitem[T-K]{11}
Ter-Krikorov A.M., {\em Optimal Control and Mathematical Economics},
Nauka, Moscow, 1977, 216 pp. (in Russian).

\bibitem[V]{5}
Vershik A.M., Several remarks on infinite-dimensional problems of
linear programming. Usp. Mat. Nauk 25:5 (1970), 117--124.

\bibitem[VCh]{6}
Vershik, A. M.; Chernyakov, A. G. Fields of convex polyhedra and
Pareto-Smale optimum. (Russian) Optimizatsiya No. 28(45) (1982),
112--145, 149. English translation in: Leifman L. (ed.) {\em
Functional analysis, optimization, and mathematical economics. A
collection of papers dedicated to the memory of Leonid Vitalevich
Kantorovich}, The Clarendon Press, Oxford University Press, New
York, 1990, 290--313

\bibitem[VCh]{22}
Vershik A., Chernyakov A., Critical points of fields of convex
polyhedra and the Pareto-Smale optimum relative to a~convex cone.
(Russian) Dokl. Akad. Nauk SSSR 266 (1982), no. 3, 529--532. English
translation: Soviet Math. Dokl. 26 (1982), no. 2, 353--356

\bibitem[VF1]{1}
Vershik A.M., Faddeev L.D. Differential geometry and Lagrangian
mechanics with constraints. Dokl. Akad. Nauk SSSR 202:3 (1972),
555--557 = Soviet Physics Doklady 17:1 (1972), 34--36.

\bibitem[VF2]{2}
Vershik A.M., Faddeev L.D., Lagrangian mechanics in an invariant
form. In: Problems of Theoretical Physics. vol. 2, Leningrad, 1975 =
Selecta Math. Sov. 1:4 (1981), 339--350.

\end{thebibl}

}

\printindex

\end{document}